\documentclass[aps,prb,12pt,onecolumn,superscriptaddress,nofootinbib,floatfix,amsmath,amssymb]{revtex4-2}

\usepackage{ragged2e}
\usepackage[utf8]{inputenc}
\usepackage{amsmath}
\usepackage{slashed}
\usepackage{graphicx}
\usepackage[caption=false]{subfig}
\usepackage{xcolor}
\usepackage[pdftex]{hyperref}
 
\begin{document}

\title{Analogue simulations of quantum gravity with fluids}
\author{Samuel L. Braunstein}
\email{sam.braunstein@york.ac.uk}
\affiliation{\scriptsize{Computer Science, University of York, York Y010 5GH, United Kingdom.}}
\author{Mir Faizal}
\email{mirfaizalmir@googlemail.com}
\affiliation{\scriptsize{Irving K. Barber School of Arts and Sciences,
University of British Columbia - Okanagan, Kelowna, British Columbia V1V 1V7, Canada}}
\affiliation{\scriptsize{Canadian Quantum Research Center 204-3002  32 Ave Vernon, BC V1T 2L7 Canada}}
\author{Lawrence M. Krauss}
\email{lawrence@originsproject.org}
\affiliation{\scriptsize{The Origins Project Foundation, Phoenix AZ 85020 USA}}

	\author{Francesco Marino}
	\email{francesco.marino@ino.cnr.it}
		\affiliation{\scriptsize{CNR-Istituto Nazionale di Ottica, Via Sansone 1, I-50019 Sesto Fiorentino (FI), Italy.}}
\affiliation{\scriptsize{INFN, Sezione di Firenze, Via Sansone 1, I-50019 Sesto Fiorentino (FI), Italy.}}

	\author{Naveed A. Shah}
	\email{naveed179755@st.jmi.ac.in}
\affiliation{\scriptsize{Department of Physics, Jamia Millia Islamia, New Delhi - 110025, India} }

\begin{abstract}
The recent technological advances in controlling and manipulating fluids have enabled the experimental realization of acoustic analogues of gravitational black holes. A flowing fluid provides an effective curved spacetime on which sound waves can propagate, allowing the simulation of gravitational geometries and related phenomena. The last decade has witnessed a variety of hydrodynamic experiments testing disparate aspects of black hole physics culminating in the recent experimental evidence of Hawking radiation and Penrose superradiance. In this Perspective, we discuss the potential use of analogue hydrodynamic systems beyond classical general relativity towards the exploration of quantum gravitational effects. These include possible insights into the information-loss paradox, black hole physics with Planck-scale quantum corrections, emergent gravity scenarios and the regularization of curvature singularities. We aim at bridging the gap between the non-overlapping communities of experimentalists working with classical and quantum fluids and quantum-gravity theorists, illustrating the opportunities made possible by the latest experimental and theoretical developments in these important areas of research.

\end{abstract}

\maketitle

\section{Curved spacetime in hydrodynamics}

In 1974, applying quantum mechanics to curved space, Hawking predicted that a form of thermal radiation should be emitted at a black hole's horizon that would eventually cause its complete evaporation \cite{hawking1,hawking2}. Previously, only rotating black holes were known to release energy. Classically, Penrose described a detailed mechanism to mine a black hole's rotational energy \cite{penrose} and Misner showed that this could amplify incident waves through a phenomenon known as superradiance \cite{misner}. Quantum mechanically, the same processes would trigger the spontaneous emission of quantum particles \cite{zeldovich,staro1,staro2}. Nevertheless, Hawking's result was a complete surprise as it demonstrated for any kind of black hole, rotating or not, that the radiation would be purely thermal with a temperature depending only on the black hole's mass, angular momentum and charge. All these processes rely on negative energy being carried into the black hole. For Hawking's process, each particle of the thermal radiation emitted is entangled with a negative-energy partner that falls into the horizon. A related phenomenon was predicted by Unruh, who showed that uniformly accelerating observers would also see thermal radiation, with a temperature proportional to their acceleration \cite{PhysRevD.14.870}.
 
The difficulty of directly observing these phenomena for astrophysical black holes has motivated the search for alternative systems that may allow one to investigate similar physics in a controlled laboratory environment. One particularly successful approach involves analogue simulators of gravity. The dawn of analogue gravity dates back to Unruh’s seminal paper in 1981 \cite{unruh}. The general idea is that, under appropriate conditions, collective excitations in condensed-matter systems evolve as fields on a curved spacetime, where this analogue spacetime is itself induced by the medium in which these fields propagate \cite{rev,barcelorev,facciorev}. The paradigmatic example is provided by sound waves in an inhomogeneous flowing fluid \cite{white,unruh,visser1,visser2} (see also the recent \cite{almeida2022analogue}). Although the waves always travel at the speed of sound relative to the flow, in the laboratory frame their velocity becomes a function of time and spatial coordinates. As a result, the waves experience an effective curved spacetime the geometry of which is specified by the acoustic metric (see Box 1).  

\begin{figure*}
\hrule
\vspace{2mm}
\section*{{Box 1: Sound cones and sonic horizons}}
\begin{minipage}{0.44\textwidth}
\linespread{1.0}\selectfont
\justify
{Light cones describe how light travels away from an event in a diagram showing both space (horizontal axes) and time (vertical axis). Only events within the cone's volume can be affected by the event occurred at $P$. In a curved spacetime the light cones are tilted towards the mass generating the curvature. 
Inside a black hole, the cones are tilted past the vertical, implying that nothing can escape from that region without traveling faster than light. The border of such a region is known as an event horizon. Sound cones similarly describe how sound propagates from a sonic event and are tilted by the underlying fluid flow.}

\end{minipage}
\begin{minipage}{0.55\textwidth}
\begin{flushright}
\includegraphics[width = 1.0\textwidth]{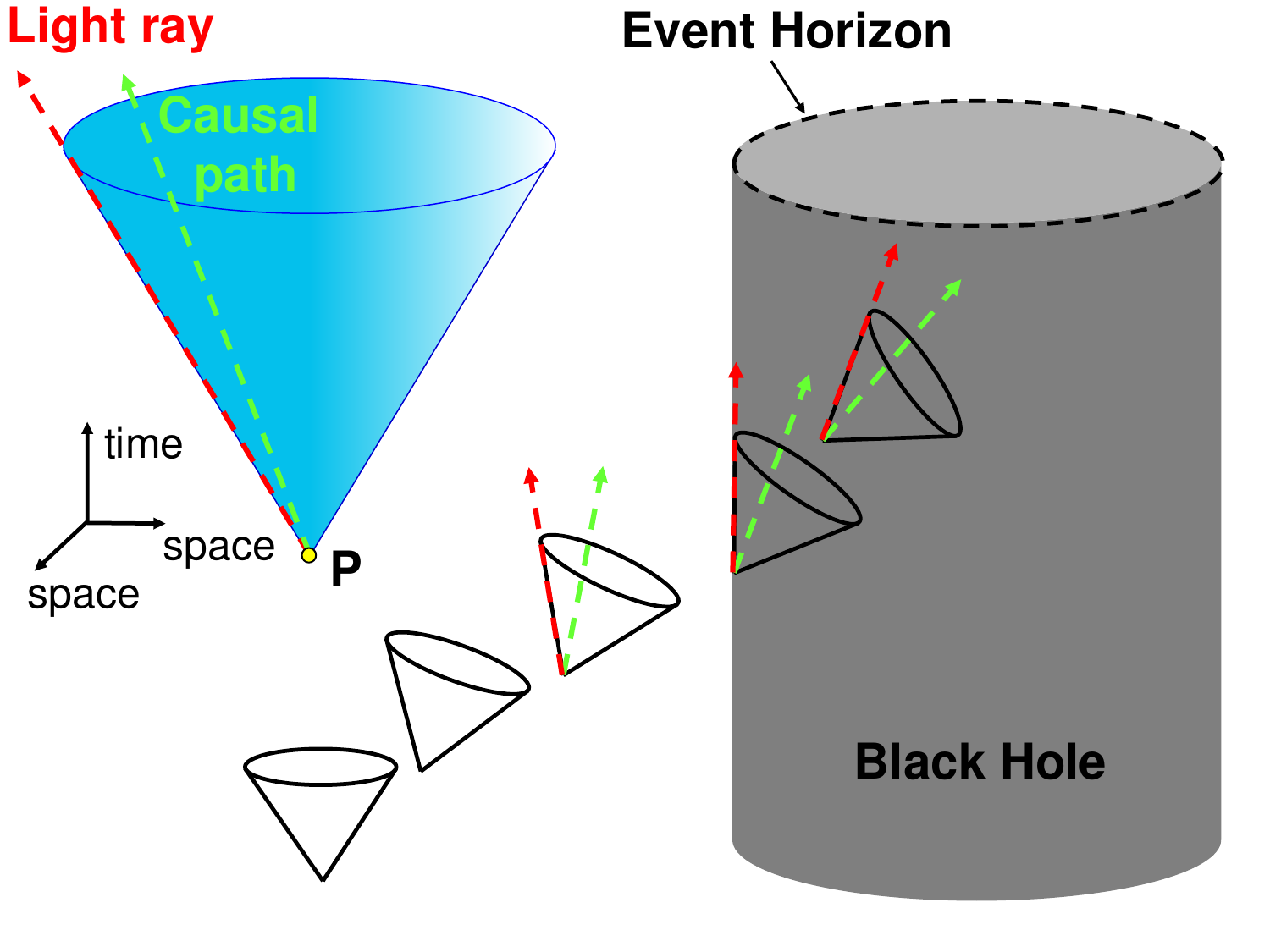}
\end{flushright}
\end{minipage}

\vspace{2mm}

\noindent
\begin{minipage}{0.5\textwidth}
\begin{flushleft}
\includegraphics[width = 1.0\textwidth]{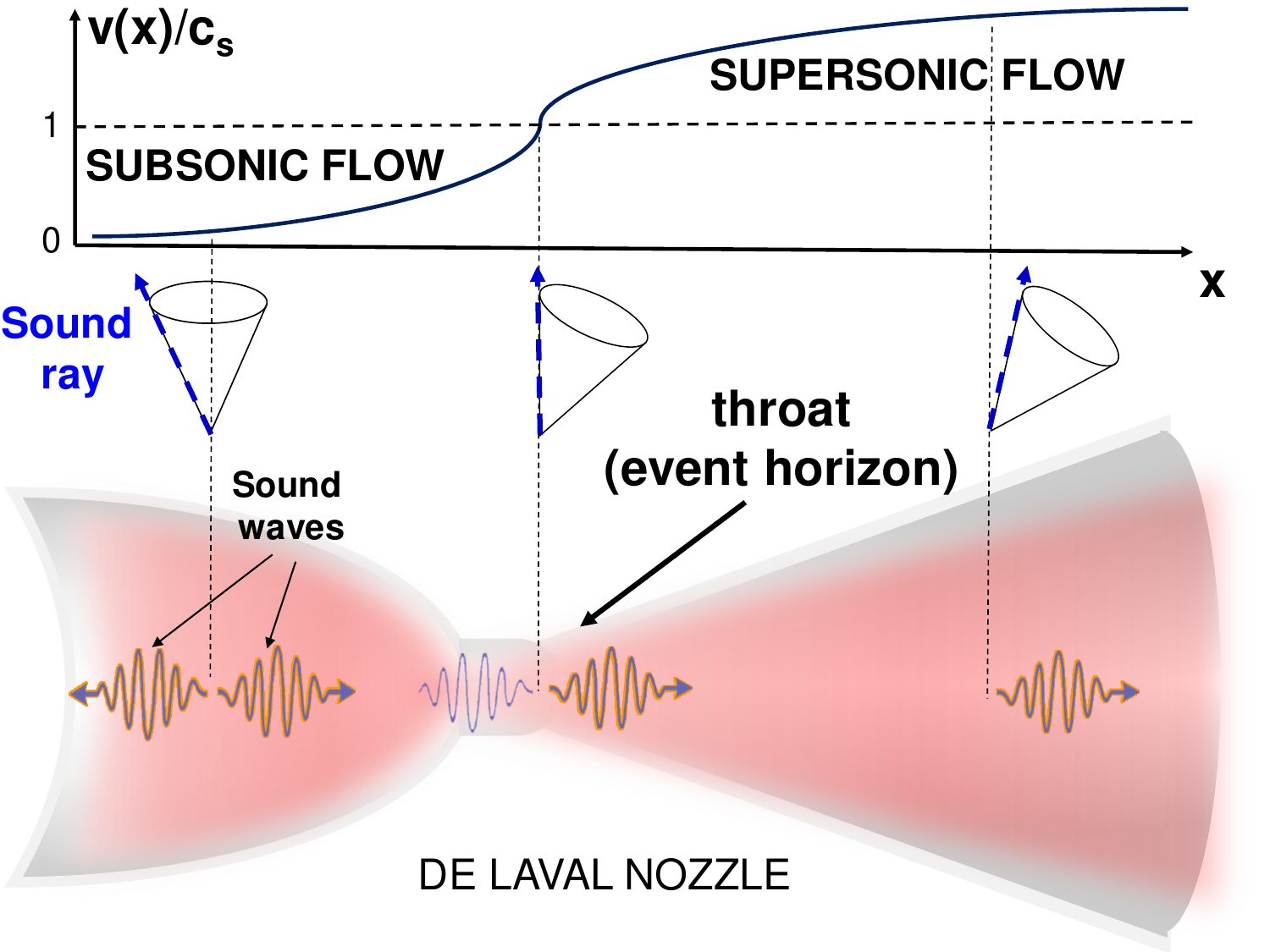}
\end{flushleft}
\end{minipage}
\begin{minipage}{0.49\textwidth}
\linespread{1.0}\selectfont
\justify
{The sound cones are governed by the quadratic equation $-c_s^2 dt^2 + (d{\bf x} - {\bf v}dt)^2 = 0$, 
where ${\bf v}$ is the fluid velocity and $c_s$ is the local speed of sound. If the flow is inhomogeneous the sound cones' tilts depend on their location in the flow, thus simulating an effective curved spacetime. The geometry is then specified by a Lorentzian metric tensor, the acoustic metric, which has the general form
\begin{equation}
g_{\mu\nu} = \left(\begin{array}{cc}
  -c_s^2 + v^2  &  -{\bf v^T} \\
  -{\bf v}  &  {\bf I} \\
\end{array}
\right)
\label{metric}
\end{equation}
where $v^2= \vert {\bf v} \vert^2$, ${\bf v^T}$ is the transpose of ${\bf v}$, and ${\bf I}$ indicates the identity matrix \cite{rev}.}
\end{minipage}

\vspace{2mm}

\begin{minipage}{0.49\textwidth}
\linespread{1.0}\selectfont
\justify
{In the region where the flow goes supersonic the sound cones are tilted past the vertical. The boundary of this region is therefore the sonic analogue of an event horizon. Rocket engine nozzles are systems where sonic horizons form (see above). In the subsonic region sound waves can propagate in all directions. At the throat the flow becomes supersonic and sound waves are dragged inward.
Similarly, horizons for surface waves can form close to waterfalls, where the flow exceeds the wave's velocity (see right).}

\end{minipage}
\begin{minipage}{0.5\textwidth}
\begin{flushright}
\includegraphics[width = 1.0\textwidth]{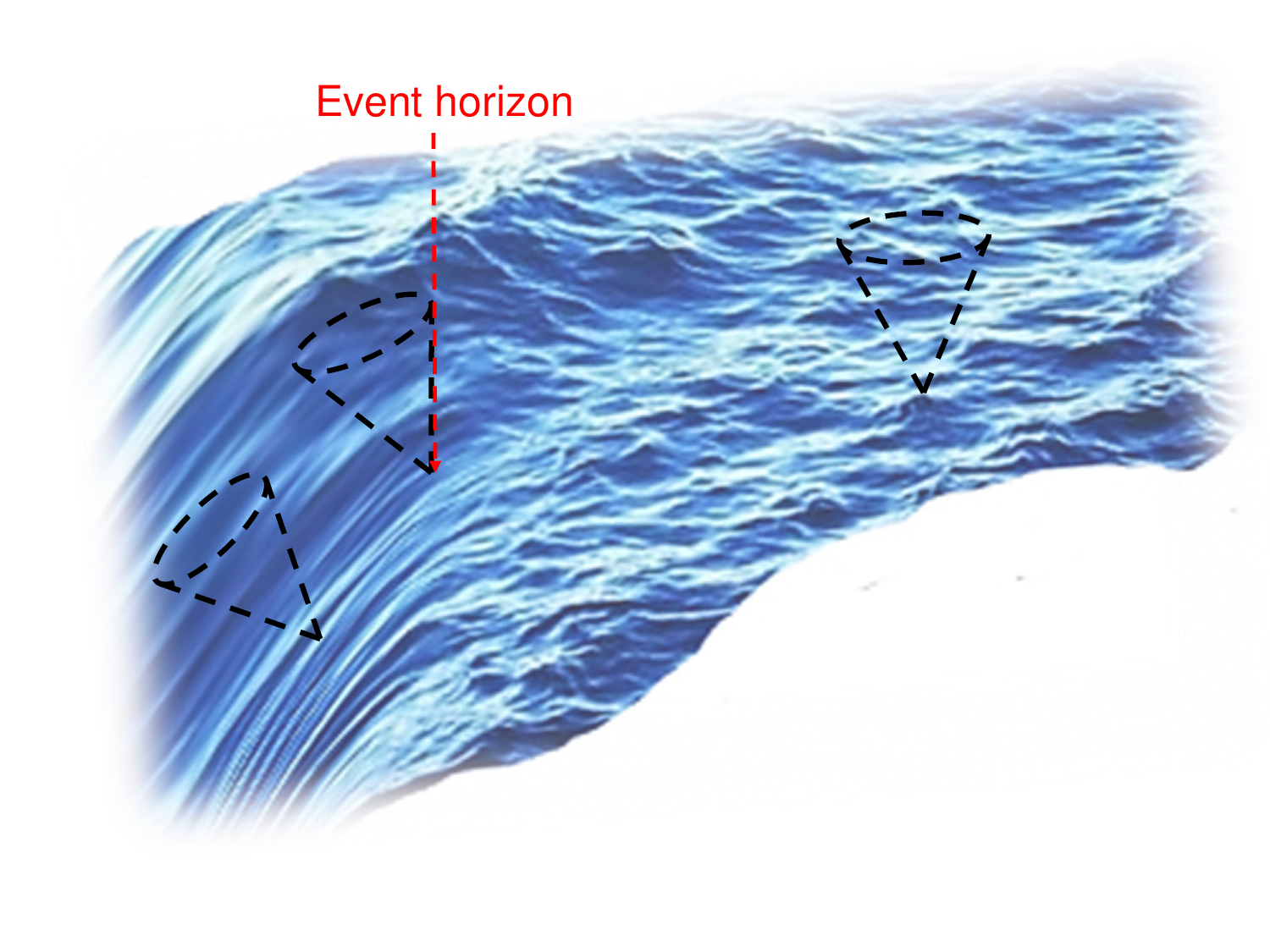}
\end{flushright}
\end{minipage}
\vspace{2mm}
\hrule
\end{figure*}

Metrics of this form allow for the simulation of several scenarios of interest in general relativity. For example, in any region where the fluid is rotating at supersonic speed, an acoustic observer (an observer whose observations are bound by the sound cones, see Box 1) is forced to co-rotate with the flow due to the supersonic dragging of inertial frames. A similar situation is found within the ergosphere surrounding rotating black holes, a spacetime region where the dragging is superluminal relative to an observer at infinity. Also, an event horizon for sound forms on a surface where the normal component of the fluid velocity passes from subsonic to supersonic: no sound waves can propagate through this surface against the flow. This sonic horizon acts as a boundary in the effective spacetime that causally disconnects the interior from the exterior, in analogy to the event horizon that general relativity predicts for black holes (the notion of causally disconnected events applies only to events bound by the sound cones, see Box 1. Signals other than sound do not experience any horizon). 

In real hydrodynamic systems, the analogy is valid only at low energies, or equivalently, large length scales, where the description in terms of macroscopic fluid variables holds (the hydrodynamic limit). In this limit, sound-like fluctuations exhibit an effective Lorentz invariance, with the velocity of sound playing the role of the locally invariant speed of light \cite{rev,unruh,visser1,visser2}. However, the above geometrical description holds only for the fluctuations, while the background flow (i.e. the metric) is governed by nonlinear hydrodynamics. As such, hydrodynamic systems allow the simulation of kinematic aspects (all phenomena related to field propagation on curved spacetime), but not of the full dynamics of Einstein gravity (although, as we shall see, a kind of gravitational dynamics may also be mimicked under specific conditions).

The last decade has witnessed a wealth of hydrodynamic experiments simulating different analogue gravitational scenarios. Event horizons in ($1+1$)-dimensions (one space plus time) have been experimentally realized in water tanks \cite{rousseaux2008,rousseaux2010}, Bose-Einstein condensates (BECs) \cite{lahav2010} quantum fluids of light \cite{elazar,nguyen} and superfluid Helium \cite{skyba}. Experiments with surface waves in water allowed for the first observation of horizon-related phenomena, such as the generation of negative-frequency modes \cite{rousseaux2008,rousseaux2010}, stimulated Hawking emission \cite{weinfurtner2011}, first measured in a weakly-nonlinear regime (as argued in \cite{euve2021nonlinear}) and then in the linear regime together with both non-stimulated and stimulated Hawking-like correlations \cite{euve2016}, and wave-scattering processes using Hawking channels of opposite energies \cite{euve2020} (see also \cite{PhysRevD.105.085022} for signals of the classical Hawking process taking place at a white hole horizon controlled by dissipation contrary to \cite{euve2016} where the white hole was controlled by the geometry.)

Quantum signatures of Hawking radiation have been reported in BECs, with the direct measurement of quantum correlations between outgoing and ingoing phonon pairs at the horizon \cite{steinhauer2016}. This result has been achieved by extracting the density-density correlations between points on opposite sides of the horizon and direcly measuring the entanglement of the Hawking and partner particles, a measurement only possible in analogue systems. Subsequent experiments in the same system verified the thermal nature of the Hawking spectrum with a temperature related to the horizon surface gravity \cite{deNova}, and the stationary nature of the process \cite{kolobov}. Quantum simulations of Unruh radiation have been achieved in parametrically modulated BECs, where the modulation has the same effect as boosting the sample to an accelerating reference frame \cite{hu2019quantum}.
Rotating black holes in ($2+1$)-dimensions have been created in quantum fluids of light \cite{vocke2018} and water tank experiments \cite{torres,torres2020}, where evidence of Penrose superradiance was first found in the dispersive regime \cite{patrick2021rotational}. Superradiance has also been observed for acoustic waves scattered by a rotating absorbing object \cite{cromb} and, more recently, for modes amplified in the sonic ergoregion generated in a photon-fluid \cite{braidotti2022,braidottiavs}. Beyond black hole physics, hydrodynamic experiments mimicking cosmological redshift and the Hubble friction associated with an expanding Universe \cite{eckel} and cosmological particle production have also been realized \cite{glorieux2022}.
After more than a decade of successful experiments testing different aspects of classical and semi-classical gravity, the extent to which the analogy can be pushed toward the domain of quantum gravity is yet to be fully explored. 
In this Perspective, we will focus on some paradigmatic examples of quantum gravitational effects and their possible hydrodynamic representations. We will illustrate new ideas and experimental proposals to gain insight into the black hole information paradox. We will describe how the spacetime geometry emerging from the microscopic dynamics of fluids could be exploited to simulate quantum gravitational corrections at high energies and other possible facets of black hole physics at the Planck scale. Finally, we will discuss the formation of curvature singularities in classical fluids and how they are naturally removed by quantum effects, giving rise to a fully regular spacetime geometry.

\section{The information paradox: a fluid dynamical approach}

The black hole information paradox represents one of the deepest puzzles in physics, exposing a potential inconsistency between quantum mechanics and general relativity. The emission of Hawking radiation at the horizon gradually leads to its complete evaporation. The final outcome of the process (a thermal bath of particles on a flat spacetime) would retain information only about the mass, electric charge and angular momentum of the original black hole, leading to the almost complete loss of information about the initial state of the matter that made up the black hole \cite{hawking3}. 
However, any loss of information is in direct contradiction to the unitary evolution of quantum systems. Simply put, a classical black hole evaporates via Hawking's process; which in turn implies the black hole is shrinking. This interplay between the emission of quantum particles and the evolution of the classical black hole, known as backreaction, forms the basis of the information paradox.

It has been suggested that this apparent inconsistency might be solved by a quantum theory of gravitation, where along with the emitted particles, the black holes too would be treated quantum mechanically \cite{RAJU20221,PhysRevD.76.024005}. Interestingly, analogue gravity in quantum fluids supports this view. If we consider for instance the complex order parameter $\psi$ describing the behaviour of a generic quantum fluid, its evolution is naturally unitary, as prescribed by quantum mechanics. However, unitarity is generally not preserved if we factorize the system dynamics into a mean-field (classical) background flow $\psi_0$ and quantum phonon fluctuations $\psi_1$ propagating on it, such that $\psi=\psi_0 + \epsilon \psi_1 + \mathcal{O}(\epsilon^2)$, where $\epsilon \ll 1$ is a perturbation parameter. If the flow associated with $\psi_0$, which in analogue models provides the background spacetime, is such that an acoustic black hole is created, the information on the quantum state of a phonon crossing the horizon would cease to be accessible to any acoustic observer, leading to an apparent loss of unitarity.  

This elementary (and oversimplified) argument suggests that the paradox may result from an incomplete knowledge of the system: the classical background geometry $\psi_0$ and its linear excitations provide only an approximate description of the full evolution accounted by $\psi$. The missing information is encoded in the neglected $\mathcal{O}(\epsilon^2)$-nonlinear terms, which take into account the quantum backreaction, i.e. the effects of the quantum phonon fluctuations over the spacetime geometry $\psi_0$ (see e.g. \cite{PhysRevD.72.105005,PhysRevA.106.053319} for the derivation of quantum backreaction from the microscopic physics of dilute BECs). Without any possibility of probing this backreaction, an acoustic observer would only have access to the classical geometry and its linear quantum fluctuations observing an apparent non-unitary evolution.

In the framework of BECs the backreaction results in quantum correlations between the background geometry and the phonon excitations, producing an entangled state \cite{liberati19}. Similarly, in the black hole evaporation process one can hypothesize the existence of correlations between the Hawking quanta and the quantum spacetime degrees of freedom inside the black hole, preserving unitarity of the system as a whole \cite{perez}. Recently, numerical simulations in a fluid of light with realistic experimental parameters have shown that phonon quantum fluctuations near a sonic horizon not only yield the aforementioned correlated emission related to the Hawking effect, but also cause quantum excitations of the acoustic black hole (quasi-normal modes) \cite{jacquet23}. Signatures of this process are found both in the correlations of density fluctuations and in the Hawking spectrum. This result provides a pristine example of quantum backreaction between phonons and background geometry, along with its effects on the Hawking emission. Experiments in BECs have already reached enough sensitivity to detect the spontaneous emission of entangled Hawking pairs at the horizon \cite{steinhauer2016}. Future experiments in such systems might soon be able to extract correlations between the condensate atomic states (which result in the global geometry) and their quantum collective excitations, thus paving the way for exploring the physics of event horizons that are purely quantum mechanical in nature.

The above ideas relate to the possibility of storing information in a kind of gravitational memory and, in turn, to the breakdown of the No-Hair Theorem beyond classical general relativity. This theorem states that stationary black holes cannot be distinguished apart from their mass, electric charge and angular momentum. Consequently, all other characteristics and details about the composition of the collapsed object or of the matter within the horizon (i.e. the potential hair) would be unavoidably lost.
While this theorem holds for classical black holes, the situation might be different if the gravitational field is quantized \cite{calmet}. Infalling particles would alter the geometry near the black hole, encoding the information about their state in its gravitational field and thus providing a possible solution to the information paradox \cite{calmet2,cheng}. These tiny modifications -- quantum bits of information, or hair -- might provide a unique quantum fingerprint of any black hole: two otherwise identical black holes in the classical theory, would then show tiny differences in their gravitational fields at the quantum level.

Quantum hair, independent of geometry, might also manifest itself through quantum geometric phases that are invisible classically but which could be observable in long-range global Aharonov-Bohm type scattering experiments \cite{krausswilczek,colemanwilczek,krausspreskill}. In fluid experiments quantum hair could be mimicked by fields of purely quantum origin generated by topological defects (vortices) and acting on phononic collective excitations with finite topological charge, similar to the Aharonov-Bohm interaction between an electrically charged particle and the electromagnetic potential. Classical signatures of this phenomenon (i.e. the phase shift of a classical wave with finite topological charge due to its interaction with a vortex has been detected in water wave experiments \cite{torres,berry,vivanco}. The detection of the same effect in superfluids at the level of quantum excitations interacting with quantized vortices has not been achieved so far. However, recent experiments have demonstrated the ability of imprinting on-demand vortex configurations in ultracold atomic systems and monitoring their interactions with phonons (see e.g. \cite{Kwon_2021}). It might therefore be possible to search for quantum hairs directly by studying the scattering between collective excitations, or phonons, with finite topological charge and analogue rotating black holes imprinted into the condensate, something that would be essentially impossible in astrophysical systems. Interestingly, along with the collective excitations, the vortex superflow simulating the rotating spacetime geometry would itself have a quantized angular momentum \cite{vocke2018}.

It is also of interest that non-classical hair-like states can originate from the (minimal) coupling of a matter field with rotating (Kerr or Kerr-Newmann) black holes \cite{herdeiro2014,benone14}. The gravitational interaction between a black hole and the field gives rise to the formation of quantized states in the vicinity of the horizon characterized by a discrete set of complex eigenfrequencies $\Omega$. Such states decay exponentially towards spatial infinity, but they can grow or decay in time depending on whether they oscillate slower or faster than the black hole rotational frequency $\Omega_{bh}$. In the first case (i.e $\mathrm{Re} (\Omega) < \Omega_{bh}$) they will be amplified via Penrose superradiance. Since they remain confined close to the horizon due to the gravitational interaction, their energy will grow exponentially in time, triggering the so-called black-hole bomb instability \cite{bhbomb}. On the other hand, modes with $\mathrm{Re} (\Omega) > \Omega_{bh}$ will eventually fall inside the horizon, leading to their complete disappearence. At the boundary between these two regimes, $\mathrm{Re} (\Omega) = \Omega_{bh}$, the matter field is in equilibrium with the black hole, so $\mathrm{Im} (\Omega)$=0, giving rise to bound states in synchronous rotation with the horizon. These states are known as scalar clouds \cite{hod1,hod2,benone14} that backreact to modify the spacetime geometry, such that it remains stationary, regular and asymptotically flat \cite{herdeiro2014}.

The above conditions clarify the minimal elements needed to construct stable black hole and external field configurations, i.e. a rotating black-hole geometry and a massive field coupled to it. Rotating black holes in ($2+1$)-dimensions have been realized by creating vortex flows in water tank experiments \cite{torres,torres2020} and quantized-vortices in superfluids of light \cite{vocke2018}. The superradiant scattering of waves at the ergosphere has been directly observed \cite{torres,braidotti2022}, although in both of these systems the elementary fluctuations have no mass and thus cannot mimic a matter field. On the other hand, a few hydrodynamic systems actually support massive phonon-like excitations. These include modified BEC condensates \cite{eg2}, quantum fluids with local and nonlocal interactions \cite{marino2019} and two-component superfluids \cite{PhysRevA.70.063615,silke-visser}, where recently the coexistence of massless and massive phonon excitations have been experimentally observed \cite{ferrari22}.
In these systems a massive phonon field on rotating spacetime geometries like those created in \cite{torres} and \cite{vocke2018} is expected to form stationary modes (neither growing nor decaying in time) in synchronous rotation with the horizon \cite{marino2021}. Such states show remarkable similarities with proposed astrophysical scalar clouds around Kerr black-holes \cite{marino2021,hod3,hod4}.
Although in fluids (and, generally, in all analogue models) the nonlinear dynamics is governed by a set of equations which greatly differ from Einstein's equations, these systems nevertheless offer a promising platform to explore the weakly nonlinear regime where Kerr-like geometries are deformed by the first-order corrections due to the cloud backreaction, thus creating the analogue of a rotating black hole with scalar hair.

\section{Microscopic spacetime structure: simulating quantum gravity with fluids?}

Soon after the prediction of black-hole evaporation it was realized that the process would involve the physics of fields at very high energies at which the semiclassical approximation itself may not hold \cite{PhysRevD.14.870}. Due to the high gravitational red shift near the horizon, the outgoing Hawking particles should originate from extremely high-frequency modes (in the free-fall reference frame of the black hole). This observation poses the question of whether any new physics at the Planck scale would modify the Hawking spectrum or even prevent the evaporation process \cite{pchen}.
 
In the absence of a self-consistent theory of quantum gravity, most studies have adopted a phenomenological approach to test the possible implications of quantum gravitational effects. One of these approaches involves a modification of the standard energy-momentum dispersion relation, $E^2= m^2c^4+p^2c^2$, to include higher-order corrections \cite{amelino98,garay-stf,amelino-doubly,magueijo}. These corrections have the effect of breaking the Lorentz symmetry at high energies/small distances, often associated with the Planck scale. The possible role of modified dispersion relations has been analysed in astrophysical observations (for a review see \cite{liberati-test}) and in the thermodynamics of black holes \cite{amelino2004,ling,nozari}.

The breaking of Lorentz symmetry in quantum gravity phenomenology is usually related to a minimal measurable length scale below which the very concepts of space and time break down. A minimal length seems incompatible with exact Lorentz invariance, because different inertial observers would not agree on such a minimal length (although, some theories are able to reconcile these two concepts \cite{speziale, Fontanini:2005ik}).  
The existence of such a minimal scale implies that spacetime, with its associated Lorentz symmetry, is not fundamental. It is in fact now generally accepted that spacetime should emerge from the microscopic dynamics of some elementary quantum objects as a collective large-scale phenomenon \cite{hu,bombelli,konopka}. In this sense, gravity would be more akin to thermodynamics \cite{jaco}, i.e. a macroscopic statistical behavior that emerges from some underlying complex many-body physics, rather than a fundamental force. The idea that general relativity emerges from quantum field theory in roughly the same sense that hydrodynamics emerges from molecular physics was already put forward by Sakharov in 1968 \cite{sakha} (see \cite{visser2002sakharov} for a review).
Interesting approaches in this direction include the causal set proposal \cite{bombelli}, quantum \textit{graphity} models \cite{konopka} (where points in spacetime are represented by linked nodes on a graph), entropic gravity \cite{Verlinde_2011}, and the group field theoretical approach to quantum gravity \cite{oriti2009group}.

A further emergent gravity scenario that has been inspired by results from black hole physics and string theory is the so-called holographic principle. The idea is that the curvature of spacetime (i.e. gravity itself) originates from some collective quantum dynamics occurring on the flat boundary of that spacetime \cite{ads1, ads2}. The boundary would then act as a kind of holographic plate with the gravitational dynamics being the result of a holographic projection.
 
Holography is usually studied on Anti-de Sitter (AdS) spacetime, a geometry which has a constant negative scalar curvature (other approaches in asymptotically flat spacetimes are the so-called \textit{celestial} \cite{pasterski2021celestial} and \textit{local} \cite{fre20,fre21} holography). Even though such a spacetime does not physically correspond to our Universe, it has been used as a toy model to understand the relationship between curved spacetime geometry and quantum theory. It is in the volume of such a spacetime that effective gravitational physics takes place. On the boundary, the related quantum dynamics is described in terms of a Conformal Field Theory (CFT), which is a quantum field theory invariant under conformal transformations (locally preserving angles, but not lengths). This conjectured relation between the gravitational dynamics in the volume and the quantum physics on the boundary is known as the AdS/CFT correspondence.

The holographic principle has been used to demonstrate that the geometrical structure in AdS generally emerges from properties of a CFT \cite{ep1,ep2}. This result has been extended by replacing the CFT by qubits, demonstrating that a classical geometry can emerge from quantum information \cite{bits}. This is suggestive of an old idea by Wheeler, who described it as `it from bit not bit from it', i.e. substance from information not information from substance \cite{itfrombit}. 

Remarkably, a spacetime geometry along with an effective Lorentz symmetry that gradually emerges at larger scales and lower energies is a common feature of phonon dynamics in both classical and quantum fluids. At large length scales, i.e. within the hydrodynamic limit, the classical expectation value of the mean fluid flow provides an effective curved spacetime on which elementary excitations can propagate (as already described in Box 1). At microscopic scales the description in terms of continuous variables no longer holds and the underlying geometric structure becomes increasingly less well-defined (see Box 2). 
In the following, we show how hydrodynamic systems may be exploited to test many of the aforementioned ideas related to black-hole physics at the Planck scale and emergent gravity scenarios.

\subsection{Modified dispersion relations}

Similar to the modified dispersion relation in quantum gravity \cite{amelino98,garay-stf,amelino-doubly,magueijo}, 
dispersion relations in fluids contain Lorentz-breaking momentum-dependent terms (see Box 2). The surface wave dispersion relation in water tanks \cite{Rousseaux_2013} and the Bogoliubov dispersion relation in atomic \cite{bogoliubov,reviewBEC} and photonic Bose gases \cite{fontaine18} are well-known examples.

The scale at which the effective Lorentz symmetry is broken is related to the underlying microscopic physics and is analogous to the Planck scale in quantum gravity phenomenology. In ideal classical inviscid fluids such a scale corresponds to the mean intermolecular distance at which the description in terms of continuous fluid variables no longer holds. Similarly, in superfluids the breakdown occurs at the the so-called coherence length, the minimal length over which the quantum fluid maintains its coherence (see Box 2). In viscous fluids, kinematic viscosity modifies the standard phonon dispersion relation introducing a dissipative term and a dispersive term breaking Lorentz invariance at short distances \cite{visser2}, although much larger than the atomic scale. Something similar occurs even for surface waves, where the analogue Planck scale  is determined by the depth of the channel in which they are propagating and by surface tension effects \cite{Rousseaux_2013}. In dipolar BECs, dipole-dipole interactions break the convexity of the dispersion relation leading to the so-called roton minimum (a local minimum in the dispersion relation corresponding to a new quasiparticle) and the breaking of Lorentz symmetry occurs at lengths of the order of the roton wavelength. Such a minimum for example has been shown to lead to analogue violations of cosmic scale invariance \cite{PhysRevLett.118.130404}.

In most fluids the excitations are massless. However, systems based on coupled-condensates \cite{PhysRevA.70.063615,silke-visser} or involving suitably shaped interactions \cite{eg2,marino2019} can also admit massive excitations and even a coexistence between massive and massless modes, as recently observed in a BEC experiment \cite{ferrari22}. As massive modes can mimic matter degrees of freedom, such systems offer a promising platform to test many ideas related to black-hole physics at Planckian scales.

\begin{figure*}
\hrule
\vspace{2mm}
\section*{{Box 2: Modified dispersion relations: quantum gravity and fluids}}
\begin{minipage}{1.0\textwidth}
\begin{flushright}
\includegraphics[width = 1.0\textwidth]{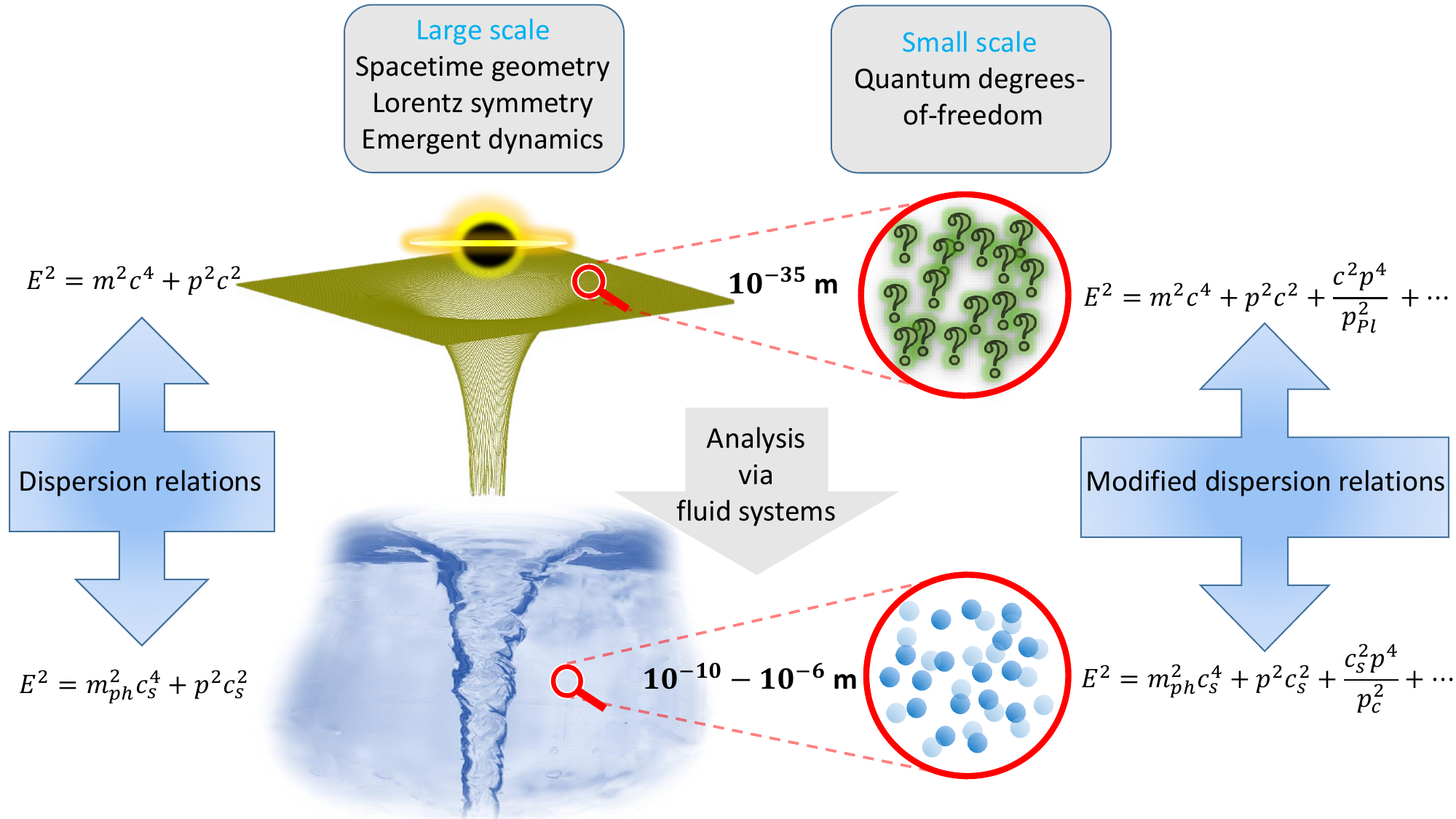}
\end{flushright}
\end{minipage}

\vspace{2mm}

\begin{minipage}{1.0\textwidth}
\linespread{1.0}\selectfont
\justify

{In fluids, the effective geometry breaks down at scales where the continuous description in terms of macroscopic variables no longer holds, in close analogy to the breakdown of spacetime geometry due to quantum gravitational effects. This scale is manifest by a breakdown of Lorentz symmetry as evidenced by the above modified phonon dispersion relations (see figure, right). There, $m_{ph}$ is the mass of phonon excitations which is the analogue of the rest mass $m$, with energy and momentum being denoted by $E$ and $p$ for either system. For $m_{ph}=0$ and truncating the expansion at the fourth-order in $p$ the above expression reduces to the well-known Bogoliubov dispersion relation in a Bose gas \cite{bogoliubov,reviewBEC} or in a circular hydraulic jump \cite{Volovik_2005,PhysRevE.83.056312}. The critical momentum $p_c = M c_s$, that here plays the role of the Planck momentum $p^{~}_{Pl}$, depends on the mass $M$ of the particles forming the condensate and is inversely proportional to the coherence length $\xi=\hbar/(M c_s)$. When $p \ll p_c$, the excitations follow the standard relativistic energy-momentum relation.}

\end{minipage}

\vspace{2mm}
\hrule
\end{figure*}
 
Phonon dispersion relations have been used to explore various issues associated with Lorentz-violating effective field theories \cite{eg1} and the effects of high-energy modes on the black hole evaporation process \cite{jacobson,unruh95,Corley}. The analysis presented in these works supported the viewpoint that Planck scale modes would not prevent the existence of black-hole radiation, but could significantly affect the emitted spectrum for high-energy modes \cite{jacob93}.

By contrast, in the low-energy regime theoretical and numerical studies in BECs predict the thermal nature of phononic Hawking radiation. This prediction was recently confirmed experimentally \cite{deNova}. These studies, among many others (see e.g., \cite{carusotto08}), demonstrated the robustness of the thermal nature of Hawking emission against short-scale modifications of field modes, thus supporting its interpretation as a low-energy effect. 

Most experiments have been conceived to operate well within the hydrodynamic limit, precisely to minimise the effects of high-energy modes. While this was reasonable in the search for evidence of Hawking radiation, it would be now interesting to engineer acoustic black holes in a regime where short-distance physics, is expected to be more relevant, as in the demonstration of analogue wormholes based on the microscopic length-scale of capillarity in water channel experiments \cite{PhysRevD.96.064042}. These effects could leave their signatures on the radiation spectrum in the form of deviations from thermality \cite{PhysRevD.71.024028}, as already reported in \cite{euve2016}. As a further example, strong dipolar interactions lead to significant violations of exact thermality (greybody factors), which can be traced back to the existence of the aforementioned roton minimum in the dispersion relation \cite{ribeiro2022impact}. 
It is worth stressing once again that it is the perfect thermality of the Hawking spectrum (i.e. the evolution from a pure initial state to a mixed final one) that underlies the information paradox. Quantum fluids of ultracold atoms, for example, could provide one avenue for examining such deviations induced by quantum degrees of freedom in the background flow.

\subsection{Phonon interactions and emergence of gravity}

In hydrodynamics, the spacetime curvature induced by an inhomogenous flow plays the role of a gravity-like field which determines the propagation of phonons. However, at the linear level of non-interacting phonon excitations, all effects due to gravitational backreaction are neglected: i.e. the acoustic spacetime geometry is not modified by the excitations propagating on it.

Nevertheless, a kind of gravitational backreaction is naturally encoded in nonlinear acoustics. A density perturbation induces a disturbance in the background flow which, as explained, forms the effective spacetime geometry on which the perturbation propagates \cite{goulart1,goulart2,cherubini,marino2016,fisher-grav}.
This extends the analogue gravity program, as it encodes in a geometric framework  the phenomenon of backreaction, i.e. the dynamical interplay between the phonons and the acoustic metric. As we will discuss in the next section, such a geometrical description can also be exploited to address the problem of curvature singularities. 

Retaining the first nonlinear corrections describing the backreaction, i.e., the $\mathcal{O}(\epsilon^2)$ terms, one finds that
the phonons follow the same equations of motion as in the linear regime, but with these corrections modifying the zeroth-order background flow (the acoustic metric). 

\begin{figure}
\begin{center}
\includegraphics*[width=0.8\columnwidth]{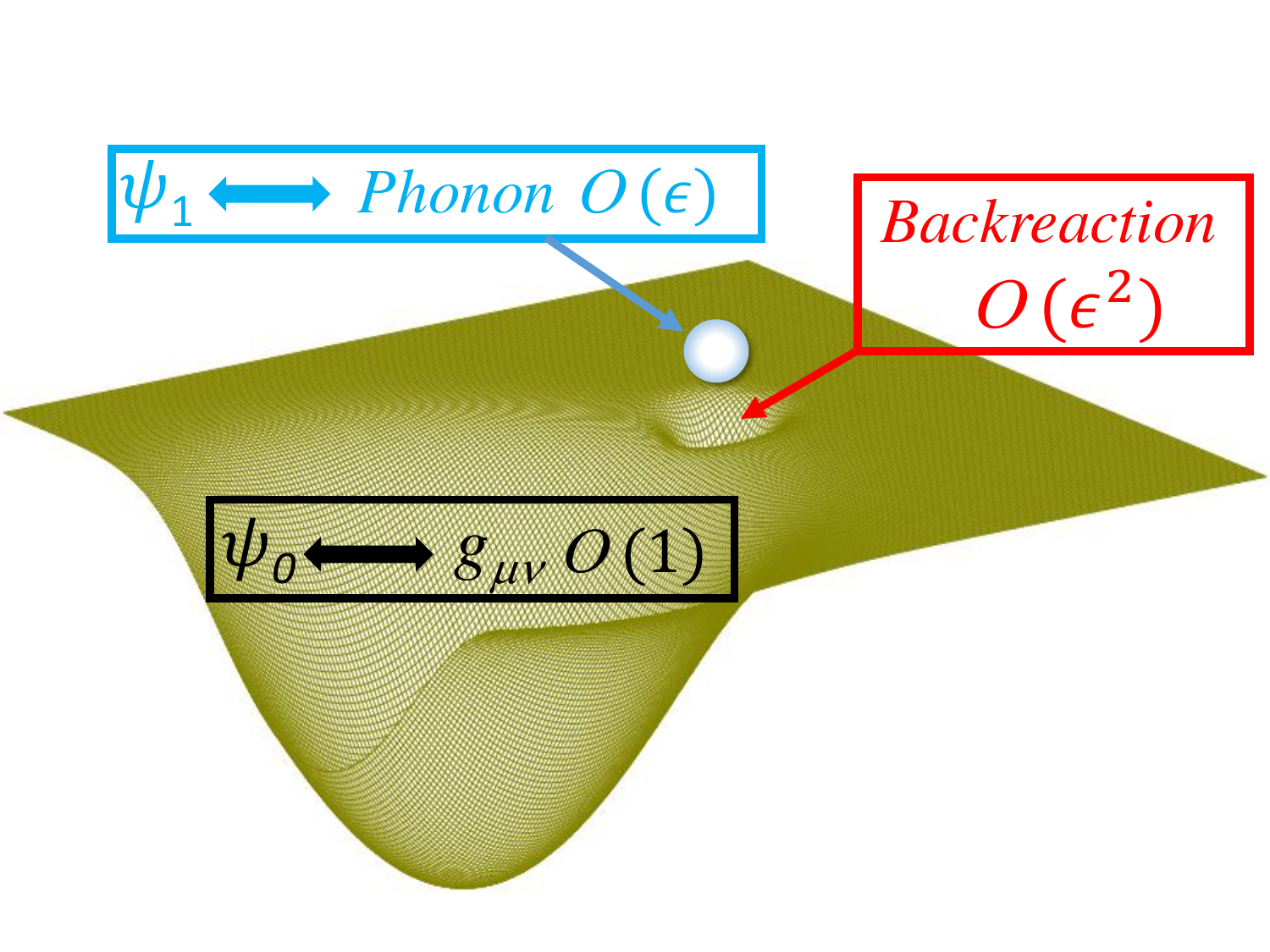}
\end{center} 
\caption{\textbf{Pictorial view of the phonon's backreaction:} a phonon propagating in an inhomogeoenous flow (i.e. an acoustic spacetime) locally modifies the background geometry.}
\label{figure2}
\end{figure}

This approach has been used to study several backreaction effects, for instance to show that acoustic black holes get cooler as they radiate phonons \cite{balbinot05}. It is also an appropriate way to show how gravitational interactions emerge at a more fundamental level. An example is provided by massive-phonon excitations in BEC systems that experience a kind of Newtonian gravity \cite{eg2,marino2019}. In these systems modified interactions between the fluid particles are engineered to alter the phonon dispersion relation and create a gap. The phonons thus acquire a rest mass. Moreover, a gravitational potential is produced by the phononic mass-density distribution, which here plays the role of matter. Hence the phonons behave as self-gravitating quantum particles. Although the analogy with standard Newtonian gravity is limited, the differences are themselves of some interest. The resulting emergent gravitational interaction has finite range (i.e. it is mediated by massive gravitons) and a non-zero cosmological constant is also present due to the backreaction of atoms which are not part of the condensate \cite{eg3}. Recently, experimental measurements of the dispersion relation of collective excitations in coupled condensates realized via parametric modulation of the trap demonstrated the coexistence of massive and massless phonon modes \cite{ferrari22}. Similar measurements in the weakly-nonlinear regime may provide information on quasiparticle interactions and back-reaction phenomena paving the way for the observation of emergent gravitational interactions in the near future.

While the non-relativistic equations of ordinary fluids do not obey the full set of symmetries associated with a geometric theory like general relativity (see e.g. \cite{eg2}), the situation changes if one considers relativistic fluids. For instance, in relativistic BEC models \cite{belenchia14} the acoustic spacetime is governed by a fully geometrical theory that takes the familiar form of the Nordstr\"{o}m scalar theory of gravitation. Although this theory is only a scalar theory of gravity, like general relativity, it satisfies the strong equivalence principle \cite{deruelle}.
The spacetime curvature is determined by the expectation value of the stress-energy of the quasiparticles, while the Newton and cosmological constants are functions of the fundamental scales of the microscopic system. This is the first example in analogue gravity in which a Lorentz invariant, geometric theory of semiclassical gravity emerges from an underlying quantum theory of matter in flat spacetime. It is possible that more complex relativistic condensates (e.g. involving two-component supefluids \cite{silke-visser} or condensates with suitably manipulated interactions \cite{marino2019}) might provide testbeds for something even closer to an emergent Einstein-like theory. 

Such systems could also enable experimental tests of several features associated with holography. Planar AdS black-hole solutions can be exactly mimicked using a nonrelativistic BEC in ($2+1$) dimensions \cite{anti1,anti4}. The advantage of engineering such solutions is to enable experimental tests of both the evolution of excitations in the volume and the correlated dynamics at the boundary. In fact, an analogue gravity system relating a fluid on the flat boundary of AdS spacetime to one in the volume has been investigated \cite{sabine15,sabine16}. The latter mimics a projection of the dynamics at the boundary and generates an effective metric that reproduces the asymptotic AdS geometry. The next step, to construct a complete analogy, would be the study of the backreaction to show whether, or under what conditions, some kind of gravitational phonon dynamics emerges. To this end, a promising development of these proposals could be their generalization to relativistic BECs, where the field equations for Nordstr\"{o}m gravity have been shown to emerge \cite{belenchia14}. Such an extension could allow one to directly probe the relation between the emergent gravitational physics and the quantum hydrodynamic degrees of freedom at the boundary \cite{dey}. In the framework of holography, it has been shown that the Einstein equations in the volume can be obtained from a relativistic fluid dynamics on the boundary \cite{ei12, ei14}. However, the AdS solution representing the gravitational dynamics in such situations only represents a mathematical device for calculating the behavior of the physical boundary fluid \cite{hc12, hc14}. On the other hand, in analogue experiments both sides of the holographic correspondence could be explicitly probed.

\section{Acoustic singularities and quantum effects} 

Curvature singularities, where the concepts of space and time break down catastrophically, are one of the most dramatic features of classical general relativity \cite{si12, si14}. 
The physical reality of these spacetime singularities has been long debated, and it is generally assumed that a quantum theory describing gravity at short distances might remove them.

Singularities abound in the equations of fluid mechanics, where they typically arise at scales where a given model fails to fully describe the system and some new physical effects come into play. While the processes of singularity formation in fluids and Einstein gravity are distinct, the study of the regularization mechanism of hydrodynamic singularities in which the curvature of the acoustic metric is somehow counterbalanced by quantum effects could provide new ideas for modelling similar phenomena in gravity theories. 

\begin{figure}
\begin{center}
\includegraphics*[width=0.7\columnwidth]{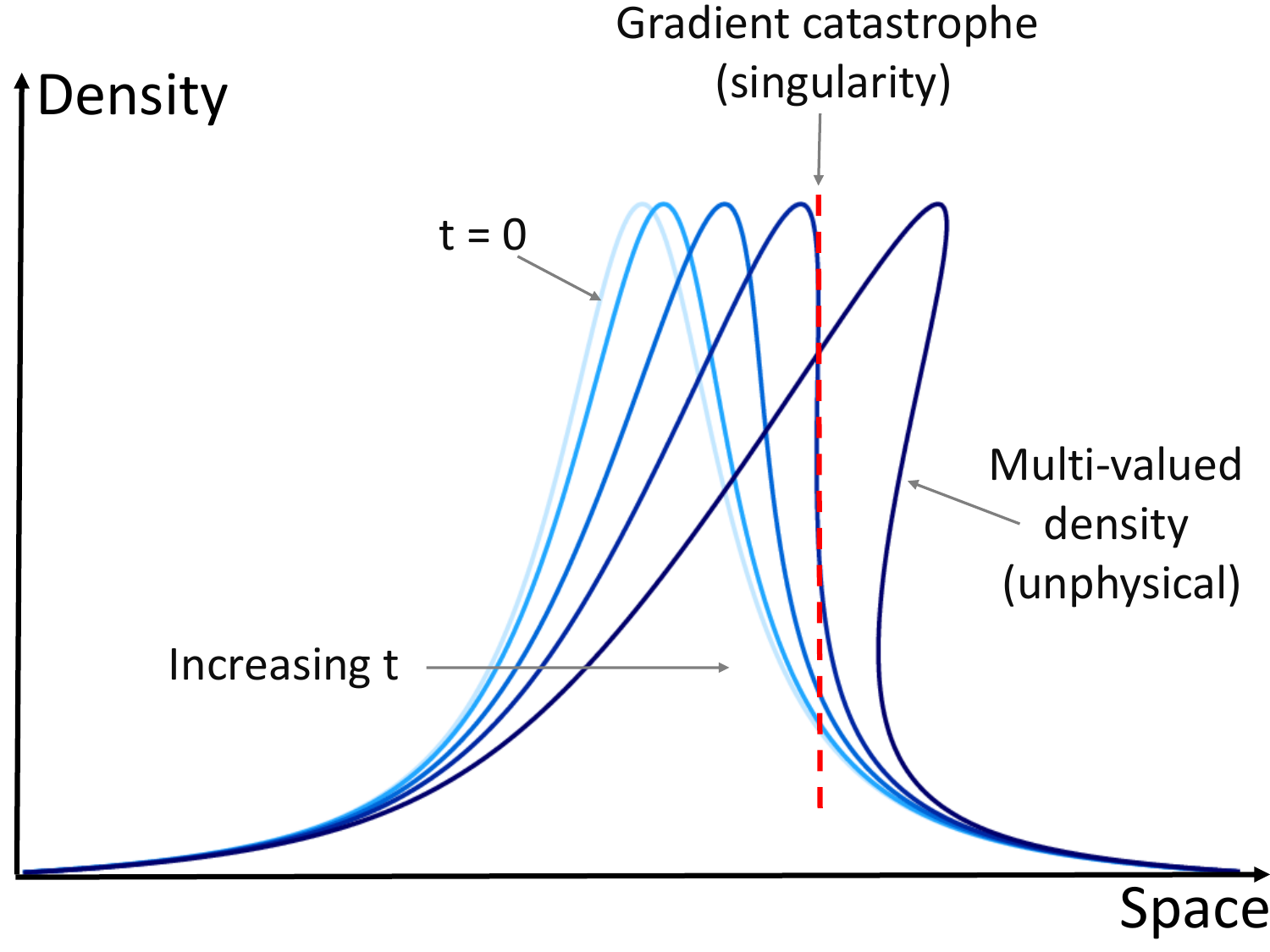}
\end{center} 
\caption{\textbf{Schematic of the self-steepening of a density wave.} Plots of density profiles in the comoving frame at different times. As time progresses, the wavefront in the propagation direction becomes increasingly steeper until it becomes vertical (gradient catastrophe).}
\label{fig4}
\end{figure}

In classical inviscid fluids a density perturbation cannot propagate forever, as a flow discontinuity is spontaneously created in a finite time. Owing to the nonlinear coupling between the density and flow velocity, a propagating density front 
induces a background flow which, in turn, modifies its propagation. Different points along the wave profile propagate at different velocities leading to self-steepening of the wavefront. In a finite time the wavefront will tip past the vertical, which would correspond to an unphysical multivalued solution of the fluid equations, yielding the so-called gradient catastrophe \cite{landau} (see Fig. \ref{fig4}). Theoretical studies of this phenomenon date back to Riemann’s seminal work on discontinuous flows in 1860 \cite{riemann}. 

Analogue gravity allows for a fully geometric description of the above self-interaction, where the gradient catastrophe can be associated with a curvature singularity of an emergent acoustic metric. As the wave density profile changes in time, it continuously modifies the underlying flow and thus the associated spacetime geometry. The latter, in turn, affects the density profile. Unlike the linear case, where the acoustic geometry is shaped by suitable external forces, here a curved spacetime geometry originates from the backreaction of a density perturbation over the background flow. In this way an initially negligible curvature/steepening will increase in time without bound \cite{marino2016}. A correspondence between the gradient catastrophe and curvature singularity of the emergent metric has been explicitly shown by direct computation of suitable curvature invariants \cite{marino2016}. Note however that such singularities are \textit{naked}, i.e. are not hidden to external observers by an event horizon.

In the framework of BECs and superfluids, quantum effects, due to the microscopic structure of these systems, completely change the picture. The condensate exhibits a certain stiffness against spatial variations of its density, given by the so-called Bohm quantum potential \cite{bohmqp,reviewBEC} that depends on the curvature of the amplitude of the wave function and goes to zero in the classical limit as the de Broglie wavelength also vanishes. When the density curvature becomes sufficiently strong, the quantum stiffness gradually comes into play to counterbalance the curvature. As one approaches the critical point where the gradient catastrophe would otherwise take place, rapid spatial oscillations of density and velocity develop \cite{fischer-cens} leading to the formation of a dispersive shock wave. These waves make the spacetime structure completely regular, therefore heading off the formation of a naked singularity. A similar behaviour is found for the quantum gravitational collapse of spherically symmetric pressureless dust \cite{PhysRevLett.128.121301}, where the infalling dust bounces when the spacetime curvature reaches the Planck scale and forms an outgoing shock wave. As a result the final black hole is nonsingular \cite{PhysRevLett.128.121301}. The emission of a shock wave is probably a very common mechanism for nonlinear systems to smear the effects of nonlinearities that concentrate energy in small regions of space. Similar scenarios could be investigated in surface wave experiments, with the surface tension eventually playing the same regularizing role as the Bohm quantum potential.
Dispersive shock waves have been observed in different settings, ranging from water waves platforms \cite{ROUSSEAUX201631}, BECs \cite{dutton} and, in particular, photon superfluids \cite{wan,xu,bienaime,Bendahmane}. From an analogue gravity perspective, then, these systems are interesting candidates to explore how quantum effects might remove singularities that would otherwise be present.  

\section{Conclusions and perspectives}

Analogue gravity systems have been demonstrated to offer a useful testbed for classical and semi-classical general relativity. Beyond providing experimental evidence of fundamental phenomena that might have not otherwise been directly observed, these experiments have also deepened our theoretical understanding of them. For instance, they have shown that Hawking radiation arises for any test field on any Lorentzian geometry containing an event horizon, regardless of whether or not that geometry has a gravitational origin, and that Penrose superradiance does not require an event horizon to occur. Analogue gravity systems have thus surpassed their initial aim and scope, providing a tool for advancing gravitational physics. After 40 years of theoretical proposals, and more than a decade of successful experiments mainly focused on the observation of the Hawking effect, the field is now mature enough to face new challenges and to push the analogy towards its own limits, approaching the domain of quantum gravity. The recent progress in controlling and manipulating classical and quantum fluids will open novel opportunities to engineer accessible analogue gravity scenarios. The next generation of analogue experiments should focus on short-distance measurements involving the manipulation and control of atomic degrees of freedom to clarify how they affect the properties of the macroscopic emergent geometry. The emergence of large-scale collective phenomena have often very similar properties in spite of the physical differences between the systems in which they arise. So we expect that these new experiments may provide useful insights into the emergence of spacetime geometry from quantum gravitational degrees of freedom. At the same time, as discussed in our perspective, recent works aimed at unraveling a full quantum theory of gravity suggest numerous effects that cannot be tested directly in astrophysical or cosmological contexts, but which might be amenable to being investigated in laboratory conditions. These phenomena range from modified dispersion relations and minimal measurable lengths, resembling those found in quantum gravity phenomenology, to the emergence of spacetime and gravity from the collective quantum dynamics of microscopic degrees of freedom. The latter can have profound consequences for not only shedding light on the information paradox and the resolution of curvature singularities, but also inspire new directions in the development of a quantum theory of gravity.

\subsection{Acknowledgements}

The authors thank M. Ciszak, P. Hoodbhoy, B.S. Kay, R. Norte, K.A. Peacock, D.J. Smith, T.S. Tsun and W.G. Unruh for valuable comments on the manuscript. The authors also thank S. Bahamonde for sharing many ideas on the information paradox and holography in analogue gravity and G. Roati for fruitful discussions on BECs experiments.

\bibliography{reference}

\begin{thebibliography}{145}%
\makeatletter
\providecommand \@ifxundefined [1]{%
 \@ifx{#1\undefined}
}%
\providecommand \@ifnum [1]{%
 \ifnum #1\expandafter \@firstoftwo
 \else \expandafter \@secondoftwo
 \fi
}%
\providecommand \@ifx [1]{%
 \ifx #1\expandafter \@firstoftwo
 \else \expandafter \@secondoftwo
 \fi
}%
\providecommand \natexlab [1]{#1}%
\providecommand \enquote  [1]{``#1''}%
\providecommand \bibnamefont  [1]{#1}%
\providecommand \bibfnamefont [1]{#1}%
\providecommand \citenamefont [1]{#1}%
\providecommand \href@noop [0]{\@secondoftwo}%
\providecommand \href [0]{\begingroup \@sanitize@url \@href}%
\providecommand \@href[1]{\@@startlink{#1}\@@href}%
\providecommand \@@href[1]{\endgroup#1\@@endlink}%
\providecommand \@sanitize@url [0]{\catcode `\\12\catcode `\$12\catcode
  `\&12\catcode `\#12\catcode `\^12\catcode `\_12\catcode `\%12\relax}%
\providecommand \@@startlink[1]{}%
\providecommand \@@endlink[0]{}%
\providecommand \url  [0]{\begingroup\@sanitize@url \@url }%
\providecommand \@url [1]{\endgroup\@href {#1}{\urlprefix }}%
\providecommand \urlprefix  [0]{URL }%
\providecommand \Eprint [0]{\href }%
\providecommand \doibase [0]{https://doi.org/}%
\providecommand \selectlanguage [0]{\@gobble}%
\providecommand \bibinfo  [0]{\@secondoftwo}%
\providecommand \bibfield  [0]{\@secondoftwo}%
\providecommand \translation [1]{[#1]}%
\providecommand \BibitemOpen [0]{}%
\providecommand \bibitemStop [0]{}%
\providecommand \bibitemNoStop [0]{.\EOS\space}%
\providecommand \EOS [0]{\spacefactor3000\relax}%
\providecommand \BibitemShut  [1]{\csname bibitem#1\endcsname}%
\let\auto@bib@innerbib\@empty
\bibitem [{\citenamefont {Hawking}(1974)}]{hawking1}%
  \BibitemOpen
  \bibfield  {author} {\bibinfo {author} {\bibfnamefont {S.~W.}\ \bibnamefont
  {Hawking}},\ }\bibfield  {title} {\bibinfo {title} {{Black hole
  explosions?}},\ }\href {https://doi.org/10.1038/248030a0} {\bibfield
  {journal} {\bibinfo  {journal} {Nature}\ }\textbf {\bibinfo {volume} {248}},\
  \bibinfo {pages} {30} (\bibinfo {year} {1974})}\BibitemShut {NoStop}%
\bibitem [{\citenamefont {Hawking}(1975)}]{hawking2}%
  \BibitemOpen
  \bibfield  {author} {\bibinfo {author} {\bibfnamefont {S.~W.}\ \bibnamefont
  {Hawking}},\ }\bibfield  {title} {\bibinfo {title} {{Particle creation by
  black holes}},\ }\href {https://doi.org/10.1007/BF02345020} {\bibfield
  {journal} {\bibinfo  {journal} {Commun. Math. Phys.}\ }\textbf {\bibinfo
  {volume} {43}},\ \bibinfo {pages} {199} (\bibinfo {year} {1975})},\ \bibinfo
  {note} {[Erratum: Commun.Math.Phys. 46, 206 (1976)]}\BibitemShut {NoStop}%
\bibitem [{\citenamefont {Penrose}\ and\ \citenamefont
  {Floyd}(1971)}]{penrose}%
  \BibitemOpen
  \bibfield  {author} {\bibinfo {author} {\bibfnamefont {R.}~\bibnamefont
  {Penrose}}\ and\ \bibinfo {author} {\bibfnamefont {R.~M.}\ \bibnamefont
  {Floyd}},\ }\bibfield  {title} {\bibinfo {title} {{Extraction of rotational
  energy from a black hole}},\ }\href {https://doi.org/10.1038/physci229177a0}
  {\bibfield  {journal} {\bibinfo  {journal} {Nature}\ }\textbf {\bibinfo
  {volume} {229}},\ \bibinfo {pages} {177} (\bibinfo {year}
  {1971})}\BibitemShut {NoStop}%
\bibitem [{\citenamefont {Misner}(1972)}]{misner}%
  \BibitemOpen
  \bibfield  {author} {\bibinfo {author} {\bibfnamefont {C.}~\bibnamefont
  {Misner}},\ }\bibfield  {title} {\bibinfo {title} {{Stability of Kerr black
  holes against scalar perturbations}},\ }\href
  {https://doi.org/10.1038/physci229177a0} {\bibfield  {journal} {\bibinfo
  {journal} {Bull. Amer. Phys. Soc.}\ }\textbf {\bibinfo {volume} {17}},\
  \bibinfo {pages} {472} (\bibinfo {year} {1972})}\BibitemShut {NoStop}%
\bibitem [{\citenamefont {{Zel'dovich}}(1971)}]{zeldovich}%
  \BibitemOpen
  \bibfield  {author} {\bibinfo {author} {\bibfnamefont {Y.~B.}\ \bibnamefont
  {{Zel'dovich}}},\ }\bibfield  {title} {\bibinfo {title} {{Generation of
  {w}aves by a {r}otating {b}ody}},\ }\href@noop {} {\bibfield  {journal}
  {\bibinfo  {journal} {Soviet Journal of Experimental and Theoretical Physics
  Letters}\ }\textbf {\bibinfo {volume} {14}},\ \bibinfo {pages} {180}
  (\bibinfo {year} {1971})}\BibitemShut {NoStop}%
\bibitem [{\citenamefont {{Starobinski{\v{i}}}}(1973)}]{staro1}%
  \BibitemOpen
  \bibfield  {author} {\bibinfo {author} {\bibfnamefont {A.~A.}\ \bibnamefont
  {{Starobinski{\v{i}}}}},\ }\bibfield  {title} {\bibinfo {title}
  {{Amplification of waves during reflection from a rotating ``black hole''}},\
  }\href@noop {} {\bibfield  {journal} {\bibinfo  {journal} {Soviet Journal of
  Experimental and Theoretical Physics}\ }\textbf {\bibinfo {volume} {37}},\
  \bibinfo {pages} {28} (\bibinfo {year} {1973})}\BibitemShut {NoStop}%
\bibitem [{\citenamefont {{Starobinski{\v{i}}}}\ and\ \citenamefont
  {{Churilov}}(1974)}]{staro2}%
  \BibitemOpen
  \bibfield  {author} {\bibinfo {author} {\bibfnamefont {A.~A.}\ \bibnamefont
  {{Starobinski{\v{i}}}}}\ and\ \bibinfo {author} {\bibfnamefont {S.~M.}\
  \bibnamefont {{Churilov}}},\ }\bibfield  {title} {\bibinfo {title}
  {{Amplification of electromagnetic and gravitational waves scattered by a
  rotating ``black hole''}},\ }\href@noop {} {\bibfield  {journal} {\bibinfo
  {journal} {Soviet Journal of Experimental and Theoretical Physics}\ }\textbf
  {\bibinfo {volume} {38}},\ \bibinfo {pages} {1} (\bibinfo {year}
  {1974})}\BibitemShut {NoStop}%
\bibitem [{\citenamefont {Unruh}(1976)}]{PhysRevD.14.870}%
  \BibitemOpen
  \bibfield  {author} {\bibinfo {author} {\bibfnamefont {W.~G.}\ \bibnamefont
  {Unruh}},\ }\bibfield  {title} {\bibinfo {title} {Notes on black-hole
  evaporation},\ }\href {https://doi.org/10.1103/PhysRevD.14.870} {\bibfield
  {journal} {\bibinfo  {journal} {Phys. Rev. D}\ }\textbf {\bibinfo {volume}
  {14}},\ \bibinfo {pages} {870} (\bibinfo {year} {1976})}\BibitemShut
  {NoStop}%
\bibitem [{\citenamefont {Unruh}(1981)}]{unruh}%
  \BibitemOpen
  \bibfield  {author} {\bibinfo {author} {\bibfnamefont {W.~G.}\ \bibnamefont
  {Unruh}},\ }\bibfield  {title} {\bibinfo {title} {Experimental black-hole
  evaporation?},\ }\href {https://doi.org/10.1103/PhysRevLett.46.1351}
  {\bibfield  {journal} {\bibinfo  {journal} {Phys. Rev. Lett.}\ }\textbf
  {\bibinfo {volume} {46}},\ \bibinfo {pages} {1351} (\bibinfo {year}
  {1981})}\BibitemShut {NoStop}%
\bibitem [{\citenamefont {Barcel{\'o}}\ \emph {et~al.}()\citenamefont
  {Barcel{\'o}}, \citenamefont {Liberati},\ and\ \citenamefont {Visser}}]{rev}%
  \BibitemOpen
  \bibfield  {author} {\bibinfo {author} {\bibfnamefont {C.}~\bibnamefont
  {Barcel{\'o}}}, \bibinfo {author} {\bibfnamefont {S.}~\bibnamefont
  {Liberati}},\ and\ \bibinfo {author} {\bibfnamefont {M.}~\bibnamefont
  {Visser}},\ }\bibfield  {title} {\bibinfo {title} {Analogue gravity},\
  }\bibfield  {journal} {\bibinfo  {journal} {Living reviews in relativity}\
  }\href {https://doi.org/https://doi.org/10.12942/lrr-2011-3}
  {https://doi.org/10.12942/lrr-2011-3}\BibitemShut {NoStop}%
\bibitem [{\citenamefont {Barcel\'o}(2019)}]{barcelorev}%
  \BibitemOpen
  \bibfield  {author} {\bibinfo {author} {\bibfnamefont {C.}~\bibnamefont
  {Barcel\'o}},\ }\bibfield  {title} {\bibinfo {title} {{Analogue black-hole
  horizons}},\ }\href {https://doi.org/10.1038/s41567-018-0367-6} {\bibfield
  {journal} {\bibinfo  {journal} {Nature Phys.}\ }\textbf {\bibinfo {volume}
  {15}},\ \bibinfo {pages} {210} (\bibinfo {year} {2019})}\BibitemShut
  {NoStop}%
\bibitem [{\citenamefont {Faccio}\ \emph {et~al.}(2013)\citenamefont {Faccio},
  \citenamefont {Belgiorno}, \citenamefont {Cacciatori}, \citenamefont
  {Gorini}, \citenamefont {Liberati},\ and\ \citenamefont
  {Moschella}}]{facciorev}%
  \BibitemOpen
  \bibinfo {editor} {\bibfnamefont {D.}~\bibnamefont {Faccio}}, \bibinfo
  {editor} {\bibfnamefont {F.}~\bibnamefont {Belgiorno}}, \bibinfo {editor}
  {\bibfnamefont {S.}~\bibnamefont {Cacciatori}}, \bibinfo {editor}
  {\bibfnamefont {V.}~\bibnamefont {Gorini}}, \bibinfo {editor} {\bibfnamefont
  {S.}~\bibnamefont {Liberati}},\ and\ \bibinfo {editor} {\bibfnamefont
  {U.}~\bibnamefont {Moschella}},\ eds.,\ \href
  {https://doi.org/10.1007/978-3-319-00266-8} {\emph {\bibinfo {title}
  {{Analogue Gravity Phenomenology}}}},\ Vol.\ \bibinfo {volume} {870}\
  (\bibinfo {year} {2013})\BibitemShut {NoStop}%
\bibitem [{\citenamefont {White}(1973)}]{white}%
  \BibitemOpen
  \bibfield  {author} {\bibinfo {author} {\bibfnamefont {R.~W.}\ \bibnamefont
  {White}},\ }\bibfield  {title} {\bibinfo {title} {Acoustic ray tracing in
  moving inhomogeneous fluids},\ }\href {https://doi.org/10.1121/1.1913522}
  {\bibfield  {journal} {\bibinfo  {journal} {The Journal of the Acoustical
  Society of America}\ }\textbf {\bibinfo {volume} {53}},\ \bibinfo {pages}
  {1700} (\bibinfo {year} {1973})},\ \Eprint
  {https://arxiv.org/abs/https://doi.org/10.1121/1.1913522}
  {https://doi.org/10.1121/1.1913522} \BibitemShut {NoStop}%
\bibitem [{\citenamefont {Visser}(1998{\natexlab{a}})}]{visser1}%
  \BibitemOpen
  \bibfield  {author} {\bibinfo {author} {\bibfnamefont {M.}~\bibnamefont
  {Visser}},\ }\bibfield  {title} {\bibinfo {title} {Hawking radiation without
  black hole entropy},\ }\href {https://doi.org/10.1103/PhysRevLett.80.3436}
  {\bibfield  {journal} {\bibinfo  {journal} {Phys. Rev. Lett.}\ }\textbf
  {\bibinfo {volume} {80}},\ \bibinfo {pages} {3436} (\bibinfo {year}
  {1998}{\natexlab{a}})}\BibitemShut {NoStop}%
\bibitem [{\citenamefont {Visser}(1998{\natexlab{b}})}]{visser2}%
  \BibitemOpen
  \bibfield  {author} {\bibinfo {author} {\bibfnamefont {M.}~\bibnamefont
  {Visser}},\ }\bibfield  {title} {\bibinfo {title} {Acoustic black holes:
  horizons, ergospheres and {H}awking radiation},\ }\href
  {https://doi.org/10.1088/0264-9381/15/6/024} {\bibfield  {journal} {\bibinfo
  {journal} {Classical and Quantum Gravity}\ }\textbf {\bibinfo {volume}
  {15}},\ \bibinfo {pages} {1767} (\bibinfo {year}
  {1998}{\natexlab{b}})}\BibitemShut {NoStop}%
\bibitem [{\citenamefont {Almeida}\ and\ \citenamefont
  {Jacquet}(2022)}]{almeida2022analogue}%
  \BibitemOpen
  \bibfield  {author} {\bibinfo {author} {\bibfnamefont {C.~R.}\ \bibnamefont
  {Almeida}}\ and\ \bibinfo {author} {\bibfnamefont {M.~J.}\ \bibnamefont
  {Jacquet}},\ }\href@noop {} {\bibinfo {title} {Analogue gravity and the
  {H}awking effect: historical perspective and literature review}} (\bibinfo
  {year} {2022}),\ \Eprint {https://arxiv.org/abs/2212.08838} {arXiv:2212.08838
  [physics.hist-ph]} \BibitemShut {NoStop}%
\bibitem [{\citenamefont {Rousseaux}\ \emph {et~al.}(2008)\citenamefont
  {Rousseaux}, \citenamefont {Mathis}, \citenamefont {Maïssa}, \citenamefont
  {Philbin},\ and\ \citenamefont {Leonhardt}}]{rousseaux2008}%
  \BibitemOpen
  \bibfield  {author} {\bibinfo {author} {\bibfnamefont {G.}~\bibnamefont
  {Rousseaux}}, \bibinfo {author} {\bibfnamefont {C.}~\bibnamefont {Mathis}},
  \bibinfo {author} {\bibfnamefont {P.}~\bibnamefont {Maïssa}}, \bibinfo
  {author} {\bibfnamefont {T.~G.}\ \bibnamefont {Philbin}},\ and\ \bibinfo
  {author} {\bibfnamefont {U.}~\bibnamefont {Leonhardt}},\ }\bibfield  {title}
  {\bibinfo {title} {Observation of negative-frequency waves in a water tank: a
  classical analogue to the {H}awking effect?},\ }\href
  {https://doi.org/10.1088/1367-2630/10/5/053015} {\bibfield  {journal}
  {\bibinfo  {journal} {New Journal of Physics}\ }\textbf {\bibinfo {volume}
  {10}},\ \bibinfo {pages} {053015} (\bibinfo {year} {2008})}\BibitemShut
  {NoStop}%
\bibitem [{\citenamefont {Rousseaux}\ \emph {et~al.}(2010)\citenamefont
  {Rousseaux}, \citenamefont {Maïssa}, \citenamefont {Mathis}, \citenamefont
  {Coullet}, \citenamefont {Philbin},\ and\ \citenamefont
  {Leonhardt}}]{rousseaux2010}%
  \BibitemOpen
  \bibfield  {author} {\bibinfo {author} {\bibfnamefont {G.}~\bibnamefont
  {Rousseaux}}, \bibinfo {author} {\bibfnamefont {P.}~\bibnamefont {Maïssa}},
  \bibinfo {author} {\bibfnamefont {C.}~\bibnamefont {Mathis}}, \bibinfo
  {author} {\bibfnamefont {P.}~\bibnamefont {Coullet}}, \bibinfo {author}
  {\bibfnamefont {T.~G.}\ \bibnamefont {Philbin}},\ and\ \bibinfo {author}
  {\bibfnamefont {U.}~\bibnamefont {Leonhardt}},\ }\bibfield  {title} {\bibinfo
  {title} {Horizon effects with surface waves on moving water},\ }\href
  {https://doi.org/10.1088/1367-2630/12/9/095018} {\bibfield  {journal}
  {\bibinfo  {journal} {New Journal of Physics}\ }\textbf {\bibinfo {volume}
  {12}},\ \bibinfo {pages} {095018} (\bibinfo {year} {2010})}\BibitemShut
  {NoStop}%
\bibitem [{\citenamefont {Lahav}\ \emph {et~al.}(2010)\citenamefont {Lahav},
  \citenamefont {Itah}, \citenamefont {Blumkin}, \citenamefont {Gordon},
  \citenamefont {Rinott}, \citenamefont {Zayats},\ and\ \citenamefont
  {Steinhauer}}]{lahav2010}%
  \BibitemOpen
  \bibfield  {author} {\bibinfo {author} {\bibfnamefont {O.}~\bibnamefont
  {Lahav}}, \bibinfo {author} {\bibfnamefont {A.}~\bibnamefont {Itah}},
  \bibinfo {author} {\bibfnamefont {A.}~\bibnamefont {Blumkin}}, \bibinfo
  {author} {\bibfnamefont {C.}~\bibnamefont {Gordon}}, \bibinfo {author}
  {\bibfnamefont {S.}~\bibnamefont {Rinott}}, \bibinfo {author} {\bibfnamefont
  {A.}~\bibnamefont {Zayats}},\ and\ \bibinfo {author} {\bibfnamefont
  {J.}~\bibnamefont {Steinhauer}},\ }\bibfield  {title} {\bibinfo {title}
  {Realization of a sonic black hole analog in a {B}ose-{E}instein
  condensate},\ }\href {https://doi.org/10.1103/PhysRevLett.105.240401}
  {\bibfield  {journal} {\bibinfo  {journal} {Phys. Rev. Lett.}\ }\textbf
  {\bibinfo {volume} {105}},\ \bibinfo {pages} {240401} (\bibinfo {year}
  {2010})}\BibitemShut {NoStop}%
\bibitem [{\citenamefont {Elazar}\ \emph {et~al.}(2012)\citenamefont {Elazar},
  \citenamefont {Fleurov},\ and\ \citenamefont {Bar-Ad}}]{elazar}%
  \BibitemOpen
  \bibfield  {author} {\bibinfo {author} {\bibfnamefont {M.}~\bibnamefont
  {Elazar}}, \bibinfo {author} {\bibfnamefont {V.}~\bibnamefont {Fleurov}},\
  and\ \bibinfo {author} {\bibfnamefont {S.}~\bibnamefont {Bar-Ad}},\
  }\bibfield  {title} {\bibinfo {title} {All-optical event horizon in an
  optical analog of a laval nozzle},\ }\href
  {https://doi.org/10.1103/PhysRevA.86.063821} {\bibfield  {journal} {\bibinfo
  {journal} {Phys. Rev. A}\ }\textbf {\bibinfo {volume} {86}},\ \bibinfo
  {pages} {063821} (\bibinfo {year} {2012})}\BibitemShut {NoStop}%
\bibitem [{\citenamefont {Nguyen}\ \emph {et~al.}(2015)\citenamefont {Nguyen},
  \citenamefont {Gerace}, \citenamefont {Carusotto}, \citenamefont {Sanvitto},
  \citenamefont {Galopin}, \citenamefont {Lema\^{\i}tre}, \citenamefont
  {Sagnes}, \citenamefont {Bloch},\ and\ \citenamefont {Amo}}]{nguyen}%
  \BibitemOpen
  \bibfield  {author} {\bibinfo {author} {\bibfnamefont {H.~S.}\ \bibnamefont
  {Nguyen}}, \bibinfo {author} {\bibfnamefont {D.}~\bibnamefont {Gerace}},
  \bibinfo {author} {\bibfnamefont {I.}~\bibnamefont {Carusotto}}, \bibinfo
  {author} {\bibfnamefont {D.}~\bibnamefont {Sanvitto}}, \bibinfo {author}
  {\bibfnamefont {E.}~\bibnamefont {Galopin}}, \bibinfo {author} {\bibfnamefont
  {A.}~\bibnamefont {Lema\^{\i}tre}}, \bibinfo {author} {\bibfnamefont
  {I.}~\bibnamefont {Sagnes}}, \bibinfo {author} {\bibfnamefont
  {J.}~\bibnamefont {Bloch}},\ and\ \bibinfo {author} {\bibfnamefont
  {A.}~\bibnamefont {Amo}},\ }\bibfield  {title} {\bibinfo {title} {Acoustic
  black hole in a stationary hydrodynamic flow of microcavity polaritons},\
  }\href {https://doi.org/10.1103/PhysRevLett.114.036402} {\bibfield  {journal}
  {\bibinfo  {journal} {Phys. Rev. Lett.}\ }\textbf {\bibinfo {volume} {114}},\
  \bibinfo {pages} {036402} (\bibinfo {year} {2015})}\BibitemShut {NoStop}%
\bibitem [{\citenamefont {\ifmmode \check{C}\else
  \v{C}\fi{}love\ifmmode~\check{c}\else \v{c}\fi{}ko}\ \emph
  {et~al.}(2019)\citenamefont {\ifmmode \check{C}\else
  \v{C}\fi{}love\ifmmode~\check{c}\else \v{c}\fi{}ko}, \citenamefont
  {Ga\ifmmode~\check{z}\else \v{z}\fi{}o}, \citenamefont {Kupka},\ and\
  \citenamefont {Skyba}}]{skyba}%
  \BibitemOpen
  \bibfield  {author} {\bibinfo {author} {\bibfnamefont {M.}~\bibnamefont
  {\ifmmode \check{C}\else \v{C}\fi{}love\ifmmode~\check{c}\else
  \v{c}\fi{}ko}}, \bibinfo {author} {\bibfnamefont {E.}~\bibnamefont
  {Ga\ifmmode~\check{z}\else \v{z}\fi{}o}}, \bibinfo {author} {\bibfnamefont
  {M.}~\bibnamefont {Kupka}},\ and\ \bibinfo {author} {\bibfnamefont
  {P.}~\bibnamefont {Skyba}},\ }\bibfield  {title} {\bibinfo {title} {Magnonic
  analog of black- and white-hole horizons in superfluid
  $^{3}\mathrm{He}\text{\ensuremath{-}}b$},\ }\href
  {https://doi.org/10.1103/PhysRevLett.123.161302} {\bibfield  {journal}
  {\bibinfo  {journal} {Phys. Rev. Lett.}\ }\textbf {\bibinfo {volume} {123}},\
  \bibinfo {pages} {161302} (\bibinfo {year} {2019})}\BibitemShut {NoStop}%
\bibitem [{\citenamefont {Weinfurtner}\ \emph {et~al.}(2011)\citenamefont
  {Weinfurtner}, \citenamefont {Tedford}, \citenamefont {Penrice},
  \citenamefont {Unruh},\ and\ \citenamefont {Lawrence}}]{weinfurtner2011}%
  \BibitemOpen
  \bibfield  {author} {\bibinfo {author} {\bibfnamefont {S.}~\bibnamefont
  {Weinfurtner}}, \bibinfo {author} {\bibfnamefont {E.~W.}\ \bibnamefont
  {Tedford}}, \bibinfo {author} {\bibfnamefont {M.~C.~J.}\ \bibnamefont
  {Penrice}}, \bibinfo {author} {\bibfnamefont {W.~G.}\ \bibnamefont {Unruh}},\
  and\ \bibinfo {author} {\bibfnamefont {G.~A.}\ \bibnamefont {Lawrence}},\
  }\bibfield  {title} {\bibinfo {title} {Measurement of stimulated {H}awking
  emission in an analogue system},\ }\href
  {https://doi.org/10.1103/PhysRevLett.106.021302} {\bibfield  {journal}
  {\bibinfo  {journal} {Phys. Rev. Lett.}\ }\textbf {\bibinfo {volume} {106}},\
  \bibinfo {pages} {021302} (\bibinfo {year} {2011})}\BibitemShut {NoStop}%
\bibitem [{\citenamefont {Euvé}\ and\ \citenamefont
  {Rousseaux}(2021)}]{euve2021nonlinear}%
  \BibitemOpen
  \bibfield  {author} {\bibinfo {author} {\bibfnamefont {L.-P.}\ \bibnamefont
  {Euvé}}\ and\ \bibinfo {author} {\bibfnamefont {G.}~\bibnamefont
  {Rousseaux}},\ }\href@noop {} {\bibinfo {title} {Non-linear processes and
  stimulated {H}awking radiation in hydrodynamics for decelerating subcritical
  free surface flows with a subluminal dispersion relation}} (\bibinfo {year}
  {2021}),\ \Eprint {https://arxiv.org/abs/2112.12504} {arXiv:2112.12504
  [gr-qc]} \BibitemShut {NoStop}%
\bibitem [{\citenamefont {Euv\'e}\ \emph {et~al.}(2016)\citenamefont {Euv\'e},
  \citenamefont {Michel}, \citenamefont {Parentani}, \citenamefont {Philbin},\
  and\ \citenamefont {Rousseaux}}]{euve2016}%
  \BibitemOpen
  \bibfield  {author} {\bibinfo {author} {\bibfnamefont {L.-P.}\ \bibnamefont
  {Euv\'e}}, \bibinfo {author} {\bibfnamefont {F.}~\bibnamefont {Michel}},
  \bibinfo {author} {\bibfnamefont {R.}~\bibnamefont {Parentani}}, \bibinfo
  {author} {\bibfnamefont {T.~G.}\ \bibnamefont {Philbin}},\ and\ \bibinfo
  {author} {\bibfnamefont {G.}~\bibnamefont {Rousseaux}},\ }\bibfield  {title}
  {\bibinfo {title} {Observation of noise correlated by the {H}awking effect in
  a water tank},\ }\href {https://doi.org/10.1103/PhysRevLett.117.121301}
  {\bibfield  {journal} {\bibinfo  {journal} {Phys. Rev. Lett.}\ }\textbf
  {\bibinfo {volume} {117}},\ \bibinfo {pages} {121301} (\bibinfo {year}
  {2016})}\BibitemShut {NoStop}%
\bibitem [{\citenamefont {Euv\'e}\ \emph {et~al.}(2020)\citenamefont {Euv\'e},
  \citenamefont {Robertson}, \citenamefont {James}, \citenamefont {Fabbri},\
  and\ \citenamefont {Rousseaux}}]{euve2020}%
  \BibitemOpen
  \bibfield  {author} {\bibinfo {author} {\bibfnamefont {L.-P.}\ \bibnamefont
  {Euv\'e}}, \bibinfo {author} {\bibfnamefont {S.}~\bibnamefont {Robertson}},
  \bibinfo {author} {\bibfnamefont {N.}~\bibnamefont {James}}, \bibinfo
  {author} {\bibfnamefont {A.}~\bibnamefont {Fabbri}},\ and\ \bibinfo {author}
  {\bibfnamefont {G.}~\bibnamefont {Rousseaux}},\ }\bibfield  {title} {\bibinfo
  {title} {Scattering of co-current surface waves on an analogue black hole},\
  }\href {https://doi.org/10.1103/PhysRevLett.124.141101} {\bibfield  {journal}
  {\bibinfo  {journal} {Phys. Rev. Lett.}\ }\textbf {\bibinfo {volume} {124}},\
  \bibinfo {pages} {141101} (\bibinfo {year} {2020})}\BibitemShut {NoStop}%
\bibitem [{\citenamefont {Fourdrinoy}\ \emph {et~al.}(2022)\citenamefont
  {Fourdrinoy}, \citenamefont {Robertson}, \citenamefont {James}, \citenamefont
  {Fabbri},\ and\ \citenamefont {Rousseaux}}]{PhysRevD.105.085022}%
  \BibitemOpen
  \bibfield  {author} {\bibinfo {author} {\bibfnamefont {J.}~\bibnamefont
  {Fourdrinoy}}, \bibinfo {author} {\bibfnamefont {S.}~\bibnamefont
  {Robertson}}, \bibinfo {author} {\bibfnamefont {N.}~\bibnamefont {James}},
  \bibinfo {author} {\bibfnamefont {A.}~\bibnamefont {Fabbri}},\ and\ \bibinfo
  {author} {\bibfnamefont {G.}~\bibnamefont {Rousseaux}},\ }\bibfield  {title}
  {\bibinfo {title} {Correlations on weakly time-dependent transcritical
  white-hole flows},\ }\href {https://doi.org/10.1103/PhysRevD.105.085022}
  {\bibfield  {journal} {\bibinfo  {journal} {Phys. Rev. D}\ }\textbf {\bibinfo
  {volume} {105}},\ \bibinfo {pages} {085022} (\bibinfo {year}
  {2022})}\BibitemShut {NoStop}%
\bibitem [{\citenamefont {Steinhauer}(2016)}]{steinhauer2016}%
  \BibitemOpen
  \bibfield  {author} {\bibinfo {author} {\bibfnamefont {J.}~\bibnamefont
  {Steinhauer}},\ }\bibfield  {title} {\bibinfo {title} {Observation of quantum
  {H}awking radiation and its entanglement in an analogue black hole},\ }\href
  {https://doi.org/10.1038/nphys3863} {\bibfield  {journal} {\bibinfo
  {journal} {Nature Physics}\ }\textbf {\bibinfo {volume} {12}},\ \bibinfo
  {pages} {959} (\bibinfo {year} {2016})}\BibitemShut {NoStop}%
\bibitem [{\citenamefont {de~Nova}\ \emph {et~al.}(2019)\citenamefont
  {de~Nova}, \citenamefont {Golubkov}, \citenamefont {Kolobov},\ and\
  \citenamefont {Steinhauer}}]{deNova}%
  \BibitemOpen
  \bibfield  {author} {\bibinfo {author} {\bibfnamefont {J.~R.~M.}\
  \bibnamefont {de~Nova}}, \bibinfo {author} {\bibfnamefont {K.}~\bibnamefont
  {Golubkov}}, \bibinfo {author} {\bibfnamefont {V.~I.}\ \bibnamefont
  {Kolobov}},\ and\ \bibinfo {author} {\bibfnamefont {J.}~\bibnamefont
  {Steinhauer}},\ }\bibfield  {title} {\bibinfo {title} {Observation of thermal
  {H}awking radiation and its temperature in an analogue black hole},\ }\href
  {https://doi.org/10.1038/s41586-019-1241-0} {\bibfield  {journal} {\bibinfo
  {journal} {Nature}\ }\textbf {\bibinfo {volume} {569}},\ \bibinfo {pages}
  {688} (\bibinfo {year} {2019})}\BibitemShut {NoStop}%
\bibitem [{\citenamefont {Kolobov}\ \emph {et~al.}(2021)\citenamefont
  {Kolobov}, \citenamefont {Golubkov}, \citenamefont {de~Nova},\ and\
  \citenamefont {Steinhauer}}]{kolobov}%
  \BibitemOpen
  \bibfield  {author} {\bibinfo {author} {\bibfnamefont {V.~I.}\ \bibnamefont
  {Kolobov}}, \bibinfo {author} {\bibfnamefont {K.}~\bibnamefont {Golubkov}},
  \bibinfo {author} {\bibfnamefont {J.~R.~M.}\ \bibnamefont {de~Nova}},\ and\
  \bibinfo {author} {\bibfnamefont {J.}~\bibnamefont {Steinhauer}},\ }\bibfield
   {title} {\bibinfo {title} {Observation of stationary spontaneous {H}awking
  radiation and the time evolution of an analogue black hole},\ }\href
  {https://doi.org/10.1038/s41567-020-01076-0} {\bibfield  {journal} {\bibinfo
  {journal} {Nature Physics}\ }\textbf {\bibinfo {volume} {17}},\ \bibinfo
  {pages} {362} (\bibinfo {year} {2021})}\BibitemShut {NoStop}%
\bibitem [{\citenamefont {Hu}\ \emph {et~al.}()\citenamefont {Hu},
  \citenamefont {Feng}, \citenamefont {Zhang},\ and\ \citenamefont
  {Chin}}]{hu2019quantum}%
  \BibitemOpen
  \bibfield  {author} {\bibinfo {author} {\bibfnamefont {J.}~\bibnamefont
  {Hu}}, \bibinfo {author} {\bibfnamefont {L.}~\bibnamefont {Feng}}, \bibinfo
  {author} {\bibfnamefont {Z.}~\bibnamefont {Zhang}},\ and\ \bibinfo {author}
  {\bibfnamefont {C.}~\bibnamefont {Chin}},\ }\bibfield  {title} {\bibinfo
  {title} {Quantum simulation of {U}nruh radiation},\ }\bibfield  {journal}
  {\bibinfo  {journal} {Nature Physics}\ }\href
  {https://doi.org/https://doi.org/10.1038/s41567-019-0537-1}
  {https://doi.org/10.1038/s41567-019-0537-1}\BibitemShut {NoStop}%
\bibitem [{\citenamefont {Vocke}\ \emph {et~al.}(2018)\citenamefont {Vocke},
  \citenamefont {Maitland}, \citenamefont {Prain}, \citenamefont {Wilson},
  \citenamefont {Biancalana}, \citenamefont {Wright}, \citenamefont {Marino},\
  and\ \citenamefont {Faccio}}]{vocke2018}%
  \BibitemOpen
  \bibfield  {author} {\bibinfo {author} {\bibfnamefont {D.}~\bibnamefont
  {Vocke}}, \bibinfo {author} {\bibfnamefont {C.}~\bibnamefont {Maitland}},
  \bibinfo {author} {\bibfnamefont {A.}~\bibnamefont {Prain}}, \bibinfo
  {author} {\bibfnamefont {K.~E.}\ \bibnamefont {Wilson}}, \bibinfo {author}
  {\bibfnamefont {F.}~\bibnamefont {Biancalana}}, \bibinfo {author}
  {\bibfnamefont {E.~M.}\ \bibnamefont {Wright}}, \bibinfo {author}
  {\bibfnamefont {F.}~\bibnamefont {Marino}},\ and\ \bibinfo {author}
  {\bibfnamefont {D.}~\bibnamefont {Faccio}},\ }\bibfield  {title} {\bibinfo
  {title} {Rotating black hole geometries in a two-dimensional photon
  superfluid},\ }\href {https://doi.org/10.1364/OPTICA.5.001099} {\bibfield
  {journal} {\bibinfo  {journal} {Optica}\ }\textbf {\bibinfo {volume} {5}},\
  \bibinfo {pages} {1099} (\bibinfo {year} {2018})}\BibitemShut {NoStop}%
\bibitem [{\citenamefont {Torres}\ \emph {et~al.}(2017)\citenamefont {Torres},
  \citenamefont {Patrick}, \citenamefont {Coutant}, \citenamefont {Richartz},
  \citenamefont {Tedford},\ and\ \citenamefont {Weinfurtner}}]{torres}%
  \BibitemOpen
  \bibfield  {author} {\bibinfo {author} {\bibfnamefont {T.}~\bibnamefont
  {Torres}}, \bibinfo {author} {\bibfnamefont {S.}~\bibnamefont {Patrick}},
  \bibinfo {author} {\bibfnamefont {A.}~\bibnamefont {Coutant}}, \bibinfo
  {author} {\bibfnamefont {M.}~\bibnamefont {Richartz}}, \bibinfo {author}
  {\bibfnamefont {E.~W.}\ \bibnamefont {Tedford}},\ and\ \bibinfo {author}
  {\bibfnamefont {S.}~\bibnamefont {Weinfurtner}},\ }\bibfield  {title}
  {\bibinfo {title} {Rotational superradiant scattering in a vortex flow},\
  }\href {https://doi.org/10.1038/nphys4151} {\bibfield  {journal} {\bibinfo
  {journal} {Nature Physics}\ }\textbf {\bibinfo {volume} {13}},\ \bibinfo
  {pages} {833} (\bibinfo {year} {2017})}\BibitemShut {NoStop}%
\bibitem [{\citenamefont {Torres}\ \emph {et~al.}(2020)\citenamefont {Torres},
  \citenamefont {Patrick}, \citenamefont {Richartz},\ and\ \citenamefont
  {Weinfurtner}}]{torres2020}%
  \BibitemOpen
  \bibfield  {author} {\bibinfo {author} {\bibfnamefont {T.}~\bibnamefont
  {Torres}}, \bibinfo {author} {\bibfnamefont {S.}~\bibnamefont {Patrick}},
  \bibinfo {author} {\bibfnamefont {M.}~\bibnamefont {Richartz}},\ and\
  \bibinfo {author} {\bibfnamefont {S.}~\bibnamefont {Weinfurtner}},\
  }\bibfield  {title} {\bibinfo {title} {Quasinormal mode oscillations in an
  analogue black hole experiment},\ }\href
  {https://doi.org/10.1103/PhysRevLett.125.011301} {\bibfield  {journal}
  {\bibinfo  {journal} {Phys. Rev. Lett.}\ }\textbf {\bibinfo {volume} {125}},\
  \bibinfo {pages} {011301} (\bibinfo {year} {2020})}\BibitemShut {NoStop}%
\bibitem [{\citenamefont {Patrick}()}]{patrick2021rotational}%
  \BibitemOpen
  \bibfield  {author} {\bibinfo {author} {\bibfnamefont {S.}~\bibnamefont
  {Patrick}},\ }\bibfield  {title} {\bibinfo {title} {Rotational superradiance
  with {B}ogoliubov dispersion},\ }\bibfield  {journal} {\bibinfo  {journal}
  {Classical and Quantum Gravity}\ }\textbf {\bibinfo {volume} {38}},\ \href
  {https://doi.org/10.1088/1361-6382/abf1fc}
  {10.1088/1361-6382/abf1fc}\BibitemShut {NoStop}%
\bibitem [{\citenamefont {Cromb}\ \emph {et~al.}(2020)\citenamefont {Cromb},
  \citenamefont {Gibson}, \citenamefont {Toninelli}, \citenamefont {Padgett},
  \citenamefont {Wright},\ and\ \citenamefont {Faccio}}]{cromb}%
  \BibitemOpen
  \bibfield  {author} {\bibinfo {author} {\bibfnamefont {M.}~\bibnamefont
  {Cromb}}, \bibinfo {author} {\bibfnamefont {G.~M.}\ \bibnamefont {Gibson}},
  \bibinfo {author} {\bibfnamefont {E.}~\bibnamefont {Toninelli}}, \bibinfo
  {author} {\bibfnamefont {M.~J.}\ \bibnamefont {Padgett}}, \bibinfo {author}
  {\bibfnamefont {E.~M.}\ \bibnamefont {Wright}},\ and\ \bibinfo {author}
  {\bibfnamefont {D.}~\bibnamefont {Faccio}},\ }\bibfield  {title} {\bibinfo
  {title} {Amplification of waves from a rotating body},\ }\href
  {https://doi.org/10.1038/s41567-020-0944-3} {\bibfield  {journal} {\bibinfo
  {journal} {Nature Physics}\ }\textbf {\bibinfo {volume} {16}},\ \bibinfo
  {pages} {1069} (\bibinfo {year} {2020})}\BibitemShut {NoStop}%
\bibitem [{\citenamefont {Braidotti}\ \emph
  {et~al.}(2022{\natexlab{a}})\citenamefont {Braidotti}, \citenamefont
  {Prizia}, \citenamefont {Maitland}, \citenamefont {Marino}, \citenamefont
  {Prain}, \citenamefont {Starshynov}, \citenamefont {Westerberg},
  \citenamefont {Wright},\ and\ \citenamefont {Faccio}}]{braidotti2022}%
  \BibitemOpen
  \bibfield  {author} {\bibinfo {author} {\bibfnamefont {M.~C.}\ \bibnamefont
  {Braidotti}}, \bibinfo {author} {\bibfnamefont {R.}~\bibnamefont {Prizia}},
  \bibinfo {author} {\bibfnamefont {C.}~\bibnamefont {Maitland}}, \bibinfo
  {author} {\bibfnamefont {F.}~\bibnamefont {Marino}}, \bibinfo {author}
  {\bibfnamefont {A.}~\bibnamefont {Prain}}, \bibinfo {author} {\bibfnamefont
  {I.}~\bibnamefont {Starshynov}}, \bibinfo {author} {\bibfnamefont
  {N.}~\bibnamefont {Westerberg}}, \bibinfo {author} {\bibfnamefont {E.~M.}\
  \bibnamefont {Wright}},\ and\ \bibinfo {author} {\bibfnamefont
  {D.}~\bibnamefont {Faccio}},\ }\bibfield  {title} {\bibinfo {title}
  {Measurement of {P}enrose superradiance in a photon superfluid},\ }\href
  {https://doi.org/10.1103/PhysRevLett.128.013901} {\bibfield  {journal}
  {\bibinfo  {journal} {Phys. Rev. Lett.}\ }\textbf {\bibinfo {volume} {128}},\
  \bibinfo {pages} {013901} (\bibinfo {year} {2022}{\natexlab{a}})}\BibitemShut
  {NoStop}%
\bibitem [{\citenamefont {Braidotti}\ \emph
  {et~al.}(2022{\natexlab{b}})\citenamefont {Braidotti}, \citenamefont
  {Marino}, \citenamefont {Wright},\ and\ \citenamefont
  {Faccio}}]{braidottiavs}%
  \BibitemOpen
  \bibfield  {author} {\bibinfo {author} {\bibfnamefont {M.~C.}\ \bibnamefont
  {Braidotti}}, \bibinfo {author} {\bibfnamefont {F.}~\bibnamefont {Marino}},
  \bibinfo {author} {\bibfnamefont {E.~M.}\ \bibnamefont {Wright}},\ and\
  \bibinfo {author} {\bibfnamefont {D.}~\bibnamefont {Faccio}},\ }\bibfield
  {title} {\bibinfo {title} {The {P}enrose process in nonlinear optics},\
  }\href {https://doi.org/10.1116/5.0073218} {\bibfield  {journal} {\bibinfo
  {journal} {AVS Quantum Science}\ }\textbf {\bibinfo {volume} {4}},\ \bibinfo
  {pages} {010501} (\bibinfo {year} {2022}{\natexlab{b}})},\ \Eprint
  {https://arxiv.org/abs/https://doi.org/10.1116/5.0073218}
  {https://doi.org/10.1116/5.0073218} \BibitemShut {NoStop}%
\bibitem [{\citenamefont {Eckel}\ \emph {et~al.}(2018)\citenamefont {Eckel},
  \citenamefont {Kumar}, \citenamefont {Jacobson}, \citenamefont {Spielman},\
  and\ \citenamefont {Campbell}}]{eckel}%
  \BibitemOpen
  \bibfield  {author} {\bibinfo {author} {\bibfnamefont {S.}~\bibnamefont
  {Eckel}}, \bibinfo {author} {\bibfnamefont {A.}~\bibnamefont {Kumar}},
  \bibinfo {author} {\bibfnamefont {T.}~\bibnamefont {Jacobson}}, \bibinfo
  {author} {\bibfnamefont {I.~B.}\ \bibnamefont {Spielman}},\ and\ \bibinfo
  {author} {\bibfnamefont {G.~K.}\ \bibnamefont {Campbell}},\ }\bibfield
  {title} {\bibinfo {title} {A rapidly expanding {B}ose-{E}instein condensate:
  An expanding universe in the lab},\ }\href
  {https://doi.org/10.1103/PhysRevX.8.021021} {\bibfield  {journal} {\bibinfo
  {journal} {Phys. Rev. X}\ }\textbf {\bibinfo {volume} {8}},\ \bibinfo {pages}
  {021021} (\bibinfo {year} {2018})}\BibitemShut {NoStop}%
\bibitem [{\citenamefont {Steinhauer}\ \emph {et~al.}(2022)\citenamefont
  {Steinhauer}, \citenamefont {Abuzarli}, \citenamefont {Aladjidi},
  \citenamefont {Bienaim\'{e}}, \citenamefont {Piekarski}, \citenamefont {Liu},
  \citenamefont {Giacobino}, \citenamefont {Bramati},\ and\ \citenamefont
  {Glorieux}}]{glorieux2022}%
  \BibitemOpen
  \bibfield  {author} {\bibinfo {author} {\bibfnamefont {J.}~\bibnamefont
  {Steinhauer}}, \bibinfo {author} {\bibfnamefont {M.}~\bibnamefont
  {Abuzarli}}, \bibinfo {author} {\bibfnamefont {T.}~\bibnamefont {Aladjidi}},
  \bibinfo {author} {\bibfnamefont {T.}~\bibnamefont {Bienaim\'{e}}}, \bibinfo
  {author} {\bibfnamefont {C.}~\bibnamefont {Piekarski}}, \bibinfo {author}
  {\bibfnamefont {W.}~\bibnamefont {Liu}}, \bibinfo {author} {\bibfnamefont
  {E.}~\bibnamefont {Giacobino}}, \bibinfo {author} {\bibfnamefont
  {A.}~\bibnamefont {Bramati}},\ and\ \bibinfo {author} {\bibfnamefont
  {Q.}~\bibnamefont {Glorieux}},\ }\bibfield  {title} {\bibinfo {title}
  {Analogue cosmological particle creation in an ultracold quantum fluid of
  light},\ }\href {https://doi.org/https://doi.org/10.1038/s41467-022-30603-1}
  {\bibfield  {journal} {\bibinfo  {journal} {Nature Comm.}\ }\textbf {\bibinfo
  {volume} {13}},\ \bibinfo {pages} {2890} (\bibinfo {year}
  {2022})}\BibitemShut {NoStop}%
\bibitem [{\citenamefont {Hawking}(1976)}]{hawking3}%
  \BibitemOpen
  \bibfield  {author} {\bibinfo {author} {\bibfnamefont {S.~W.}\ \bibnamefont
  {Hawking}},\ }\bibfield  {title} {\bibinfo {title} {Breakdown of
  predictability in gravitational collapse},\ }\href
  {https://doi.org/10.1103/PhysRevD.14.2460} {\bibfield  {journal} {\bibinfo
  {journal} {Phys. Rev. D}\ }\textbf {\bibinfo {volume} {14}},\ \bibinfo
  {pages} {2460} (\bibinfo {year} {1976})}\BibitemShut {NoStop}%
\bibitem [{\citenamefont {Raju}(2022)}]{RAJU20221}%
  \BibitemOpen
  \bibfield  {author} {\bibinfo {author} {\bibfnamefont {S.}~\bibnamefont
  {Raju}},\ }\bibfield  {title} {\bibinfo {title} {Lessons from the information
  paradox},\ }\href
  {https://doi.org/https://doi.org/10.1016/j.physrep.2021.10.001} {\bibfield
  {journal} {\bibinfo  {journal} {Physics Reports}\ }\textbf {\bibinfo {volume}
  {943}},\ \bibinfo {pages} {1} (\bibinfo {year} {2022})}\BibitemShut {NoStop}%
\bibitem [{\citenamefont {Vachaspati}\ \emph {et~al.}(2007)\citenamefont
  {Vachaspati}, \citenamefont {Stojkovic},\ and\ \citenamefont
  {Krauss}}]{PhysRevD.76.024005}%
  \BibitemOpen
  \bibfield  {author} {\bibinfo {author} {\bibfnamefont {T.}~\bibnamefont
  {Vachaspati}}, \bibinfo {author} {\bibfnamefont {D.}~\bibnamefont
  {Stojkovic}},\ and\ \bibinfo {author} {\bibfnamefont {L.~M.}\ \bibnamefont
  {Krauss}},\ }\bibfield  {title} {\bibinfo {title} {Observation of incipient
  black holes and the information loss problem},\ }\href
  {https://doi.org/10.1103/PhysRevD.76.024005} {\bibfield  {journal} {\bibinfo
  {journal} {Phys. Rev. D}\ }\textbf {\bibinfo {volume} {76}},\ \bibinfo
  {pages} {024005} (\bibinfo {year} {2007})}\BibitemShut {NoStop}%
\bibitem [{\citenamefont {Sch\"utzhold}\ \emph {et~al.}(2005)\citenamefont
  {Sch\"utzhold}, \citenamefont {Uhlmann}, \citenamefont {Xu},\ and\
  \citenamefont {Fischer}}]{PhysRevD.72.105005}%
  \BibitemOpen
  \bibfield  {author} {\bibinfo {author} {\bibfnamefont {R.}~\bibnamefont
  {Sch\"utzhold}}, \bibinfo {author} {\bibfnamefont {M.}~\bibnamefont
  {Uhlmann}}, \bibinfo {author} {\bibfnamefont {Y.}~\bibnamefont {Xu}},\ and\
  \bibinfo {author} {\bibfnamefont {U.~R.}\ \bibnamefont {Fischer}},\
  }\bibfield  {title} {\bibinfo {title} {Quantum backreaction in dilute
  {B}ose-{E}instein condensates},\ }\href
  {https://doi.org/10.1103/PhysRevD.72.105005} {\bibfield  {journal} {\bibinfo
  {journal} {Phys. Rev. D}\ }\textbf {\bibinfo {volume} {72}},\ \bibinfo
  {pages} {105005} (\bibinfo {year} {2005})}\BibitemShut {NoStop}%
\bibitem [{\citenamefont {Baak}\ \emph {et~al.}(2022)\citenamefont {Baak},
  \citenamefont {Ribeiro},\ and\ \citenamefont
  {Fischer}}]{PhysRevA.106.053319}%
  \BibitemOpen
  \bibfield  {author} {\bibinfo {author} {\bibfnamefont {S.-S.}\ \bibnamefont
  {Baak}}, \bibinfo {author} {\bibfnamefont {C.~C.~H.}\ \bibnamefont
  {Ribeiro}},\ and\ \bibinfo {author} {\bibfnamefont {U.~R.}\ \bibnamefont
  {Fischer}},\ }\bibfield  {title} {\bibinfo {title} {Number-conserving
  solution for dynamical quantum backreaction in a {B}ose-{E}instein
  condensate},\ }\href {https://doi.org/10.1103/PhysRevA.106.053319} {\bibfield
   {journal} {\bibinfo  {journal} {Phys. Rev. A}\ }\textbf {\bibinfo {volume}
  {106}},\ \bibinfo {pages} {053319} (\bibinfo {year} {2022})}\BibitemShut
  {NoStop}%
\bibitem [{\citenamefont {Liberati}\ \emph {et~al.}(2019)\citenamefont
  {Liberati}, \citenamefont {Tricella},\ and\ \citenamefont
  {Trombettoni}}]{liberati19}%
  \BibitemOpen
  \bibfield  {author} {\bibinfo {author} {\bibfnamefont {S.}~\bibnamefont
  {Liberati}}, \bibinfo {author} {\bibfnamefont {G.}~\bibnamefont {Tricella}},\
  and\ \bibinfo {author} {\bibfnamefont {A.}~\bibnamefont {Trombettoni}},\
  }\bibfield  {title} {\bibinfo {title} {The information loss problem: An
  analogue gravity perspective},\ }\href {https://doi.org/10.3390/e21100940}
  {\bibfield  {journal} {\bibinfo  {journal} {Entropy}\ }\textbf {\bibinfo
  {volume} {21}},\ \bibinfo {pages} {940} (\bibinfo {year} {2019})}\BibitemShut
  {NoStop}%
\bibitem [{\citenamefont {Perez}(2015)}]{perez}%
  \BibitemOpen
  \bibfield  {author} {\bibinfo {author} {\bibfnamefont {A.}~\bibnamefont
  {Perez}},\ }\bibfield  {title} {\bibinfo {title} {No firewalls in quantum
  gravity: the role of discreteness of quantum geometry in resolving the
  information loss paradox},\ }\href
  {https://doi.org/10.1088/0264-9381/32/8/084001} {\bibfield  {journal}
  {\bibinfo  {journal} {Classical and Quantum Gravity}\ }\textbf {\bibinfo
  {volume} {32}},\ \bibinfo {pages} {084001} (\bibinfo {year}
  {2015})}\BibitemShut {NoStop}%
\bibitem [{\citenamefont {Jacquet}\ \emph {et~al.}(2023)\citenamefont
  {Jacquet}, \citenamefont {Giacomelli}, \citenamefont {Valnais}, \citenamefont
  {Joly}, \citenamefont {Claude}, \citenamefont {Giacobino}, \citenamefont
  {Glorieux}, \citenamefont {Carusotto},\ and\ \citenamefont
  {Bramati}}]{jacquet23}%
  \BibitemOpen
  \bibfield  {author} {\bibinfo {author} {\bibfnamefont {M.~J.}\ \bibnamefont
  {Jacquet}}, \bibinfo {author} {\bibfnamefont {L.}~\bibnamefont {Giacomelli}},
  \bibinfo {author} {\bibfnamefont {Q.}~\bibnamefont {Valnais}}, \bibinfo
  {author} {\bibfnamefont {M.}~\bibnamefont {Joly}}, \bibinfo {author}
  {\bibfnamefont {F.}~\bibnamefont {Claude}}, \bibinfo {author} {\bibfnamefont
  {E.}~\bibnamefont {Giacobino}}, \bibinfo {author} {\bibfnamefont
  {Q.}~\bibnamefont {Glorieux}}, \bibinfo {author} {\bibfnamefont
  {I.}~\bibnamefont {Carusotto}},\ and\ \bibinfo {author} {\bibfnamefont
  {A.}~\bibnamefont {Bramati}},\ }\bibfield  {title} {\bibinfo {title} {Quantum
  vacuum excitation of a quasinormal mode in an analog model of black hole
  spacetime},\ }\href {https://doi.org/10.1103/PhysRevLett.130.111501}
  {\bibfield  {journal} {\bibinfo  {journal} {Phys. Rev. Lett.}\ }\textbf
  {\bibinfo {volume} {130}},\ \bibinfo {pages} {111501} (\bibinfo {year}
  {2023})}\BibitemShut {NoStop}%
\bibitem [{\citenamefont {Calmet}\ \emph {et~al.}(2022)\citenamefont {Calmet},
  \citenamefont {Casadio}, \citenamefont {Hsu},\ and\ \citenamefont
  {Kuipers}}]{calmet}%
  \BibitemOpen
  \bibfield  {author} {\bibinfo {author} {\bibfnamefont {X.}~\bibnamefont
  {Calmet}}, \bibinfo {author} {\bibfnamefont {R.}~\bibnamefont {Casadio}},
  \bibinfo {author} {\bibfnamefont {S.~D.~H.}\ \bibnamefont {Hsu}},\ and\
  \bibinfo {author} {\bibfnamefont {F.}~\bibnamefont {Kuipers}},\ }\bibfield
  {title} {\bibinfo {title} {Quantum hair from gravity},\ }\href
  {https://doi.org/10.1103/PhysRevLett.128.111301} {\bibfield  {journal}
  {\bibinfo  {journal} {Phys. Rev. Lett.}\ }\textbf {\bibinfo {volume} {128}},\
  \bibinfo {pages} {111301} (\bibinfo {year} {2022})}\BibitemShut {NoStop}%
\bibitem [{\citenamefont {Calmet}\ and\ \citenamefont {Hsu}(2022)}]{calmet2}%
  \BibitemOpen
  \bibfield  {author} {\bibinfo {author} {\bibfnamefont {X.}~\bibnamefont
  {Calmet}}\ and\ \bibinfo {author} {\bibfnamefont {S.~D.}\ \bibnamefont
  {Hsu}},\ }\bibfield  {title} {\bibinfo {title} {Quantum hair and black hole
  information},\ }\href
  {https://doi.org/https://doi.org/10.1016/j.physletb.2022.136995} {\bibfield
  {journal} {\bibinfo  {journal} {Physics Letters B}\ }\textbf {\bibinfo
  {volume} {827}},\ \bibinfo {pages} {136995} (\bibinfo {year}
  {2022})}\BibitemShut {NoStop}%
\bibitem [{\citenamefont {Cheng}\ and\ \citenamefont {An}(2021)}]{cheng}%
  \BibitemOpen
  \bibfield  {author} {\bibinfo {author} {\bibfnamefont {P.}~\bibnamefont
  {Cheng}}\ and\ \bibinfo {author} {\bibfnamefont {Y.}~\bibnamefont {An}},\
  }\bibfield  {title} {\bibinfo {title} {Soft black hole information paradox:
  Page curve from {M}axwell soft hair of a black hole},\ }\href
  {https://doi.org/10.1103/PhysRevD.103.126020} {\bibfield  {journal} {\bibinfo
   {journal} {Phys. Rev. D}\ }\textbf {\bibinfo {volume} {103}},\ \bibinfo
  {pages} {126020} (\bibinfo {year} {2021})}\BibitemShut {NoStop}%
\bibitem [{\citenamefont {Krauss}\ and\ \citenamefont
  {Wilczek}(1989)}]{krausswilczek}%
  \BibitemOpen
  \bibfield  {author} {\bibinfo {author} {\bibfnamefont {L.~M.}\ \bibnamefont
  {Krauss}}\ and\ \bibinfo {author} {\bibfnamefont {F.}~\bibnamefont
  {Wilczek}},\ }\bibfield  {title} {\bibinfo {title} {Discrete gauge symmetry
  in continuum theories},\ }\href {https://doi.org/10.1103/PhysRevLett.62.1221}
  {\bibfield  {journal} {\bibinfo  {journal} {Phys. Rev. Lett.}\ }\textbf
  {\bibinfo {volume} {62}},\ \bibinfo {pages} {1221} (\bibinfo {year}
  {1989})}\BibitemShut {NoStop}%
\bibitem [{\citenamefont {Coleman}\ \emph {et~al.}(1992)\citenamefont
  {Coleman}, \citenamefont {Krauss}, \citenamefont {Preskill},\ and\
  \citenamefont {Wilczek}}]{colemanwilczek}%
  \BibitemOpen
  \bibfield  {author} {\bibinfo {author} {\bibfnamefont {S.~R.}\ \bibnamefont
  {Coleman}}, \bibinfo {author} {\bibfnamefont {L.~M.}\ \bibnamefont {Krauss}},
  \bibinfo {author} {\bibfnamefont {J.}~\bibnamefont {Preskill}},\ and\
  \bibinfo {author} {\bibfnamefont {F.}~\bibnamefont {Wilczek}},\ }\bibfield
  {title} {\bibinfo {title} {{Quantum hair and quantum gravity}},\ }\href
  {https://doi.org/10.1007/BF00756870} {\bibfield  {journal} {\bibinfo
  {journal} {Gen. Rel. Grav.}\ }\textbf {\bibinfo {volume} {24}},\ \bibinfo
  {pages} {9} (\bibinfo {year} {1992})}\BibitemShut {NoStop}%
\bibitem [{\citenamefont {Preskill}\ and\ \citenamefont
  {Krauss}(1990)}]{krausspreskill}%
  \BibitemOpen
  \bibfield  {author} {\bibinfo {author} {\bibfnamefont {J.}~\bibnamefont
  {Preskill}}\ and\ \bibinfo {author} {\bibfnamefont {L.~M.}\ \bibnamefont
  {Krauss}},\ }\bibfield  {title} {\bibinfo {title} {Local discrete symmetry
  and quantum-mechanical hair},\ }\href
  {https://doi.org/https://doi.org/10.1016/0550-3213(90)90262-C} {\bibfield
  {journal} {\bibinfo  {journal} {Nuclear Physics B}\ }\textbf {\bibinfo
  {volume} {341}},\ \bibinfo {pages} {50} (\bibinfo {year} {1990})}\BibitemShut
  {NoStop}%
\bibitem [{\citenamefont {Berry}\ \emph {et~al.}(1980)\citenamefont {Berry},
  \citenamefont {Chambers}, \citenamefont {Large}, \citenamefont {Upstill},\
  and\ \citenamefont {Walmsley}}]{berry}%
  \BibitemOpen
  \bibfield  {author} {\bibinfo {author} {\bibfnamefont {M.~V.}\ \bibnamefont
  {Berry}}, \bibinfo {author} {\bibfnamefont {R.~G.}\ \bibnamefont {Chambers}},
  \bibinfo {author} {\bibfnamefont {M.~D.}\ \bibnamefont {Large}}, \bibinfo
  {author} {\bibfnamefont {C.}~\bibnamefont {Upstill}},\ and\ \bibinfo {author}
  {\bibfnamefont {J.~C.}\ \bibnamefont {Walmsley}},\ }\bibfield  {title}
  {\bibinfo {title} {Wavefront dislocations in the {A}haronov-{B}ohm effect and
  its water wave analogue},\ }\href {https://doi.org/10.1088/0143-0807/1/3/008}
  {\bibfield  {journal} {\bibinfo  {journal} {European Journal of Physics}\
  }\textbf {\bibinfo {volume} {1}},\ \bibinfo {pages} {154} (\bibinfo {year}
  {1980})}\BibitemShut {NoStop}%
\bibitem [{\citenamefont {Vivanco}\ \emph {et~al.}(1999)\citenamefont
  {Vivanco}, \citenamefont {Melo}, \citenamefont {Coste},\ and\ \citenamefont
  {Lund}}]{vivanco}%
  \BibitemOpen
  \bibfield  {author} {\bibinfo {author} {\bibfnamefont {F.}~\bibnamefont
  {Vivanco}}, \bibinfo {author} {\bibfnamefont {F.}~\bibnamefont {Melo}},
  \bibinfo {author} {\bibfnamefont {C.}~\bibnamefont {Coste}},\ and\ \bibinfo
  {author} {\bibfnamefont {F.}~\bibnamefont {Lund}},\ }\bibfield  {title}
  {\bibinfo {title} {Surface wave scattering by a vertical vortex and the
  symmetry of the {A}haronov-{B}ohm wave function},\ }\href
  {https://doi.org/10.1103/PhysRevLett.83.1966} {\bibfield  {journal} {\bibinfo
   {journal} {Phys. Rev. Lett.}\ }\textbf {\bibinfo {volume} {83}},\ \bibinfo
  {pages} {1966} (\bibinfo {year} {1999})}\BibitemShut {NoStop}%
\bibitem [{\citenamefont {Kwon}\ \emph {et~al.}(2021)\citenamefont {Kwon},
  \citenamefont {Pace}, \citenamefont {Xhani}, \citenamefont {Galantucci},
  \citenamefont {Falconi}, \citenamefont {Inguscio}, \citenamefont {Scazza},\
  and\ \citenamefont {Roati}}]{Kwon_2021}%
  \BibitemOpen
  \bibfield  {author} {\bibinfo {author} {\bibfnamefont {W.~J.}\ \bibnamefont
  {Kwon}}, \bibinfo {author} {\bibfnamefont {G.~D.}\ \bibnamefont {Pace}},
  \bibinfo {author} {\bibfnamefont {K.}~\bibnamefont {Xhani}}, \bibinfo
  {author} {\bibfnamefont {L.}~\bibnamefont {Galantucci}}, \bibinfo {author}
  {\bibfnamefont {A.~M.}\ \bibnamefont {Falconi}}, \bibinfo {author}
  {\bibfnamefont {M.}~\bibnamefont {Inguscio}}, \bibinfo {author}
  {\bibfnamefont {F.}~\bibnamefont {Scazza}},\ and\ \bibinfo {author}
  {\bibfnamefont {G.}~\bibnamefont {Roati}},\ }\bibfield  {title} {\bibinfo
  {title} {Sound emission and annihilations in a programmable quantum vortex
  collider},\ }\href {https://doi.org/10.1038/s41586-021-04047-4} {\bibfield
  {journal} {\bibinfo  {journal} {Nature}\ }\textbf {\bibinfo {volume} {600}},\
  \bibinfo {pages} {64} (\bibinfo {year} {2021})}\BibitemShut {NoStop}%
\bibitem [{\citenamefont {Herdeiro}\ and\ \citenamefont
  {Radu}(2014)}]{herdeiro2014}%
  \BibitemOpen
  \bibfield  {author} {\bibinfo {author} {\bibfnamefont {C.~A.~R.}\
  \bibnamefont {Herdeiro}}\ and\ \bibinfo {author} {\bibfnamefont
  {E.}~\bibnamefont {Radu}},\ }\bibfield  {title} {\bibinfo {title} {Kerr black
  holes with scalar hair},\ }\href
  {https://doi.org/10.1103/PhysRevLett.112.221101} {\bibfield  {journal}
  {\bibinfo  {journal} {Phys. Rev. Lett.}\ }\textbf {\bibinfo {volume} {112}},\
  \bibinfo {pages} {221101} (\bibinfo {year} {2014})}\BibitemShut {NoStop}%
\bibitem [{\citenamefont {Benone}\ \emph {et~al.}(2014)\citenamefont {Benone},
  \citenamefont {Crispino}, \citenamefont {Herdeiro},\ and\ \citenamefont
  {Radu}}]{benone14}%
  \BibitemOpen
  \bibfield  {author} {\bibinfo {author} {\bibfnamefont {C.~L.}\ \bibnamefont
  {Benone}}, \bibinfo {author} {\bibfnamefont {L.~C.~B.}\ \bibnamefont
  {Crispino}}, \bibinfo {author} {\bibfnamefont {C.}~\bibnamefont {Herdeiro}},\
  and\ \bibinfo {author} {\bibfnamefont {E.}~\bibnamefont {Radu}},\ }\bibfield
  {title} {\bibinfo {title} {Kerr-newman scalar clouds},\ }\href
  {https://doi.org/10.1103/PhysRevD.90.104024} {\bibfield  {journal} {\bibinfo
  {journal} {Phys. Rev. D}\ }\textbf {\bibinfo {volume} {90}},\ \bibinfo
  {pages} {104024} (\bibinfo {year} {2014})}\BibitemShut {NoStop}%
\bibitem [{\citenamefont {Press}\ and\ \citenamefont
  {Teukolsky}(1972)}]{bhbomb}%
  \BibitemOpen
  \bibfield  {author} {\bibinfo {author} {\bibfnamefont {W.~H.}\ \bibnamefont
  {Press}}\ and\ \bibinfo {author} {\bibfnamefont {S.~A.}\ \bibnamefont
  {Teukolsky}},\ }\bibfield  {title} {\bibinfo {title} {{Floating Orbits,
  Superradiant scattering and the black-hole bomb}},\ }\href
  {https://doi.org/10.1038/238211a0} {\bibfield  {journal} {\bibinfo  {journal}
  {Nature}\ }\textbf {\bibinfo {volume} {238}},\ \bibinfo {pages} {211}
  (\bibinfo {year} {1972})}\BibitemShut {NoStop}%
\bibitem [{\citenamefont {Hod}(2012)}]{hod1}%
  \BibitemOpen
  \bibfield  {author} {\bibinfo {author} {\bibfnamefont {S.}~\bibnamefont
  {Hod}},\ }\bibfield  {title} {\bibinfo {title} {Stationary scalar clouds
  around rotating black holes},\ }\href
  {https://doi.org/10.1103/PhysRevD.86.104026} {\bibfield  {journal} {\bibinfo
  {journal} {Phys. Rev. D}\ }\textbf {\bibinfo {volume} {86}},\ \bibinfo
  {pages} {104026} (\bibinfo {year} {2012})}\BibitemShut {NoStop}%
\bibitem [{hod(2013)}]{hod2}%
  \BibitemOpen
  \href {https://doi.org/10.1140/epjc/s10052-013-2378-x} {\ \textbf {\bibinfo
  {volume} {73}},\ \bibinfo {pages} {2378} (\bibinfo {year}
  {2013})}\BibitemShut {NoStop}%
\bibitem [{\citenamefont {Girelli}\ \emph {et~al.}(2008)\citenamefont
  {Girelli}, \citenamefont {Liberati},\ and\ \citenamefont {Sindoni}}]{eg2}%
  \BibitemOpen
  \bibfield  {author} {\bibinfo {author} {\bibfnamefont {F.}~\bibnamefont
  {Girelli}}, \bibinfo {author} {\bibfnamefont {S.}~\bibnamefont {Liberati}},\
  and\ \bibinfo {author} {\bibfnamefont {L.}~\bibnamefont {Sindoni}},\
  }\bibfield  {title} {\bibinfo {title} {Gravitational dynamics in
  {B}ose-{E}instein condensates},\ }\href
  {https://doi.org/10.1103/PhysRevD.78.084013} {\bibfield  {journal} {\bibinfo
  {journal} {Phys. Rev. D}\ }\textbf {\bibinfo {volume} {78}},\ \bibinfo
  {pages} {084013} (\bibinfo {year} {2008})}\BibitemShut {NoStop}%
\bibitem [{\citenamefont {Marino}(2019)}]{marino2019}%
  \BibitemOpen
  \bibfield  {author} {\bibinfo {author} {\bibfnamefont {F.}~\bibnamefont
  {Marino}},\ }\bibfield  {title} {\bibinfo {title} {Massive phonons and
  gravitational dynamics in a photon-fluid model},\ }\href
  {https://doi.org/10.1103/PhysRevA.100.063825} {\bibfield  {journal} {\bibinfo
   {journal} {Phys. Rev. A}\ }\textbf {\bibinfo {volume} {100}},\ \bibinfo
  {pages} {063825} (\bibinfo {year} {2019})}\BibitemShut {NoStop}%
\bibitem [{\citenamefont {Fischer}\ and\ \citenamefont
  {Sch\"utzhold}(2004)}]{PhysRevA.70.063615}%
  \BibitemOpen
  \bibfield  {author} {\bibinfo {author} {\bibfnamefont {U.~R.}\ \bibnamefont
  {Fischer}}\ and\ \bibinfo {author} {\bibfnamefont {R.}~\bibnamefont
  {Sch\"utzhold}},\ }\bibfield  {title} {\bibinfo {title} {Quantum simulation
  of cosmic inflation in two-component bose-einstein condensates},\ }\href
  {https://doi.org/10.1103/PhysRevA.70.063615} {\bibfield  {journal} {\bibinfo
  {journal} {Phys. Rev. A}\ }\textbf {\bibinfo {volume} {70}},\ \bibinfo
  {pages} {063615} (\bibinfo {year} {2004})}\BibitemShut {NoStop}%
\bibitem [{\citenamefont {Visser}\ and\ \citenamefont
  {Weinfurtner}(2005)}]{silke-visser}%
  \BibitemOpen
  \bibfield  {author} {\bibinfo {author} {\bibfnamefont {M.}~\bibnamefont
  {Visser}}\ and\ \bibinfo {author} {\bibfnamefont {S.}~\bibnamefont
  {Weinfurtner}},\ }\bibfield  {title} {\bibinfo {title} {Massive
  {K}lein-{G}ordon equation from a {B}ose-{E}instein-condensation-based
  analogue spacetime},\ }\href {https://doi.org/10.1103/PhysRevD.72.044020}
  {\bibfield  {journal} {\bibinfo  {journal} {Phys. Rev. D}\ }\textbf {\bibinfo
  {volume} {72}},\ \bibinfo {pages} {044020} (\bibinfo {year}
  {2005})}\BibitemShut {NoStop}%
\bibitem [{\citenamefont {Cominotti}\ \emph {et~al.}(2022)\citenamefont
  {Cominotti}, \citenamefont {Berti}, \citenamefont {Farolfi}, \citenamefont
  {Zenesini}, \citenamefont {Lamporesi}, \citenamefont {Carusotto},
  \citenamefont {Recati},\ and\ \citenamefont {Ferrari}}]{ferrari22}%
  \BibitemOpen
  \bibfield  {author} {\bibinfo {author} {\bibfnamefont {R.}~\bibnamefont
  {Cominotti}}, \bibinfo {author} {\bibfnamefont {A.}~\bibnamefont {Berti}},
  \bibinfo {author} {\bibfnamefont {A.}~\bibnamefont {Farolfi}}, \bibinfo
  {author} {\bibfnamefont {A.}~\bibnamefont {Zenesini}}, \bibinfo {author}
  {\bibfnamefont {G.}~\bibnamefont {Lamporesi}}, \bibinfo {author}
  {\bibfnamefont {I.}~\bibnamefont {Carusotto}}, \bibinfo {author}
  {\bibfnamefont {A.}~\bibnamefont {Recati}},\ and\ \bibinfo {author}
  {\bibfnamefont {G.}~\bibnamefont {Ferrari}},\ }\bibfield  {title} {\bibinfo
  {title} {Observation of massless and massive collective excitations with
  faraday patterns in a two-component superfluid},\ }\href
  {https://doi.org/10.1103/PhysRevLett.128.210401} {\bibfield  {journal}
  {\bibinfo  {journal} {Phys. Rev. Lett.}\ }\textbf {\bibinfo {volume} {128}},\
  \bibinfo {pages} {210401} (\bibinfo {year} {2022})}\BibitemShut {NoStop}%
\bibitem [{\citenamefont {Ciszak}\ and\ \citenamefont
  {Marino}(2021)}]{marino2021}%
  \BibitemOpen
  \bibfield  {author} {\bibinfo {author} {\bibfnamefont {M.}~\bibnamefont
  {Ciszak}}\ and\ \bibinfo {author} {\bibfnamefont {F.}~\bibnamefont
  {Marino}},\ }\bibfield  {title} {\bibinfo {title} {Acoustic black-hole bombs
  and scalar clouds in a photon-fluid model},\ }\href
  {https://doi.org/10.1103/PhysRevD.103.045004} {\bibfield  {journal} {\bibinfo
   {journal} {Phys. Rev. D}\ }\textbf {\bibinfo {volume} {103}},\ \bibinfo
  {pages} {045004} (\bibinfo {year} {2021})}\BibitemShut {NoStop}%
\bibitem [{\citenamefont {Hod}(2021{\natexlab{a}})}]{hod3}%
  \BibitemOpen
  \bibfield  {author} {\bibinfo {author} {\bibfnamefont {S.}~\bibnamefont
  {Hod}},\ }\bibfield  {title} {\bibinfo {title} {Stationary scalar clouds
  supported by rapidly-rotating acoustic black holes in a photon-fluid model},\
  }\href {https://doi.org/10.1103/PhysRevD.103.084003} {\bibfield  {journal}
  {\bibinfo  {journal} {Phys. Rev. D}\ }\textbf {\bibinfo {volume} {103}},\
  \bibinfo {pages} {084003} (\bibinfo {year} {2021}{\natexlab{a}})}\BibitemShut
  {NoStop}%
\bibitem [{\citenamefont {Hod}(2021{\natexlab{b}})}]{hod4}%
  \BibitemOpen
  \bibfield  {author} {\bibinfo {author} {\bibfnamefont {S.}~\bibnamefont
  {Hod}},\ }\bibfield  {title} {\bibinfo {title} {No-short scalar hair theorem
  for spinning acoustic black holes in a photon-fluid model},\ }\href
  {https://doi.org/10.1103/PhysRevD.104.104041} {\bibfield  {journal} {\bibinfo
   {journal} {Phys. Rev. D}\ }\textbf {\bibinfo {volume} {104}},\ \bibinfo
  {pages} {104041} (\bibinfo {year} {2021}{\natexlab{b}})}\BibitemShut
  {NoStop}%
\bibitem [{\citenamefont {Chen}\ \emph {et~al.}(2015)\citenamefont {Chen},
  \citenamefont {Ong},\ and\ \citenamefont {h.~Yeom}}]{pchen}%
  \BibitemOpen
  \bibfield  {author} {\bibinfo {author} {\bibfnamefont {P.}~\bibnamefont
  {Chen}}, \bibinfo {author} {\bibfnamefont {Y.}~\bibnamefont {Ong}},\ and\
  \bibinfo {author} {\bibfnamefont {D.}~\bibnamefont {h.~Yeom}},\ }\bibfield
  {title} {\bibinfo {title} {Black hole remnants and the information loss
  paradox},\ }\href {https://doi.org/10.1016/j.physrep.2015.10.007} {\bibfield
  {journal} {\bibinfo  {journal} {Physics Reports}\ }\textbf {\bibinfo {volume}
  {603}},\ \bibinfo {pages} {1} (\bibinfo {year} {2015})}\BibitemShut {NoStop}%
\bibitem [{\citenamefont {Amelino-Camelia}\ \emph {et~al.}(1998)\citenamefont
  {Amelino-Camelia}, \citenamefont {Ellis}, \citenamefont {Mavromatos},
  \citenamefont {Nanopoulos},\ and\ \citenamefont {Sarkar}}]{amelino98}%
  \BibitemOpen
  \bibfield  {author} {\bibinfo {author} {\bibfnamefont {G.}~\bibnamefont
  {Amelino-Camelia}}, \bibinfo {author} {\bibfnamefont {J.~R.}\ \bibnamefont
  {Ellis}}, \bibinfo {author} {\bibfnamefont {N.~E.}\ \bibnamefont
  {Mavromatos}}, \bibinfo {author} {\bibfnamefont {D.~V.}\ \bibnamefont
  {Nanopoulos}},\ and\ \bibinfo {author} {\bibfnamefont {S.}~\bibnamefont
  {Sarkar}},\ }\bibfield  {title} {\bibinfo {title} {{Tests of quantum gravity
  from observations of gamma-ray bursts}},\ }\href
  {https://doi.org/10.1038/31647} {\bibfield  {journal} {\bibinfo  {journal}
  {Nature}\ }\textbf {\bibinfo {volume} {393}},\ \bibinfo {pages} {763}
  (\bibinfo {year} {1998})},\ \Eprint {https://arxiv.org/abs/astro-ph/9712103}
  {arXiv:astro-ph/9712103} \BibitemShut {NoStop}%
\bibitem [{\citenamefont {Garay}(1998)}]{garay-stf}%
  \BibitemOpen
  \bibfield  {author} {\bibinfo {author} {\bibfnamefont {L.~J.}\ \bibnamefont
  {Garay}},\ }\bibfield  {title} {\bibinfo {title} {Spacetime foam as a quantum
  thermal bath},\ }\href {https://doi.org/10.1103/PhysRevLett.80.2508}
  {\bibfield  {journal} {\bibinfo  {journal} {Phys. Rev. Lett.}\ }\textbf
  {\bibinfo {volume} {80}},\ \bibinfo {pages} {2508} (\bibinfo {year}
  {1998})}\BibitemShut {NoStop}%
\bibitem [{\citenamefont {Amelino-Camelia}(2002)}]{amelino-doubly}%
  \BibitemOpen
  \bibfield  {author} {\bibinfo {author} {\bibfnamefont {G.}~\bibnamefont
  {Amelino-Camelia}},\ }\bibfield  {title} {\bibinfo {title} {Special
  treatment},\ }\href {https://doi.org/10.1038/418034a} {\bibfield  {journal}
  {\bibinfo  {journal} {Nature}\ }\textbf {\bibinfo {volume} {418}},\ \bibinfo
  {pages} {34} (\bibinfo {year} {2002})}\BibitemShut {NoStop}%
\bibitem [{\citenamefont {Magueijo}\ and\ \citenamefont
  {Smolin}(2003)}]{magueijo}%
  \BibitemOpen
  \bibfield  {author} {\bibinfo {author} {\bibfnamefont {J.}~\bibnamefont
  {Magueijo}}\ and\ \bibinfo {author} {\bibfnamefont {L.}~\bibnamefont
  {Smolin}},\ }\bibfield  {title} {\bibinfo {title} {Generalized {L}orentz
  invariance with an invariant energy scale},\ }\href
  {https://doi.org/10.1103/PhysRevD.67.044017} {\bibfield  {journal} {\bibinfo
  {journal} {Phys. Rev. D}\ }\textbf {\bibinfo {volume} {67}},\ \bibinfo
  {pages} {044017} (\bibinfo {year} {2003})}\BibitemShut {NoStop}%
\bibitem [{\citenamefont {Liberati}(2013)}]{liberati-test}%
  \BibitemOpen
  \bibfield  {author} {\bibinfo {author} {\bibfnamefont {S.}~\bibnamefont
  {Liberati}},\ }\bibfield  {title} {\bibinfo {title} {Tests of {L}orentz
  invariance: a 2013 update},\ }\href
  {https://doi.org/10.1088/0264-9381/30/13/133001} {\bibfield  {journal}
  {\bibinfo  {journal} {Classical and Quantum Gravity}\ }\textbf {\bibinfo
  {volume} {30}},\ \bibinfo {pages} {133001} (\bibinfo {year}
  {2013})}\BibitemShut {NoStop}%
\bibitem [{\citenamefont {Amelino-Camelia}\ \emph {et~al.}(2004)\citenamefont
  {Amelino-Camelia}, \citenamefont {Arzano},\ and\ \citenamefont
  {Procaccini}}]{amelino2004}%
  \BibitemOpen
  \bibfield  {author} {\bibinfo {author} {\bibfnamefont {G.}~\bibnamefont
  {Amelino-Camelia}}, \bibinfo {author} {\bibfnamefont {M.}~\bibnamefont
  {Arzano}},\ and\ \bibinfo {author} {\bibfnamefont {A.}~\bibnamefont
  {Procaccini}},\ }\bibfield  {title} {\bibinfo {title} {Severe constraints on
  the loop-quantum-gravity energy-momentum dispersion relation from the
  black-hole area-entropy law},\ }\href
  {https://doi.org/10.1103/PhysRevD.70.107501} {\bibfield  {journal} {\bibinfo
  {journal} {Phys. Rev. D}\ }\textbf {\bibinfo {volume} {70}},\ \bibinfo
  {pages} {107501} (\bibinfo {year} {2004})}\BibitemShut {NoStop}%
\bibitem [{\citenamefont {Ling}\ \emph {et~al.}(2006)\citenamefont {Ling},
  \citenamefont {Hu},\ and\ \citenamefont {Li}}]{ling}%
  \BibitemOpen
  \bibfield  {author} {\bibinfo {author} {\bibfnamefont {Y.}~\bibnamefont
  {Ling}}, \bibinfo {author} {\bibfnamefont {B.}~\bibnamefont {Hu}},\ and\
  \bibinfo {author} {\bibfnamefont {X.}~\bibnamefont {Li}},\ }\bibfield
  {title} {\bibinfo {title} {Modified dispersion relations and black hole
  physics},\ }\href {https://doi.org/10.1103/PhysRevD.73.087702} {\bibfield
  {journal} {\bibinfo  {journal} {Phys. Rev. D}\ }\textbf {\bibinfo {volume}
  {73}},\ \bibinfo {pages} {087702} (\bibinfo {year} {2006})}\BibitemShut
  {NoStop}%
\bibitem [{\citenamefont {Nozari}\ and\ \citenamefont
  {Sefidgar}(2006)}]{nozari}%
  \BibitemOpen
  \bibfield  {author} {\bibinfo {author} {\bibfnamefont {K.}~\bibnamefont
  {Nozari}}\ and\ \bibinfo {author} {\bibfnamefont {A.}~\bibnamefont
  {Sefidgar}},\ }\bibfield  {title} {\bibinfo {title} {Comparison of approaches
  to quantum correction of black hole thermodynamics},\ }\href
  {https://doi.org/https://doi.org/10.1016/j.physletb.2006.02.043} {\bibfield
  {journal} {\bibinfo  {journal} {Physics Letters B}\ }\textbf {\bibinfo
  {volume} {635}},\ \bibinfo {pages} {156} (\bibinfo {year}
  {2006})}\BibitemShut {NoStop}%
\bibitem [{\citenamefont {Rovelli}\ and\ \citenamefont
  {Speziale}(2003)}]{speziale}%
  \BibitemOpen
  \bibfield  {author} {\bibinfo {author} {\bibfnamefont {C.}~\bibnamefont
  {Rovelli}}\ and\ \bibinfo {author} {\bibfnamefont {S.}~\bibnamefont
  {Speziale}},\ }\bibfield  {title} {\bibinfo {title} {Reconcile {P}lanck-scale
  discreteness and the {L}orentz-{F}itzgerald contraction},\ }\href
  {https://doi.org/10.1103/PhysRevD.67.064019} {\bibfield  {journal} {\bibinfo
  {journal} {Phys. Rev. D}\ }\textbf {\bibinfo {volume} {67}},\ \bibinfo
  {pages} {064019} (\bibinfo {year} {2003})}\BibitemShut {NoStop}%
\bibitem [{\citenamefont {Fontanini}\ \emph {et~al.}(2006)\citenamefont
  {Fontanini}, \citenamefont {Spallucci},\ and\ \citenamefont
  {Padmanabhan}}]{Fontanini:2005ik}%
  \BibitemOpen
  \bibfield  {author} {\bibinfo {author} {\bibfnamefont {M.}~\bibnamefont
  {Fontanini}}, \bibinfo {author} {\bibfnamefont {E.}~\bibnamefont
  {Spallucci}},\ and\ \bibinfo {author} {\bibfnamefont {T.}~\bibnamefont
  {Padmanabhan}},\ }\bibfield  {title} {\bibinfo {title} {Zero-point length
  from string fluctuations},\ }\href
  {https://doi.org/https://doi.org/10.1016/j.physletb.2005.12.039} {\bibfield
  {journal} {\bibinfo  {journal} {Physics Letters B}\ }\textbf {\bibinfo
  {volume} {633}},\ \bibinfo {pages} {627} (\bibinfo {year}
  {2006})}\BibitemShut {NoStop}%
\bibitem [{\citenamefont {Hu}(2005)}]{hu}%
  \BibitemOpen
  \bibfield  {author} {\bibinfo {author} {\bibfnamefont {B.~L.}\ \bibnamefont
  {Hu}},\ }\bibfield  {title} {\bibinfo {title} {Can spacetime be a
  condensate?},\ }\href {https://doi.org/10.1007/s10773-005-8895-0} {\bibfield
  {journal} {\bibinfo  {journal} {International Journal of Theoretical
  Physics}\ }\textbf {\bibinfo {volume} {44}},\ \bibinfo {pages} {1785}
  (\bibinfo {year} {2005})}\BibitemShut {NoStop}%
\bibitem [{\citenamefont {Bombelli}\ \emph {et~al.}(1987)\citenamefont
  {Bombelli}, \citenamefont {Lee}, \citenamefont {Meyer},\ and\ \citenamefont
  {Sorkin}}]{bombelli}%
  \BibitemOpen
  \bibfield  {author} {\bibinfo {author} {\bibfnamefont {L.}~\bibnamefont
  {Bombelli}}, \bibinfo {author} {\bibfnamefont {J.}~\bibnamefont {Lee}},
  \bibinfo {author} {\bibfnamefont {D.}~\bibnamefont {Meyer}},\ and\ \bibinfo
  {author} {\bibfnamefont {R.~D.}\ \bibnamefont {Sorkin}},\ }\bibfield  {title}
  {\bibinfo {title} {Space-time as a causal set},\ }\href
  {https://doi.org/10.1103/PhysRevLett.59.521} {\bibfield  {journal} {\bibinfo
  {journal} {Phys. Rev. Lett.}\ }\textbf {\bibinfo {volume} {59}},\ \bibinfo
  {pages} {521} (\bibinfo {year} {1987})}\BibitemShut {NoStop}%
\bibitem [{\citenamefont {Konopka}\ \emph {et~al.}(2008)\citenamefont
  {Konopka}, \citenamefont {Markopoulou},\ and\ \citenamefont
  {Severini}}]{konopka}%
  \BibitemOpen
  \bibfield  {author} {\bibinfo {author} {\bibfnamefont {T.}~\bibnamefont
  {Konopka}}, \bibinfo {author} {\bibfnamefont {F.}~\bibnamefont
  {Markopoulou}},\ and\ \bibinfo {author} {\bibfnamefont {S.}~\bibnamefont
  {Severini}},\ }\bibfield  {title} {\bibinfo {title} {Quantum graphity: A
  model of emergent locality},\ }\href
  {https://doi.org/10.1103/PhysRevD.77.104029} {\bibfield  {journal} {\bibinfo
  {journal} {Phys. Rev. D}\ }\textbf {\bibinfo {volume} {77}},\ \bibinfo
  {pages} {104029} (\bibinfo {year} {2008})}\BibitemShut {NoStop}%
\bibitem [{\citenamefont {Jacobson}(1995)}]{jaco}%
  \BibitemOpen
  \bibfield  {author} {\bibinfo {author} {\bibfnamefont {T.}~\bibnamefont
  {Jacobson}},\ }\bibfield  {title} {\bibinfo {title} {Thermodynamics of
  spacetime: The {E}instein equation of state},\ }\href
  {https://doi.org/10.1103/PhysRevLett.75.1260} {\bibfield  {journal} {\bibinfo
   {journal} {Phys. Rev. Lett.}\ }\textbf {\bibinfo {volume} {75}},\ \bibinfo
  {pages} {1260} (\bibinfo {year} {1995})}\BibitemShut {NoStop}%
\bibitem [{\citenamefont {Sakharov}(1968)}]{sakha}%
  \BibitemOpen
  \bibfield  {author} {\bibinfo {author} {\bibfnamefont {A.~D.}\ \bibnamefont
  {Sakharov}},\ }\bibfield  {title} {\bibinfo {title} {Vacuum quantum
  fluctuations in curved space and the theory of gravitation},\ }\href@noop {}
  {\bibfield  {journal} {\bibinfo  {journal} {Sov. Phys. Dokl.}\ }\textbf
  {\bibinfo {volume} {12}} (\bibinfo {year} {1968})}\BibitemShut {NoStop}%
\bibitem [{\citenamefont {Visser}(2002)}]{visser2002sakharov}%
  \BibitemOpen
  \bibfield  {author} {\bibinfo {author} {\bibfnamefont {M.}~\bibnamefont
  {Visser}},\ }\bibfield  {title} {\bibinfo {title} {Sakharov's induced
  gravity: a modern perspective},\ }\href
  {https://doi.org/10.1142/S0217732302006886} {\bibfield  {journal} {\bibinfo
  {journal} {Modern Physics Letters A}\ }\textbf {\bibinfo {volume} {17}},\
  \bibinfo {pages} {977} (\bibinfo {year} {2002})}\BibitemShut {NoStop}%
\bibitem [{\citenamefont {Verlinde}(2011)}]{Verlinde_2011}%
  \BibitemOpen
  \bibfield  {author} {\bibinfo {author} {\bibfnamefont {E.}~\bibnamefont
  {Verlinde}},\ }\bibfield  {title} {\bibinfo {title} {On the origin of gravity
  and the laws of {N}ewton},\ }\href {https://doi.org/10.1007/jhep04(2011)029}
  {\bibfield  {journal} {\bibinfo  {journal} {Journal of High Energy Physics}\
  }\textbf {\bibinfo {volume} {2011}},\ \bibinfo {pages} {4} (\bibinfo {year}
  {2011})}\BibitemShut {NoStop}%
\bibitem [{\citenamefont {Oriti}(2009)}]{oriti2009group}%
  \BibitemOpen
  \bibfield  {author} {\bibinfo {author} {\bibfnamefont {D.}~\bibnamefont
  {Oriti}},\ }\href@noop {} {\bibinfo {title} {The group field theory approach
  to quantum gravity: some recent results}} (\bibinfo {year} {2009}),\ \Eprint
  {https://arxiv.org/abs/0912.2441} {arXiv:0912.2441 [hep-th]} \BibitemShut
  {NoStop}%
\bibitem [{\citenamefont {Beisert~et al.}(2011)}]{ads1}%
  \BibitemOpen
  \bibfield  {author} {\bibinfo {author} {\bibfnamefont {N.}~\bibnamefont
  {Beisert~et al.}},\ }\bibfield  {title} {\bibinfo {title} {Review of
  {AdS}/{CFT} integrability: An overview},\ }\href
  {https://doi.org/10.1007/s11005-011-0529-2} {\bibfield  {journal} {\bibinfo
  {journal} {Letters in Mathematical Physics}\ }\textbf {\bibinfo {volume}
  {99}},\ \bibinfo {pages} {3} (\bibinfo {year} {2011})}\BibitemShut {NoStop}%
\bibitem [{\citenamefont {Maldacena}(1999)}]{ads2}%
  \BibitemOpen
  \bibfield  {author} {\bibinfo {author} {\bibfnamefont {J.}~\bibnamefont
  {Maldacena}},\ }\href {https://doi.org/10.1023/a:1026654312961} {\bibfield
  {journal} {\bibinfo  {journal} {International Journal of Theoretical
  Physics}\ }\textbf {\bibinfo {volume} {38}},\ \bibinfo {pages} {1113}
  (\bibinfo {year} {1999})}\BibitemShut {NoStop}%
\bibitem [{\citenamefont {Pasterski}\ \emph {et~al.}(2021)\citenamefont
  {Pasterski}, \citenamefont {Pate},\ and\ \citenamefont
  {Raclariu}}]{pasterski2021celestial}%
  \BibitemOpen
  \bibfield  {author} {\bibinfo {author} {\bibfnamefont {S.}~\bibnamefont
  {Pasterski}}, \bibinfo {author} {\bibfnamefont {M.}~\bibnamefont {Pate}},\
  and\ \bibinfo {author} {\bibfnamefont {A.-M.}\ \bibnamefont {Raclariu}},\
  }\href@noop {} {\bibinfo {title} {Celestial holography}} (\bibinfo {year}
  {2021}),\ \Eprint {https://arxiv.org/abs/2111.11392} {arXiv:2111.11392
  [hep-th]} \BibitemShut {NoStop}%
\bibitem [{\citenamefont {Freidel}\ \emph {et~al.}(2020)\citenamefont
  {Freidel}, \citenamefont {Geiller},\ and\ \citenamefont {Pranzetti}}]{fre20}%
  \BibitemOpen
  \bibfield  {author} {\bibinfo {author} {\bibfnamefont {L.}~\bibnamefont
  {Freidel}}, \bibinfo {author} {\bibfnamefont {M.}~\bibnamefont {Geiller}},\
  and\ \bibinfo {author} {\bibfnamefont {D.}~\bibnamefont {Pranzetti}},\
  }\bibfield  {title} {\bibinfo {title} {Edge modes of gravity. part {I}:
  {C}orner potentials and charges},\ }\href
  {https://doi.org/10.1007/JHEP11(2020)026} {\bibfield  {journal} {\bibinfo
  {journal} {JHEP}\ }\textbf {\bibinfo {volume} {11}},\ \bibinfo {pages}
  {026}}\BibitemShut {NoStop}%
\bibitem [{\citenamefont {Freidel}\ \emph {et~al.}(2021)\citenamefont
  {Freidel}, \citenamefont {Oliveri}, \citenamefont {Pranzetti},\ and\
  \citenamefont {Speziale}}]{fre21}%
  \BibitemOpen
  \bibfield  {author} {\bibinfo {author} {\bibfnamefont {L.}~\bibnamefont
  {Freidel}}, \bibinfo {author} {\bibfnamefont {R.}~\bibnamefont {Oliveri}},
  \bibinfo {author} {\bibfnamefont {D.}~\bibnamefont {Pranzetti}},\ and\
  \bibinfo {author} {\bibfnamefont {S.}~\bibnamefont {Speziale}},\ }\bibfield
  {title} {\bibinfo {title} {Extended corner symmetry, charge bracket and
  {E}instein’s equations},\ }\href {https://doi.org/10.1007/jhep04(2011)029}
  {\bibfield  {journal} {\bibinfo  {journal} {Journal of High Energy Physics}\
  }\textbf {\bibinfo {volume} {2021}},\ \bibinfo {pages} {1} (\bibinfo {year}
  {2021})}\BibitemShut {NoStop}%
\bibitem [{\citenamefont {Jafferis}\ \emph {et~al.}()\citenamefont {Jafferis},
  \citenamefont {Zlokapa}, \citenamefont {Lykken}, \citenamefont {Kolchmeyer},
  \citenamefont {Davis}, \citenamefont {Lauk}, \citenamefont {Neven},\ and\
  \citenamefont {Spiropulu}}]{ep1}%
  \BibitemOpen
  \bibfield  {author} {\bibinfo {author} {\bibfnamefont {D.}~\bibnamefont
  {Jafferis}}, \bibinfo {author} {\bibfnamefont {A.}~\bibnamefont {Zlokapa}},
  \bibinfo {author} {\bibfnamefont {J.~D.}\ \bibnamefont {Lykken}}, \bibinfo
  {author} {\bibfnamefont {D.~K.}\ \bibnamefont {Kolchmeyer}}, \bibinfo
  {author} {\bibfnamefont {S.~I.}\ \bibnamefont {Davis}}, \bibinfo {author}
  {\bibfnamefont {N.}~\bibnamefont {Lauk}}, \bibinfo {author} {\bibfnamefont
  {H.}~\bibnamefont {Neven}},\ and\ \bibinfo {author} {\bibfnamefont
  {M.}~\bibnamefont {Spiropulu}},\ }\bibfield  {title} {\bibinfo {title}
  {Traversable wormhole dynamics on a quantum processor},\ }\href
  {https://doi.org/10.1038/s41586-022-05424-3} {\bibfield  {journal} {\bibinfo
  {journal} {Nature}\ }\textbf {\bibinfo {volume} {612}},\ \bibinfo {pages}
  {51}}\BibitemShut {NoStop}%
\bibitem [{\citenamefont {Jafferis}\ and\ \citenamefont
  {Schneider}(2022)}]{ep2}%
  \BibitemOpen
  \bibfield  {author} {\bibinfo {author} {\bibfnamefont {D.~L.}\ \bibnamefont
  {Jafferis}}\ and\ \bibinfo {author} {\bibfnamefont {E.}~\bibnamefont
  {Schneider}},\ }\bibfield  {title} {\bibinfo {title} {Stringy {ER} = {EPR}},\
  }\bibfield  {journal} {\bibinfo  {journal} {Journal of High Energy Physics}\
  }\textbf {\bibinfo {volume} {2022}},\ \href
  {https://doi.org/10.1007/jhep10(2022)195} {10.1007/jhep10(2022)195} (\bibinfo
  {year} {2022})\BibitemShut {NoStop}%
\bibitem [{\citenamefont {Van~Raamsdonk}(2020)}]{bits}%
  \BibitemOpen
  \bibfield  {author} {\bibinfo {author} {\bibfnamefont {M.}~\bibnamefont
  {Van~Raamsdonk}},\ }\bibfield  {title} {\bibinfo {title} {{Spacetime from
  bits}},\ }\href {https://doi.org/10.1126/science.aay9560} {\bibfield
  {journal} {\bibinfo  {journal} {Science}\ }\textbf {\bibinfo {volume}
  {370}},\ \bibinfo {pages} {198} (\bibinfo {year} {2020})}\BibitemShut
  {NoStop}%
\bibitem [{\citenamefont {Wheeler}(1989)}]{itfrombit}%
  \BibitemOpen
  \bibfield  {author} {\bibinfo {author} {\bibfnamefont {J.~A.}\ \bibnamefont
  {Wheeler}},\ }\bibfield  {title} {\bibinfo {title} {{Information, physics,
  quantum: The search for links}},\ }in\ \href@noop {} {\emph {\bibinfo
  {booktitle} {{3rd International Symposium on Foundations of Quantum Mechanics
  in Light}}}}\ (\bibinfo {year} {1989})\BibitemShut {NoStop}%
\bibitem [{\citenamefont {Rousseaux}(2013)}]{Rousseaux_2013}%
  \BibitemOpen
  \bibfield  {author} {\bibinfo {author} {\bibfnamefont {G.}~\bibnamefont
  {Rousseaux}},\ }\bibfield  {title} {\bibinfo {title} {The basics of water
  waves theory for analogue gravity},\ }in\ \href
  {https://doi.org/10.1007/978-3-319-00266-8_5} {\emph {\bibinfo {booktitle}
  {Lecture Notes in Physics}}}\ (\bibinfo  {publisher} {Springer International
  Publishing},\ \bibinfo {year} {2013})\ pp.\ \bibinfo {pages}
  {81--107}\BibitemShut {NoStop}%
\bibitem [{\citenamefont {Bogoliubov}(1947)}]{bogoliubov}%
  \BibitemOpen
  \bibfield  {author} {\bibinfo {author} {\bibfnamefont {N.~N.}\ \bibnamefont
  {Bogoliubov}},\ }\bibfield  {title} {\bibinfo {title} {{On the theory of
  superfluidity}},\ }\href@noop {} {\bibfield  {journal} {\bibinfo  {journal}
  {J. Phys. (USSR)}\ }\textbf {\bibinfo {volume} {11}},\ \bibinfo {pages} {23}
  (\bibinfo {year} {1947})}\BibitemShut {NoStop}%
\bibitem [{\citenamefont {Dalfovo}\ \emph {et~al.}(1999)\citenamefont
  {Dalfovo}, \citenamefont {Giorgini}, \citenamefont {Pitaevskii},\ and\
  \citenamefont {Stringari}}]{reviewBEC}%
  \BibitemOpen
  \bibfield  {author} {\bibinfo {author} {\bibfnamefont {F.}~\bibnamefont
  {Dalfovo}}, \bibinfo {author} {\bibfnamefont {S.}~\bibnamefont {Giorgini}},
  \bibinfo {author} {\bibfnamefont {L.~P.}\ \bibnamefont {Pitaevskii}},\ and\
  \bibinfo {author} {\bibfnamefont {S.}~\bibnamefont {Stringari}},\ }\bibfield
  {title} {\bibinfo {title} {Theory of {B}ose-{E}instein condensation in
  trapped gases},\ }\href {https://doi.org/10.1103/RevModPhys.71.463}
  {\bibfield  {journal} {\bibinfo  {journal} {Rev. Mod. Phys.}\ }\textbf
  {\bibinfo {volume} {71}},\ \bibinfo {pages} {463} (\bibinfo {year}
  {1999})}\BibitemShut {NoStop}%
\bibitem [{\citenamefont {Fontaine}\ \emph {et~al.}(2018)\citenamefont
  {Fontaine}, \citenamefont {Bienaim\'e}, \citenamefont {Pigeon}, \citenamefont
  {Giacobino}, \citenamefont {Bramati},\ and\ \citenamefont
  {Glorieux}}]{fontaine18}%
  \BibitemOpen
  \bibfield  {author} {\bibinfo {author} {\bibfnamefont {Q.}~\bibnamefont
  {Fontaine}}, \bibinfo {author} {\bibfnamefont {T.}~\bibnamefont
  {Bienaim\'e}}, \bibinfo {author} {\bibfnamefont {S.}~\bibnamefont {Pigeon}},
  \bibinfo {author} {\bibfnamefont {E.}~\bibnamefont {Giacobino}}, \bibinfo
  {author} {\bibfnamefont {A.}~\bibnamefont {Bramati}},\ and\ \bibinfo {author}
  {\bibfnamefont {Q.}~\bibnamefont {Glorieux}},\ }\bibfield  {title} {\bibinfo
  {title} {Observation of the {B}ogoliubov dispersion in a fluid of light},\
  }\href {https://doi.org/10.1103/PhysRevLett.121.183604} {\bibfield  {journal}
  {\bibinfo  {journal} {Phys. Rev. Lett.}\ }\textbf {\bibinfo {volume} {121}},\
  \bibinfo {pages} {183604} (\bibinfo {year} {2018})}\BibitemShut {NoStop}%
\bibitem [{\citenamefont {Ch\"a}\ and\ \citenamefont
  {Fischer}(2017)}]{PhysRevLett.118.130404}%
  \BibitemOpen
  \bibfield  {author} {\bibinfo {author} {\bibfnamefont {S.-Y.}\ \bibnamefont
  {Ch\"a}}\ and\ \bibinfo {author} {\bibfnamefont {U.~R.}\ \bibnamefont
  {Fischer}},\ }\bibfield  {title} {\bibinfo {title} {Probing the scale
  invariance of the inflationary power spectrum in expanding
  quasi-two-dimensional dipolar condensates},\ }\href
  {https://doi.org/10.1103/PhysRevLett.118.130404} {\bibfield  {journal}
  {\bibinfo  {journal} {Phys. Rev. Lett.}\ }\textbf {\bibinfo {volume} {118}},\
  \bibinfo {pages} {130404} (\bibinfo {year} {2017})}\BibitemShut {NoStop}%
\bibitem [{\citenamefont {Volovik}(2005)}]{Volovik_2005}%
  \BibitemOpen
  \bibfield  {author} {\bibinfo {author} {\bibfnamefont {G.~E.}\ \bibnamefont
  {Volovik}},\ }\bibfield  {title} {\bibinfo {title} {Hydraulic jump as a white
  hole},\ }\href {https://doi.org/10.1134/1.2166908} {\bibfield  {journal}
  {\bibinfo  {journal} {Journal of Experimental and Theoretical Physics
  Letters}\ }\textbf {\bibinfo {volume} {82}},\ \bibinfo {pages} {624}
  (\bibinfo {year} {2005})}\BibitemShut {NoStop}%
\bibitem [{\citenamefont {Jannes}\ \emph {et~al.}(2011)\citenamefont {Jannes},
  \citenamefont {Piquet}, \citenamefont {Ma\"{\i}ssa}, \citenamefont {Mathis},\
  and\ \citenamefont {Rousseaux}}]{PhysRevE.83.056312}%
  \BibitemOpen
  \bibfield  {author} {\bibinfo {author} {\bibfnamefont {G.}~\bibnamefont
  {Jannes}}, \bibinfo {author} {\bibfnamefont {R.}~\bibnamefont {Piquet}},
  \bibinfo {author} {\bibfnamefont {P.}~\bibnamefont {Ma\"{\i}ssa}}, \bibinfo
  {author} {\bibfnamefont {C.}~\bibnamefont {Mathis}},\ and\ \bibinfo {author}
  {\bibfnamefont {G.}~\bibnamefont {Rousseaux}},\ }\bibfield  {title} {\bibinfo
  {title} {Experimental demonstration of the supersonic-subsonic bifurcation in
  the circular jump: A hydrodynamic white hole},\ }\href
  {https://doi.org/10.1103/PhysRevE.83.056312} {\bibfield  {journal} {\bibinfo
  {journal} {Phys. Rev. E}\ }\textbf {\bibinfo {volume} {83}},\ \bibinfo
  {pages} {056312} (\bibinfo {year} {2011})}\BibitemShut {NoStop}%
\bibitem [{\citenamefont {Liberati}\ \emph {et~al.}(2006)\citenamefont
  {Liberati}, \citenamefont {Visser},\ and\ \citenamefont {Weinfurtner}}]{eg1}%
  \BibitemOpen
  \bibfield  {author} {\bibinfo {author} {\bibfnamefont {S.}~\bibnamefont
  {Liberati}}, \bibinfo {author} {\bibfnamefont {M.}~\bibnamefont {Visser}},\
  and\ \bibinfo {author} {\bibfnamefont {S.}~\bibnamefont {Weinfurtner}},\
  }\bibfield  {title} {\bibinfo {title} {Naturalness in an emergent analogue
  spacetime},\ }\href {https://doi.org/10.1103/PhysRevLett.96.151301}
  {\bibfield  {journal} {\bibinfo  {journal} {Phys. Rev. Lett.}\ }\textbf
  {\bibinfo {volume} {96}},\ \bibinfo {pages} {151301} (\bibinfo {year}
  {2006})}\BibitemShut {NoStop}%
\bibitem [{\citenamefont {Unruh}(1995)}]{jacobson}%
  \BibitemOpen
  \bibfield  {author} {\bibinfo {author} {\bibfnamefont {W.~G.}\ \bibnamefont
  {Unruh}},\ }\bibfield  {title} {\bibinfo {title} {Sonic analogue of black
  holes and the effects of high frequencies on black hole evaporation},\ }\href
  {https://doi.org/10.1103/PhysRevD.51.2827} {\bibfield  {journal} {\bibinfo
  {journal} {Phys. Rev. D}\ }\textbf {\bibinfo {volume} {51}},\ \bibinfo
  {pages} {2827} (\bibinfo {year} {1995})}\BibitemShut {NoStop}%
\bibitem [{\citenamefont {Jacobson}(1991)}]{unruh95}%
  \BibitemOpen
  \bibfield  {author} {\bibinfo {author} {\bibfnamefont {T.}~\bibnamefont
  {Jacobson}},\ }\bibfield  {title} {\bibinfo {title} {Black-hole evaporation
  and ultrashort distances},\ }\href {https://doi.org/10.1103/PhysRevD.44.1731}
  {\bibfield  {journal} {\bibinfo  {journal} {Phys. Rev. D}\ }\textbf {\bibinfo
  {volume} {44}},\ \bibinfo {pages} {1731} (\bibinfo {year}
  {1991})}\BibitemShut {NoStop}%
\bibitem [{\citenamefont {Corley}\ and\ \citenamefont
  {Jacobson}(1996)}]{Corley}%
  \BibitemOpen
  \bibfield  {author} {\bibinfo {author} {\bibfnamefont {S.}~\bibnamefont
  {Corley}}\ and\ \bibinfo {author} {\bibfnamefont {T.}~\bibnamefont
  {Jacobson}},\ }\bibfield  {title} {\bibinfo {title} {Hawking spectrum and
  high frequency dispersion},\ }\href
  {https://doi.org/10.1103/PhysRevD.54.1568} {\bibfield  {journal} {\bibinfo
  {journal} {Phys. Rev. D}\ }\textbf {\bibinfo {volume} {54}},\ \bibinfo
  {pages} {1568} (\bibinfo {year} {1996})}\BibitemShut {NoStop}%
\bibitem [{\citenamefont {Jacobson}(1993)}]{jacob93}%
  \BibitemOpen
  \bibfield  {author} {\bibinfo {author} {\bibfnamefont {T.}~\bibnamefont
  {Jacobson}},\ }\bibfield  {title} {\bibinfo {title} {Black hole radiation in
  the presence of a short distance cutoff},\ }\href
  {https://doi.org/10.1103/PhysRevD.48.728} {\bibfield  {journal} {\bibinfo
  {journal} {Phys. Rev. D}\ }\textbf {\bibinfo {volume} {48}},\ \bibinfo
  {pages} {728} (\bibinfo {year} {1993})}\BibitemShut {NoStop}%
\bibitem [{\citenamefont {Carusotto}\ \emph {et~al.}(2008)\citenamefont
  {Carusotto}, \citenamefont {Fagnocchi}, \citenamefont {Recati}, \citenamefont
  {Balbinot},\ and\ \citenamefont {Fabbri}}]{carusotto08}%
  \BibitemOpen
  \bibfield  {author} {\bibinfo {author} {\bibfnamefont {I.}~\bibnamefont
  {Carusotto}}, \bibinfo {author} {\bibfnamefont {S.}~\bibnamefont
  {Fagnocchi}}, \bibinfo {author} {\bibfnamefont {A.}~\bibnamefont {Recati}},
  \bibinfo {author} {\bibfnamefont {R.}~\bibnamefont {Balbinot}},\ and\
  \bibinfo {author} {\bibfnamefont {A.}~\bibnamefont {Fabbri}},\ }\bibfield
  {title} {\bibinfo {title} {Numerical observation of {H}awking radiation from
  acoustic black holes in atomic {B}ose–{E}instein condensates},\ }\href
  {https://doi.org/10.1088/1367-2630/10/10/103001} {\bibfield  {journal}
  {\bibinfo  {journal} {New Journal of Physics}\ }\textbf {\bibinfo {volume}
  {10}},\ \bibinfo {pages} {103001} (\bibinfo {year} {2008})}\BibitemShut
  {NoStop}%
\bibitem [{\citenamefont {Euv\'e}\ and\ \citenamefont
  {Rousseaux}(2017)}]{PhysRevD.96.064042}%
  \BibitemOpen
  \bibfield  {author} {\bibinfo {author} {\bibfnamefont {L.-P.}\ \bibnamefont
  {Euv\'e}}\ and\ \bibinfo {author} {\bibfnamefont {G.}~\bibnamefont
  {Rousseaux}},\ }\bibfield  {title} {\bibinfo {title} {Classical analogue of
  an interstellar travel through a hydrodynamic wormhole},\ }\href
  {https://doi.org/10.1103/PhysRevD.96.064042} {\bibfield  {journal} {\bibinfo
  {journal} {Phys. Rev. D}\ }\textbf {\bibinfo {volume} {96}},\ \bibinfo
  {pages} {064042} (\bibinfo {year} {2017})}\BibitemShut {NoStop}%
\bibitem [{\citenamefont {Unruh}\ and\ \citenamefont
  {Sch\"utzhold}(2005)}]{PhysRevD.71.024028}%
  \BibitemOpen
  \bibfield  {author} {\bibinfo {author} {\bibfnamefont {W.~G.}\ \bibnamefont
  {Unruh}}\ and\ \bibinfo {author} {\bibfnamefont {R.}~\bibnamefont
  {Sch\"utzhold}},\ }\bibfield  {title} {\bibinfo {title} {Universality of the
  {H}awking effect},\ }\href {https://doi.org/10.1103/PhysRevD.71.024028}
  {\bibfield  {journal} {\bibinfo  {journal} {Phys. Rev. D}\ }\textbf {\bibinfo
  {volume} {71}},\ \bibinfo {pages} {024028} (\bibinfo {year}
  {2005})}\BibitemShut {NoStop}%
\bibitem [{\citenamefont {Ribeiro}\ and\ \citenamefont
  {Fischer}(2022)}]{ribeiro2022impact}%
  \BibitemOpen
  \bibfield  {author} {\bibinfo {author} {\bibfnamefont {C.~C.~H.}\
  \bibnamefont {Ribeiro}}\ and\ \bibinfo {author} {\bibfnamefont {U.~R.}\
  \bibnamefont {Fischer}},\ }\href@noop {} {\bibinfo {title} {Impact of
  trans-{P}lanckian excitations on black-hole radiation in dipolar
  condensates}} (\bibinfo {year} {2022}),\ \Eprint
  {https://arxiv.org/abs/2211.01243} {arXiv:2211.01243 [gr-qc]} \BibitemShut
  {NoStop}%
\bibitem [{\citenamefont {Novello}\ and\ \citenamefont
  {Goulart}(2011)}]{goulart1}%
  \BibitemOpen
  \bibfield  {author} {\bibinfo {author} {\bibfnamefont {M.}~\bibnamefont
  {Novello}}\ and\ \bibinfo {author} {\bibfnamefont {E.}~\bibnamefont
  {Goulart}},\ }\bibfield  {title} {\bibinfo {title} {Beyond analog gravity:
  the case of exceptional dynamics},\ }\href
  {https://doi.org/10.1088/0264-9381/28/14/145022} {\bibfield  {journal}
  {\bibinfo  {journal} {Classical and Quantum Gravity}\ }\textbf {\bibinfo
  {volume} {28}},\ \bibinfo {pages} {145022} (\bibinfo {year}
  {2011})}\BibitemShut {NoStop}%
\bibitem [{\citenamefont {Goulart}\ \emph {et~al.}(2011)\citenamefont
  {Goulart}, \citenamefont {Novello}, \citenamefont {Falciano},\ and\
  \citenamefont {Toniato}}]{goulart2}%
  \BibitemOpen
  \bibfield  {author} {\bibinfo {author} {\bibfnamefont {E.}~\bibnamefont
  {Goulart}}, \bibinfo {author} {\bibfnamefont {M.}~\bibnamefont {Novello}},
  \bibinfo {author} {\bibfnamefont {F.~T.}\ \bibnamefont {Falciano}},\ and\
  \bibinfo {author} {\bibfnamefont {J.~D.}\ \bibnamefont {Toniato}},\
  }\bibfield  {title} {\bibinfo {title} {Hidden geometries in nonlinear
  theories: a novel aspect of analogue gravity},\ }\href
  {https://doi.org/10.1088/0264-9381/28/24/245008} {\bibfield  {journal}
  {\bibinfo  {journal} {Classical and Quantum Gravity}\ }\textbf {\bibinfo
  {volume} {28}},\ \bibinfo {pages} {245008} (\bibinfo {year}
  {2011})}\BibitemShut {NoStop}%
\bibitem [{\citenamefont {Cherubini}\ and\ \citenamefont
  {Filippi}(2011)}]{cherubini}%
  \BibitemOpen
  \bibfield  {author} {\bibinfo {author} {\bibfnamefont {C.}~\bibnamefont
  {Cherubini}}\ and\ \bibinfo {author} {\bibfnamefont {S.}~\bibnamefont
  {Filippi}},\ }\bibfield  {title} {\bibinfo {title} {Von {M}ises' potential
  flow wave equation and nonlinear analog gravity},\ }\href
  {https://doi.org/10.1103/PhysRevD.84.124010} {\bibfield  {journal} {\bibinfo
  {journal} {Phys. Rev. D}\ }\textbf {\bibinfo {volume} {84}},\ \bibinfo
  {pages} {124010} (\bibinfo {year} {2011})}\BibitemShut {NoStop}%
\bibitem [{\citenamefont {Marino}\ \emph {et~al.}()\citenamefont {Marino},
  \citenamefont {Maitland}, \citenamefont {Vocke}, \citenamefont {Ortolan},\
  and\ \citenamefont {Faccio}}]{marino2016}%
  \BibitemOpen
  \bibfield  {author} {\bibinfo {author} {\bibfnamefont {F.}~\bibnamefont
  {Marino}}, \bibinfo {author} {\bibfnamefont {C.}~\bibnamefont {Maitland}},
  \bibinfo {author} {\bibfnamefont {D.}~\bibnamefont {Vocke}}, \bibinfo
  {author} {\bibfnamefont {A.}~\bibnamefont {Ortolan}},\ and\ \bibinfo {author}
  {\bibfnamefont {D.}~\bibnamefont {Faccio}},\ }\bibfield  {title} {\bibinfo
  {title} {Emergent geometries and nonlinear-wave dynamics in photon fluids},\
  }\href {https://doi.org/https://doi.org/10.1038/srep23282} {\bibfield
  {journal} {\bibinfo  {journal} {Scientific Reports}\ }\textbf {\bibinfo
  {volume} {6}},\ \bibinfo {pages} {23282}}\BibitemShut {NoStop}%
\bibitem [{\citenamefont {Datta}\ and\ \citenamefont
  {Fischer}(2022)}]{fisher-grav}%
  \BibitemOpen
  \bibfield  {author} {\bibinfo {author} {\bibfnamefont {S.}~\bibnamefont
  {Datta}}\ and\ \bibinfo {author} {\bibfnamefont {U.~R.}\ \bibnamefont
  {Fischer}},\ }\bibfield  {title} {\bibinfo {title} {Analogue gravitational
  field from nonlinear fluid dynamics},\ }\href
  {https://doi.org/10.1088/1361-6382/ac4828} {\bibfield  {journal} {\bibinfo
  {journal} {Classical and Quantum Gravity}\ }\textbf {\bibinfo {volume}
  {39}},\ \bibinfo {pages} {075018} (\bibinfo {year} {2022})}\BibitemShut
  {NoStop}%
\bibitem [{\citenamefont {Balbinot}\ \emph {et~al.}(2005)\citenamefont
  {Balbinot}, \citenamefont {Fagnocchi}, \citenamefont {Fabbri},\ and\
  \citenamefont {Procopio}}]{balbinot05}%
  \BibitemOpen
  \bibfield  {author} {\bibinfo {author} {\bibfnamefont {R.}~\bibnamefont
  {Balbinot}}, \bibinfo {author} {\bibfnamefont {S.}~\bibnamefont {Fagnocchi}},
  \bibinfo {author} {\bibfnamefont {A.}~\bibnamefont {Fabbri}},\ and\ \bibinfo
  {author} {\bibfnamefont {G.~P.}\ \bibnamefont {Procopio}},\ }\bibfield
  {title} {\bibinfo {title} {Backreaction in acoustic black holes},\ }\href
  {https://doi.org/10.1103/PhysRevLett.94.161302} {\bibfield  {journal}
  {\bibinfo  {journal} {Phys. Rev. Lett.}\ }\textbf {\bibinfo {volume} {94}},\
  \bibinfo {pages} {161302} (\bibinfo {year} {2005})}\BibitemShut {NoStop}%
\bibitem [{\citenamefont {Finazzi}\ \emph {et~al.}(2012)\citenamefont
  {Finazzi}, \citenamefont {Liberati},\ and\ \citenamefont {Sindoni}}]{eg3}%
  \BibitemOpen
  \bibfield  {author} {\bibinfo {author} {\bibfnamefont {S.}~\bibnamefont
  {Finazzi}}, \bibinfo {author} {\bibfnamefont {S.}~\bibnamefont {Liberati}},\
  and\ \bibinfo {author} {\bibfnamefont {L.}~\bibnamefont {Sindoni}},\
  }\bibfield  {title} {\bibinfo {title} {Cosmological constant: A lesson from
  {B}ose-{E}instein condensates},\ }\href
  {https://doi.org/10.1103/PhysRevLett.108.071101} {\bibfield  {journal}
  {\bibinfo  {journal} {Phys. Rev. Lett.}\ }\textbf {\bibinfo {volume} {108}},\
  \bibinfo {pages} {071101} (\bibinfo {year} {2012})}\BibitemShut {NoStop}%
\bibitem [{\citenamefont {Belenchia}\ \emph {et~al.}(2014)\citenamefont
  {Belenchia}, \citenamefont {Liberati},\ and\ \citenamefont
  {Mohd}}]{belenchia14}%
  \BibitemOpen
  \bibfield  {author} {\bibinfo {author} {\bibfnamefont {A.}~\bibnamefont
  {Belenchia}}, \bibinfo {author} {\bibfnamefont {S.}~\bibnamefont
  {Liberati}},\ and\ \bibinfo {author} {\bibfnamefont {A.}~\bibnamefont
  {Mohd}},\ }\bibfield  {title} {\bibinfo {title} {Emergent gravitational
  dynamics in a relativistic {B}ose-{E}instein condensate},\ }\href
  {https://doi.org/10.1103/PhysRevD.90.104015} {\bibfield  {journal} {\bibinfo
  {journal} {Phys. Rev. D}\ }\textbf {\bibinfo {volume} {90}},\ \bibinfo
  {pages} {104015} (\bibinfo {year} {2014})}\BibitemShut {NoStop}%
\bibitem [{\citenamefont {Deruelle}(2011)}]{deruelle}%
  \BibitemOpen
  \bibfield  {author} {\bibinfo {author} {\bibfnamefont {N.}~\bibnamefont
  {Deruelle}},\ }\bibfield  {title} {\bibinfo {title} {{Nordstrom's scalar
  theory of gravity and the equivalence principle}},\ }\href
  {https://doi.org/10.1007/s10714-011-1247-x} {\bibfield  {journal} {\bibinfo
  {journal} {Gen. Rel. Grav.}\ }\textbf {\bibinfo {volume} {43}},\ \bibinfo
  {pages} {3337} (\bibinfo {year} {2011})},\ \Eprint
  {https://arxiv.org/abs/1104.4608} {arXiv:1104.4608 [gr-qc]} \BibitemShut
  {NoStop}%
\bibitem [{\citenamefont {Mosna}\ \emph {et~al.}(2016)\citenamefont {Mosna},
  \citenamefont {Pitelli},\ and\ \citenamefont {Richartz}}]{anti1}%
  \BibitemOpen
  \bibfield  {author} {\bibinfo {author} {\bibfnamefont {R.~A.}\ \bibnamefont
  {Mosna}}, \bibinfo {author} {\bibfnamefont {J.~a. P.~M.}\ \bibnamefont
  {Pitelli}},\ and\ \bibinfo {author} {\bibfnamefont {M.}~\bibnamefont
  {Richartz}},\ }\bibfield  {title} {\bibinfo {title} {Analogue model for
  anti-de {S}itter as a description of point sources in fluids},\ }\href
  {https://doi.org/10.1103/PhysRevD.94.104065} {\bibfield  {journal} {\bibinfo
  {journal} {Phys. Rev. D}\ }\textbf {\bibinfo {volume} {94}},\ \bibinfo
  {pages} {104065} (\bibinfo {year} {2016})}\BibitemShut {NoStop}%
\bibitem [{\citenamefont {Aruquipa}\ \emph {et~al.}(2018)\citenamefont
  {Aruquipa}, \citenamefont {Mosna},\ and\ \citenamefont {Pitelli}}]{anti4}%
  \BibitemOpen
  \bibfield  {author} {\bibinfo {author} {\bibfnamefont {D.~Q.}\ \bibnamefont
  {Aruquipa}}, \bibinfo {author} {\bibfnamefont {R.~A.}\ \bibnamefont
  {Mosna}},\ and\ \bibinfo {author} {\bibfnamefont {J.~a. P.~M.}\ \bibnamefont
  {Pitelli}},\ }\bibfield  {title} {\bibinfo {title} {Analogue gravity and
  radial fluid flows: The case of ads and its deformations},\ }\href
  {https://doi.org/10.1103/PhysRevD.97.104056} {\bibfield  {journal} {\bibinfo
  {journal} {Phys. Rev. D}\ }\textbf {\bibinfo {volume} {97}},\ \bibinfo
  {pages} {104056} (\bibinfo {year} {2018})}\BibitemShut {NoStop}%
\bibitem [{\citenamefont {Hossenfelder}(2015)}]{sabine15}%
  \BibitemOpen
  \bibfield  {author} {\bibinfo {author} {\bibfnamefont {S.}~\bibnamefont
  {Hossenfelder}},\ }\bibfield  {title} {\bibinfo {title} {Analog systems for
  gravity duals},\ }\href {https://doi.org/10.1103/PhysRevD.91.124064}
  {\bibfield  {journal} {\bibinfo  {journal} {Phys. Rev. D}\ }\textbf {\bibinfo
  {volume} {91}},\ \bibinfo {pages} {124064} (\bibinfo {year}
  {2015})}\BibitemShut {NoStop}%
\bibitem [{\citenamefont {Hossenfelder}(2016)}]{sabine16}%
  \BibitemOpen
  \bibfield  {author} {\bibinfo {author} {\bibfnamefont {S.}~\bibnamefont
  {Hossenfelder}},\ }\bibfield  {title} {\bibinfo {title} {A relativistic
  acoustic metric for planar black holes},\ }\href
  {https://doi.org/https://doi.org/10.1016/j.physletb.2015.11.026} {\bibfield
  {journal} {\bibinfo  {journal} {Physics Letters B}\ }\textbf {\bibinfo
  {volume} {752}},\ \bibinfo {pages} {13} (\bibinfo {year} {2016})}\BibitemShut
  {NoStop}%
\bibitem [{\citenamefont {Dey}\ \emph {et~al.}(2016)\citenamefont {Dey},
  \citenamefont {Liberati},\ and\ \citenamefont {Turcati}}]{dey}%
  \BibitemOpen
  \bibfield  {author} {\bibinfo {author} {\bibfnamefont {R.}~\bibnamefont
  {Dey}}, \bibinfo {author} {\bibfnamefont {S.}~\bibnamefont {Liberati}},\ and\
  \bibinfo {author} {\bibfnamefont {R.}~\bibnamefont {Turcati}},\ }\bibfield
  {title} {\bibinfo {title} {Ads and ds black hole solutions in analogue
  gravity: The relativistic and nonrelativistic cases},\ }\href
  {https://doi.org/10.1103/PhysRevD.94.104068} {\bibfield  {journal} {\bibinfo
  {journal} {Phys. Rev. D}\ }\textbf {\bibinfo {volume} {94}},\ \bibinfo
  {pages} {104068} (\bibinfo {year} {2016})}\BibitemShut {NoStop}%
\bibitem [{\citenamefont {Bhattacharyya}\ \emph
  {et~al.}(2008{\natexlab{a}})\citenamefont {Bhattacharyya}, \citenamefont
  {Minwalla}, \citenamefont {Hubeny},\ and\ \citenamefont {Rangamani}}]{ei12}%
  \BibitemOpen
  \bibfield  {author} {\bibinfo {author} {\bibfnamefont {S.}~\bibnamefont
  {Bhattacharyya}}, \bibinfo {author} {\bibfnamefont {S.}~\bibnamefont
  {Minwalla}}, \bibinfo {author} {\bibfnamefont {V.~E.}\ \bibnamefont
  {Hubeny}},\ and\ \bibinfo {author} {\bibfnamefont {M.}~\bibnamefont
  {Rangamani}},\ }\bibfield  {title} {\bibinfo {title} {Nonlinear fluid
  dynamics from gravity},\ }\href
  {https://doi.org/10.1088/1126-6708/2008/02/045} {\bibfield  {journal}
  {\bibinfo  {journal} {Journal of High Energy Physics}\ }\textbf {\bibinfo
  {volume} {2008}},\ \bibinfo {pages} {045} (\bibinfo {year}
  {2008}{\natexlab{a}})}\BibitemShut {NoStop}%
\bibitem [{\citenamefont {Bhattacharyya}\ \emph
  {et~al.}(2008{\natexlab{b}})\citenamefont {Bhattacharyya}, \citenamefont
  {Hubeny}, \citenamefont {Loganayagam}, \citenamefont {Mandal}, \citenamefont
  {Minwalla}, \citenamefont {Morita}, \citenamefont {Rangamani},\ and\
  \citenamefont {Reall}}]{ei14}%
  \BibitemOpen
  \bibfield  {author} {\bibinfo {author} {\bibfnamefont {S.}~\bibnamefont
  {Bhattacharyya}}, \bibinfo {author} {\bibfnamefont {V.~E.}\ \bibnamefont
  {Hubeny}}, \bibinfo {author} {\bibfnamefont {R.}~\bibnamefont {Loganayagam}},
  \bibinfo {author} {\bibfnamefont {G.}~\bibnamefont {Mandal}}, \bibinfo
  {author} {\bibfnamefont {S.}~\bibnamefont {Minwalla}}, \bibinfo {author}
  {\bibfnamefont {T.}~\bibnamefont {Morita}}, \bibinfo {author} {\bibfnamefont
  {M.}~\bibnamefont {Rangamani}},\ and\ \bibinfo {author} {\bibfnamefont
  {H.~S.}\ \bibnamefont {Reall}},\ }\bibfield  {title} {\bibinfo {title} {Local
  fluid dynamical entropy from gravity},\ }\href
  {https://doi.org/10.1088/1126-6708/2008/06/055} {\bibfield  {journal}
  {\bibinfo  {journal} {Journal of High Energy Physics}\ }\textbf {\bibinfo
  {volume} {2008}},\ \bibinfo {pages} {055} (\bibinfo {year}
  {2008}{\natexlab{b}})}\BibitemShut {NoStop}%
\bibitem [{\citenamefont {Waeber}\ and\ \citenamefont {Yaffe}()}]{hc12}%
  \BibitemOpen
  \bibfield  {author} {\bibinfo {author} {\bibfnamefont {S.}~\bibnamefont
  {Waeber}}\ and\ \bibinfo {author} {\bibfnamefont {L.~G.}\ \bibnamefont
  {Yaffe}},\ }\bibfield  {title} {\bibinfo {title} {Colliding localized, lumpy
  holographic shocks with a granular nuclear structure},\ }\href
  {https://doi.org/https://doi.org/10.1007/JHEP03(2023)208} {\bibfield
  {journal} {\bibinfo  {journal} {Journal of High Energy Physics}\ }\textbf
  {\bibinfo {volume} {2023}},\ \bibinfo {pages} {1}}\BibitemShut {NoStop}%
\bibitem [{\citenamefont {Folkestad}\ \emph {et~al.}()\citenamefont
  {Folkestad}, \citenamefont {Grozdanov}, \citenamefont {Rajagopal},\ and\
  \citenamefont {van~der Schee}}]{hc14}%
  \BibitemOpen
  \bibfield  {author} {\bibinfo {author} {\bibfnamefont {A.}~\bibnamefont
  {Folkestad}}, \bibinfo {author} {\bibfnamefont {S.}~\bibnamefont
  {Grozdanov}}, \bibinfo {author} {\bibfnamefont {K.}~\bibnamefont
  {Rajagopal}},\ and\ \bibinfo {author} {\bibfnamefont {W.}~\bibnamefont
  {van~der Schee}},\ }\bibfield  {title} {\bibinfo {title} {Coupling constant
  corrections in a holographic model of heavy ion collisions with nonzero
  baryon number density},\ }\href
  {https://doi.org/https://doi.org/10.1007/JHEP12(2019)093} {\bibfield
  {journal} {\bibinfo  {journal} {Journal of High Energy Physics}\ }\textbf
  {\bibinfo {volume} {2019}},\ \bibinfo {pages} {1}}\BibitemShut {NoStop}%
\bibitem [{\citenamefont {Penrose}(1965)}]{si12}%
  \BibitemOpen
  \bibfield  {author} {\bibinfo {author} {\bibfnamefont {R.}~\bibnamefont
  {Penrose}},\ }\bibfield  {title} {\bibinfo {title} {Gravitational collapse
  and space-time singularities},\ }\href
  {https://doi.org/10.1103/PhysRevLett.14.57} {\bibfield  {journal} {\bibinfo
  {journal} {Phys. Rev. Lett.}\ }\textbf {\bibinfo {volume} {14}},\ \bibinfo
  {pages} {57} (\bibinfo {year} {1965})}\BibitemShut {NoStop}%
\bibitem [{\citenamefont {Hawking}\ and\ \citenamefont {Penrose}(1970)}]{si14}%
  \BibitemOpen
  \bibfield  {author} {\bibinfo {author} {\bibfnamefont {S.~W.}\ \bibnamefont
  {Hawking}}\ and\ \bibinfo {author} {\bibfnamefont {R.}~\bibnamefont
  {Penrose}},\ }\bibfield  {title} {\bibinfo {title} {{The singularities of
  gravitational collapse and cosmology}},\ }\href
  {https://doi.org/10.1098/rspa.1970.0021} {\bibfield  {journal} {\bibinfo
  {journal} {Proc. Roy. Soc. Lond. A}\ }\textbf {\bibinfo {volume} {314}},\
  \bibinfo {pages} {529} (\bibinfo {year} {1970})}\BibitemShut {NoStop}%
\bibitem [{\citenamefont {Landau}\ and\ \citenamefont
  {Lifshitz}(2013)}]{landau}%
  \BibitemOpen
  \bibfield  {author} {\bibinfo {author} {\bibfnamefont {L.~D.}\ \bibnamefont
  {Landau}}\ and\ \bibinfo {author} {\bibfnamefont {E.~M.}\ \bibnamefont
  {Lifshitz}},\ }\href@noop {} {\emph {\bibinfo {title} {Fluid Mechanics:
  Course of Theoretical Physics, Volume 6}}},\ Vol.~\bibinfo {volume} {6}\
  (\bibinfo  {publisher} {Elsevier},\ \bibinfo {year} {2013})\BibitemShut
  {NoStop}%
\bibitem [{\citenamefont {Riemann}(1860)}]{riemann}%
  \BibitemOpen
  \bibfield  {author} {\bibinfo {author} {\bibfnamefont {B.}~\bibnamefont
  {Riemann}},\ }\bibfield  {title} {\bibinfo {title} {{U}eber die fortpflanzung
  ebener luftwellen von endlicher schwingungsweite},\ }\href
  {http://eudml.org/doc/135717} {\bibfield  {journal} {\bibinfo  {journal}
  {Abhandlungen der Königlichen Gesellschaft der Wissenschaften in
  Göttingen}\ }\textbf {\bibinfo {volume} {8}},\ \bibinfo {pages} {43}
  (\bibinfo {year} {1860})}\BibitemShut {NoStop}%
\bibitem [{\citenamefont {Bohm}(1952)}]{bohmqp}%
  \BibitemOpen
  \bibfield  {author} {\bibinfo {author} {\bibfnamefont {D.}~\bibnamefont
  {Bohm}},\ }\bibfield  {title} {\bibinfo {title} {A suggested interpretation
  of the quantum theory in terms of "hidden" variables. i},\ }\href
  {https://doi.org/10.1103/PhysRev.85.166} {\bibfield  {journal} {\bibinfo
  {journal} {Phys. Rev.}\ }\textbf {\bibinfo {volume} {85}},\ \bibinfo {pages}
  {166} (\bibinfo {year} {1952})}\BibitemShut {NoStop}%
\bibitem [{\citenamefont {Fischer}\ and\ \citenamefont
  {Datta}(2023)}]{fischer-cens}%
  \BibitemOpen
  \bibfield  {author} {\bibinfo {author} {\bibfnamefont {U.~R.}\ \bibnamefont
  {Fischer}}\ and\ \bibinfo {author} {\bibfnamefont {S.}~\bibnamefont
  {Datta}},\ }\bibfield  {title} {\bibinfo {title} {Dispersive censor of
  acoustic spacetimes with a shock-wave singularity},\ }\href
  {https://doi.org/10.1103/PhysRevD.107.084023} {\bibfield  {journal} {\bibinfo
   {journal} {Phys. Rev. D}\ }\textbf {\bibinfo {volume} {107}},\ \bibinfo
  {pages} {084023} (\bibinfo {year} {2023})}\BibitemShut {NoStop}%
\bibitem [{\citenamefont {Husain}\ \emph {et~al.}(2022)\citenamefont {Husain},
  \citenamefont {Kelly}, \citenamefont {Santacruz},\ and\ \citenamefont
  {Wilson-Ewing}}]{PhysRevLett.128.121301}%
  \BibitemOpen
  \bibfield  {author} {\bibinfo {author} {\bibfnamefont {V.}~\bibnamefont
  {Husain}}, \bibinfo {author} {\bibfnamefont {J.~G.}\ \bibnamefont {Kelly}},
  \bibinfo {author} {\bibfnamefont {R.}~\bibnamefont {Santacruz}},\ and\
  \bibinfo {author} {\bibfnamefont {E.}~\bibnamefont {Wilson-Ewing}},\
  }\bibfield  {title} {\bibinfo {title} {Quantum gravity of dust collapse:
  Shock waves from black holes},\ }\href
  {https://doi.org/10.1103/PhysRevLett.128.121301} {\bibfield  {journal}
  {\bibinfo  {journal} {Phys. Rev. Lett.}\ }\textbf {\bibinfo {volume} {128}},\
  \bibinfo {pages} {121301} (\bibinfo {year} {2022})}\BibitemShut {NoStop}%
\bibitem [{\citenamefont {Rousseaux}\ \emph {et~al.}(2016)\citenamefont
  {Rousseaux}, \citenamefont {Mougenot}, \citenamefont {Chatellier},
  \citenamefont {David},\ and\ \citenamefont {Calluaud}}]{ROUSSEAUX201631}%
  \BibitemOpen
  \bibfield  {author} {\bibinfo {author} {\bibfnamefont {G.}~\bibnamefont
  {Rousseaux}}, \bibinfo {author} {\bibfnamefont {J.-M.}\ \bibnamefont
  {Mougenot}}, \bibinfo {author} {\bibfnamefont {L.}~\bibnamefont
  {Chatellier}}, \bibinfo {author} {\bibfnamefont {L.}~\bibnamefont {David}},\
  and\ \bibinfo {author} {\bibfnamefont {D.}~\bibnamefont {Calluaud}},\
  }\bibfield  {title} {\bibinfo {title} {A novel method to generate tidal-like
  bores in the laboratory},\ }\href
  {https://doi.org/https://doi.org/10.1016/j.euromechflu.2015.08.004}
  {\bibfield  {journal} {\bibinfo  {journal} {European Journal of Mechanics -
  B/Fluids}\ }\textbf {\bibinfo {volume} {55}},\ \bibinfo {pages} {31}
  (\bibinfo {year} {2016})}\BibitemShut {NoStop}%
\bibitem [{\citenamefont {Dutton}\ \emph {et~al.}(2001)\citenamefont {Dutton},
  \citenamefont {Budde}, \citenamefont {Slowe},\ and\ \citenamefont
  {Hau}}]{dutton}%
  \BibitemOpen
  \bibfield  {author} {\bibinfo {author} {\bibfnamefont {Z.}~\bibnamefont
  {Dutton}}, \bibinfo {author} {\bibfnamefont {M.}~\bibnamefont {Budde}},
  \bibinfo {author} {\bibfnamefont {C.}~\bibnamefont {Slowe}},\ and\ \bibinfo
  {author} {\bibfnamefont {L.~V.}\ \bibnamefont {Hau}},\ }\bibfield  {title}
  {\bibinfo {title} {Observation of quantum shock waves created with ultra-
  compressed slow light pulses in a {B}ose-{E}instein condensate},\ }\href
  {https://doi.org/10.1126/science.1062527} {\bibfield  {journal} {\bibinfo
  {journal} {Science}\ }\textbf {\bibinfo {volume} {293}},\ \bibinfo {pages}
  {663} (\bibinfo {year} {2001})}\BibitemShut {NoStop}%
\bibitem [{\citenamefont {Wan}\ \emph {et~al.}(2006)\citenamefont {Wan},
  \citenamefont {Jia},\ and\ \citenamefont {Fleischer}}]{wan}%
  \BibitemOpen
  \bibfield  {author} {\bibinfo {author} {\bibfnamefont {W.}~\bibnamefont
  {Wan}}, \bibinfo {author} {\bibfnamefont {S.}~\bibnamefont {Jia}},\ and\
  \bibinfo {author} {\bibfnamefont {J.~W.}\ \bibnamefont {Fleischer}},\
  }\bibfield  {title} {\bibinfo {title} {Dispersive superfluid-like shock waves
  in nonlinear optics},\ }\href {https://doi.org/10.1038/nphys486} {\bibfield
  {journal} {\bibinfo  {journal} {Nature Physics}\ }\textbf {\bibinfo {volume}
  {3}},\ \bibinfo {pages} {46} (\bibinfo {year} {2006})}\BibitemShut {NoStop}%
\bibitem [{\citenamefont {Xu}\ \emph {et~al.}(2015)\citenamefont {Xu},
  \citenamefont {Vocke}, \citenamefont {Faccio}, \citenamefont {Garnier},
  \citenamefont {Trillo},\ and\ \citenamefont {Picozzi}}]{xu}%
  \BibitemOpen
  \bibfield  {author} {\bibinfo {author} {\bibfnamefont {G.}~\bibnamefont
  {Xu}}, \bibinfo {author} {\bibfnamefont {D.}~\bibnamefont {Vocke}}, \bibinfo
  {author} {\bibfnamefont {D.}~\bibnamefont {Faccio}}, \bibinfo {author}
  {\bibfnamefont {J.}~\bibnamefont {Garnier}}, \bibinfo {author} {\bibfnamefont
  {S.}~\bibnamefont {Trillo}},\ and\ \bibinfo {author} {\bibfnamefont
  {A.}~\bibnamefont {Picozzi}},\ }\bibfield  {title} {\bibinfo {title} {From
  coherent shocklets to giant collective incoherent shock waves in nonlocal
  turbulent flows.},\ }\href {https://doi.org/10.1038/ncomms9131} {\bibfield
  {journal} {\bibinfo  {journal} {Nature communications}\ }\textbf {\bibinfo
  {volume} {6}},\ \bibinfo {pages} {8131} (\bibinfo {year} {2015})}\BibitemShut
  {NoStop}%
\bibitem [{\citenamefont {Bienaim\'e}\ \emph {et~al.}(2021)\citenamefont
  {Bienaim\'e}, \citenamefont {Isoard}, \citenamefont {Fontaine}, \citenamefont
  {Bramati}, \citenamefont {Kamchatnov}, \citenamefont {Glorieux},\ and\
  \citenamefont {Pavloff}}]{bienaime}%
  \BibitemOpen
  \bibfield  {author} {\bibinfo {author} {\bibfnamefont {T.}~\bibnamefont
  {Bienaim\'e}}, \bibinfo {author} {\bibfnamefont {M.}~\bibnamefont {Isoard}},
  \bibinfo {author} {\bibfnamefont {Q.}~\bibnamefont {Fontaine}}, \bibinfo
  {author} {\bibfnamefont {A.}~\bibnamefont {Bramati}}, \bibinfo {author}
  {\bibfnamefont {A.~M.}\ \bibnamefont {Kamchatnov}}, \bibinfo {author}
  {\bibfnamefont {Q.}~\bibnamefont {Glorieux}},\ and\ \bibinfo {author}
  {\bibfnamefont {N.}~\bibnamefont {Pavloff}},\ }\bibfield  {title} {\bibinfo
  {title} {Quantitative analysis of shock wave dynamics in a fluid of light},\
  }\href {https://doi.org/10.1103/PhysRevLett.126.183901} {\bibfield  {journal}
  {\bibinfo  {journal} {Phys. Rev. Lett.}\ }\textbf {\bibinfo {volume} {126}},\
  \bibinfo {pages} {183901} (\bibinfo {year} {2021})}\BibitemShut {NoStop}%
\bibitem [{\citenamefont {Bendahmane}\ \emph {et~al.}(2022)\citenamefont
  {Bendahmane}, \citenamefont {Xu}, \citenamefont {Kudlinski}, \citenamefont
  {Mussot},\ and\ \citenamefont {Trillo}}]{Bendahmane}%
  \BibitemOpen
  \bibfield  {author} {\bibinfo {author} {\bibfnamefont {A.}~\bibnamefont
  {Bendahmane}}, \bibinfo {author} {\bibfnamefont {G.}~\bibnamefont {Xu}},
  \bibinfo {author} {\bibfnamefont {A.}~\bibnamefont {Kudlinski}}, \bibinfo
  {author} {\bibfnamefont {A.}~\bibnamefont {Mussot}},\ and\ \bibinfo {author}
  {\bibfnamefont {S.}~\bibnamefont {Trillo}},\ }\bibfield  {title} {\bibinfo
  {title} {From coherent shocklets to giant collective incoherent shock waves
  in nonlocal turbulent flows.},\ }\href
  {https://doi.org/10.1038/s41467-022-30734-5} {\bibfield  {journal} {\bibinfo
  {journal} {Nature communications}\ }\textbf {\bibinfo {volume} {13}},\
  \bibinfo {pages} {3137} (\bibinfo {year} {2022})}\BibitemShut {NoStop}%
\end{thebibliography}%
\end{document}